%% file: lawther_mrk590_xray_paper.tex
\documentclass[fleqn,usenatbib,useAMS]{mnras}
\usepackage{deluxetable}
\usepackage{epic,eepic}
\usepackage{graphicx}
\usepackage[usenames,dvipsnames]{color}
\usepackage{wasysym}

\usepackage{amsmath, amssymb}

\usepackage[english]{babel}

\newcommand{\Rom}[1]
    {\MakeUppercase{\romannumeral #1}}

\title[Analysis of X-ray Flares in Mrk 590]{Flares in the Changing Look AGN Mrk 590. \Rom{2}: Deep X-ray observations reveal a Comptonizing inner accretion flow}

\author[Lawther, D., Vestergaard, M., Raimundo, S., Fan, X., Koay, J.Y.]{Lawther, D.,$^1$ Vestergaard, M.,$^{1,2}$ Raimundo, S.,$^{2,3}$, Fan, X.,$^{1}$, Koay, J.Y.$^{4,5}$ \\
    $^1$ Steward Observatory, University of Arizona, 933 N. Cherry Avenue, 85721 Tucson, AZ, USA\\
	$^2$ DARK, Niels Bohr Institute, University of Copenhagen, Denmark.\\
    $^3$ University of Southampton, University Road,  Southampton, United Kingdom.\\
    $^4$ Niigata University, 8050 Ikarashi-nino-cho, Nishi-ku, Niigata 950-2181, Japan \\
	$^5$ Institute of Astronomy and Astrophysics, Academia Sinica,  Roosevelt Rd,  Taipei 10617,  Taiwan,  R.O.C.}

\begin{document}
	
\maketitle

\begin{abstract}
Mrk 590 is a Changing Look AGN currently in an unusual repeat X-ray and UV flaring state. Here, we report on deep X-ray observations with \emph{XMM-Newton}, \emph{NuSTAR}, and \emph{NICER}, obtained at a range of X-ray flux levels. We detect a prominent soft excess below 2 keV; its flux is tightly correlated with that of both the X-ray and UV continuum, and it persists at the lowest flux levels captured. Our Bayesian model comparison strongly favors inverse Comptonization as the origin of this soft excess, instead of blurred reflection. We find only weak reflection features, with $R\apprle0.4$ assuming Compton-thick reflection. Most of this reprocessing occurs at least $\sim$800 gravitational radii (roughly three light-days) from the continuum source. Relativistically broadened emission is weak or absent, suggesting the lack of a standard `thin disk' at small radii. We confirm that the predicted broad-band emission due to Comptonization is roughly consistent with the observed UV--optical photometry. This implies an optically thick, warm ($kT_e\sim0.3$ keV) scattering region that extends to at least $\sim10^3$ gravitational radii, reprocessing any UV thermal emission. The lack of a standard `thin disk' may also explain the puzzling $\sim3$-day X-ray to UV delay previously measured for Mrk 590. Overall, we find that the X-ray spectral changes in Mrk 590 are minimal, despite substantial luminosity changes. Other well-studied changing look AGN display more dramatic spectral evolution, \emph{e.g.}, disappearing continuum or soft excess. This suggests that a diversity of physical mechanisms in the inner accretion flow may produce a UV--optical changing-look event.
\end{abstract}

% Considering the previously detected  $\sim3$-day X-ray to UV delay, this may indicate that the outer disk beyond the Comptonization region, or the inner broad-line region, produces the lion's share of the observed UV--optical flux.


\input{sections/sec_introduction}

\input{sections/sec_data}

\input{sections/sec_xray_analysis}

\input{sections/sec_discussion}

\input{sections/conclusion}

\bibliographystyle{mnras}
\bibliography{bibliography}

%\newpage
 
\onecolumn

\input{tables/xray_fit_tables}

%\input{tables/model_comparison_table}

\twocolumn
\appendix

\input{appendixA/appendixA}

\clearpage
\input{appendixB/appendixB}

\clearpage
\input{appendixC/appendixC}

\clearpage
\input{appendixD/appendixD}

\end{document}

%% file: sections/sec_introduction.tex
\section{Introduction}\label{sec:introduction}

\defcitealias{Lawther2023}{Paper 1}

Active Galactic Nuclei (AGN) emit copiously over the entire electromagnetic spectrum \citep[\emph{e.g.},][]{Sanders1989, Elvis1994, Richards2006}. They are powered by the release of gravitational energy as gas from their host galaxy forms an accretion flow and feeds the central supermassive black hole \citep[\emph{e.g.},][]{Shields1978, Malkan1982, Malkan1983, StorchiBergman2019}. The standard theory of accretion onto black holes is presented in the seminal works of \citet{Shakura1973} and \citet{Novikov1973}, hereafter referred to as `thin-disk models'. The observed UV--optical AGN continuum \citep[`big blue bump', e.g.,][]{Shields1978, Malkan1982, Sanders1989, Siemiginowska1995, Grupe1998, Scott2014} is often attributed to emission from a thermal disk. However, the applicability of thin-disk models to AGN is debated, as they fail to predict the observed variability timescales and spectral turnovers \citep{Antonucci2015, Lawrence2018}.

Typically, AGN display stochastic UV--optical flux variations of order $\sim10\%$ on rest-frame timescales of days to months \citep[\emph{e.g.},][]{VandenBerk2004, Kelly2009, Rumbaugh2018, Chanchaiworawit2024}. This level of variability may be due to thermal instabilities in the disk \citep[\emph{e.g.},][]{Kelly2009}, plus a high-frequency contribution from reprocessing of intrinsic X-ray variability \citep[the `lamp-post model', \emph{e.g.}][]{Cackett2007, Edelson2019, Guo2022}. In recent years a class of Changing Look AGN (CLAGN) has been identified, displaying more extreme variability behavior. In CLAGN, the UV--optical continuum and broad emission lines (dis)appear on timescales of months to years \citep[\emph{e.g.},][]{Penston1984, Shappee2014, Denney2014, LaMassa2015, Runnoe2016, MacLeod2016,  LaMassa2017, Rumbaugh2018, MacLeod2019}. In broad terms, this behavior may be due to variable \emph{obscuration} of the central engine, or to strong variability of the ionizing continuum. \citet{Ricci2023} classify these types as \emph{changing-obscuration} and \emph{changing-state} AGN, respectively. Analytical thin-disk models predict significant continuum variability on long timescales, $\sim10^3$--$10^5$ years \citep[\emph{e.g.},][]{LaMassa2015, MacLeod2016, Noda2018, Lawrence2018}, inconsistent with the changing-state sources. Some numerical simulations of accretion disks may capture additional variability mechanisms; \citet{Jiang2018} and \citet{Jiang2020} find strong variability on timescales of a few years, consistent with observations. Alternatively, the accretion state of the disk may transition out of the `thin-disk' mode and into a radiatively inefficient, advective state \citep[\emph{e.g.},][]{Narayan1994, Yuan2014}, when the Eddington luminosity ratio drops below $\sim1\%$ \citep[\emph{e.g.},][]{Noda2018, Veronese2024}. However, the mechanism driving such a state transition (and thus, the predicted timescale) is largely unknown. Clearly, `changing-state' CLAGN are important case studies to advance our understanding of AGN accretion.

Many CLAGN display extreme X-ray variability \citep[\emph{e.g.,}][]{Denney2014, Noda2018, Ricci2020, Kollatschny2023, Lawther2023, Saha2023}, presumably linked to processes very near the central black hole and inner accretion flow. In this work, to harness the potential of X-ray observations to probe the inner accretion flow, we study a series of deep X-ray observations of the CLAGN Mrk 590 ($z=0.026385$) obtained since 2002. A \emph{bona fide} AGN in the 1980s-1990s, this target `turned off' around year 2010 \citep{Denney2014} and regained its broad emission lines during 2017 \citep{Raimundo2019}. Since then, it displays X-ray and UV--optical flare-ups once or twice a year, during which the X-ray flux increases by a factor $\sim5$--10 for a few weeks. \citet{Lawther2023} (hereafter, \citetalias{Lawther2023}) find that the X-rays lead the UV variability by $\sim3$ days during these flares. This is much longer than the predicted X-ray to UV delay for a `thin disk' with a central X-ray source. In this work, we further investigate the X-ray emission that drives this flaring activity. Our primary goals are as follows: \emph{1)} to document variability in the X-ray emission features, and determine their relationship to continuum variability; \emph{2)} to find the most likely physical model of the X-ray emission in Mrk 590, and infer the accretion flow properties based thereon. To provide context for this study, we now briefly outline the typical X-ray emission features in AGN.

\emph{Primary continuum:} AGN display a power law-like X-ray continuum extending from $\sim0.2$ keV or below, with a high-energy cutoff around $200$ keV \citep{Zdziarski1995, Tortosa2018}. The predicted emission from a `thin disk' does not extend to such high energies. Instead, the X-ray continuum is thought to be due to inverse Compton scattering of UV seed photons in a hot ($kT\sim100$ keV), optically thin plasma \citep{Sunyaev1980, Haardt1993, Zdziarski1994, Zdziarski1995, Zdziarski1996}. The required seed photons may originate in a `thin disk' or other thermal source in the accretion flow. The size of this region - the so-called hot corona - is constrained to a few gravitational radii, due to its rapid coherent variability \citep[\emph{e.g.},][]{Guilbert1983, Barr1986, Fabian2015}. The geometry of the hot corona is largely unknown \citep[\emph{e.g.},][]{Tortosa2018}, although X-ray polarimetry results favor deviations from spherical symmetry \citep{Gianolli2023, Ingram2023}. 

\emph{Soft excess:} An excess of soft X-rays below rest-frame 2 keV, relative to the primary continuum, is detected in around $80$\% of low-obscuration AGN, and may be intrinsically near-ubiquitous  \citep{Bianchi2009}. The soft excess is not consistent with thermal `thin disk' emission \citep[\emph{e.g.},][]{Laor1997, Leighly1999, Done2012}. Instead, two production mechanisms are typically considered: \emph{1)} the inverse Compton scattering of a thermal distribution of UV seed photons in a warm ($kT\sim0.2$ keV), optically thick medium \citep[\emph{e.g.},][]{Czerny1987, Magdziarz1998, Petrucci2004, Done2012, Middei2020}. We refer to this type of soft excess as `warm-Comptonized emission' throughout. \citet{Petrucci2018} present one possible physical model, where the warm Comptonizing region is a moderately optically thick atmosphere `sandwiching' a cool disk; the atmosphere is heated by direct dissipation of accretion energy, \emph{e.g.}, via the magnetic dynamo mechanism \citep{Begelman2015}. \emph{2)} Alternatively, relativistically blurred  atomic-line reprocessing (`reflection') can produce a featureless soft excess feature; this mechanism requires a dense accretion flow extending to small radii, such that the individual lines are rotationally blurred out \citep[\emph{e.g.},][]{Ross2005, Crummy2006, Walton2013, Dauser2014}. Thus, in either case, study of the soft excess can probe the nature of the inner accretion flow.

\emph{Reflection features:} Most AGN display some evidence of iron K-shell emission \citep{Nandra2007, Miller2007}. These fluorescent lines are due to X-ray illumination of atomic gas. If a `thin disk' extends to the innermost stable orbit, a relativistically broadened component is expected \citep[\emph{e.g.},][]{Fabian1989, MeyerHofmeister2011}. Broad Fe K is challenging to confirm observationally, but is detected in some local AGN \citep[\emph{e.g.},][]{Tanaka1995, Turner2002, Mason2003, Porquet2003, Nandra2007, Bhayani2011}. X-ray illumination also produces soft X-ray line emission, along with a broad Compton hump peaking at $\sim$25 keV \citep[\emph{e.g.},][]{Reynolds1999, Fabian2009}. Collectively, these are the most prominent features of the so-called reflection spectrum. The full reflection spectrum can be simulated self-consistently, for a given geometry, by several different reflection models \citep[\emph{e.g.},][]{Garcia2014, Balokovic2018, Dauser2022}.

\paragraph*{Outline of this work:} In \S \ref{sec:observations} we describe the \emph{XMM-Newton}, \emph{NuSTAR}, and \emph{NICER} observations used in this work, along with our basic data processing procedures. There, we also present \emph{Swift} X-ray lightcurves to document new flaring activity since June 2021. In \S \ref{sec:analysis} we describe our analysis of the X-ray data, including a variability study (\S \ref{sec:analysis_bbody}), and physical modeling in low- and high-flux states (\S \ref{sec:analysis_joint}). We discuss our results in \S \ref{sec:discussion}, and conclude in \S \ref{sec:conclusion}.

%% file: sections/sec_data.tex
\section{Observations and Data Processing}\label{sec:observations}

\subsection{\emph{Swift}}

\begin{figure*}
    \centering
    \includegraphics[scale=0.495]{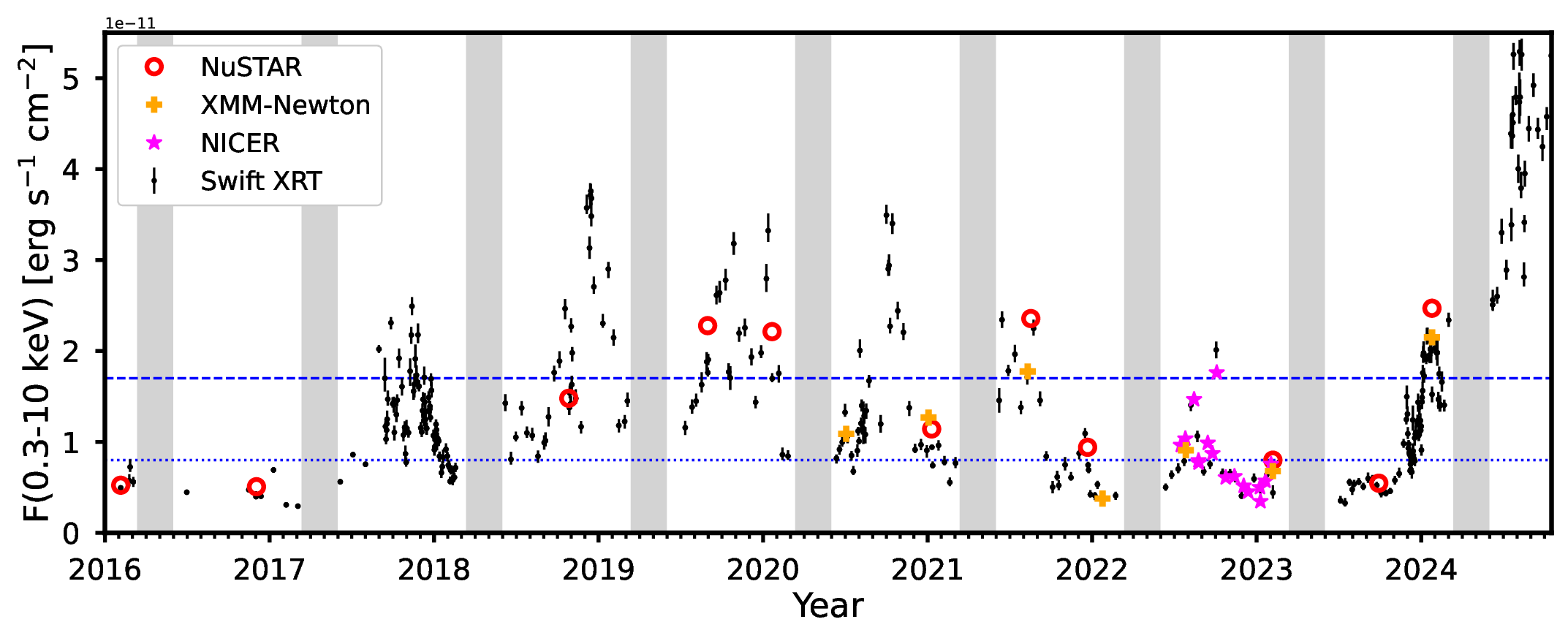}
    \vskip -5.5pt

    \caption{\emph{Swift} XRT 0.3--10 keV lightcurve (black circles with error bars) for Mrk 590 since June 2015. Orange crosses indicate the 0.3--10 keV flux measured with \emph{XMM-Newton}; red squares represent extrapolations of the \emph{NuSTAR} observations (3-79 keV) to the 0.3--10 keV regime. Magenta stars indicate \emph{NICER} observations with usable data (\S \ref{sec:observations}). All fluxes are extracted using a power-law model with Galactic absorption. The blue dotted and dashed horizontal lines indicate the maximum flux level for inclusion in the `low state' joint data set, and the minimum for inclusion in the `high state' (\S \ref{sec:analysis_joint}), respectively. Mrk 590 is behind the Sun during early March -- mid-June (gray shaded regions). Data obtained since June 2021 are not included in \citetalias{Lawther2023}. The prolonged, powerful high-flux state during Summer 2024 will be addressed in future work (\emph{Lawther et al., in prep.}).}\label{fig:xrt_lightcurve}
\end{figure*}

The Neil Gehrels \emph{Swift} Observatory \citep{Burrows2005} has been monitoring Mrk 590 since 2013 \citepalias{Lawther2023}. In this work we include new \emph{Swift} XRT observations obtained between June 2021 and September 2024 (Swift Cycle 18, Program 1821134, PI: Lawther; VLBA/Swift Joint Proposal, Program VLBA/22A-217, PI: Vestergaard; Swift Cycle 19, Program 1922187, PI: Lawther; VLBA/Swift Joint Proposal, Program VLBA/24A-374, PI: Vestergaard). These new data are processed and analyzed following the same methods as \citetalias{Lawther2023}; we use version 6.32.1 of the \textsc{HEASoft} pipeline. We present the \emph{Swift} XRT lightcurve since June 2015, including the new data collected since 2021, in Figure \ref{fig:xrt_lightcurve}. Based on the most recent data (Programs 1922187 and VLBA/24A-374), we report that Mrk 590 entered a new, prolonged bright state in Summer 2024, which we will address in future work (\emph{Lawther et al., in prep.}). Based on the \emph{Swift} monitoring, we obtained several observations with \emph{XMM-Newton} (\S \ref{sec:data_xmm}), \emph{NuSTAR} (\S \ref{sec:data_nustar}) and \emph{NICER} (\S \ref{sec:data_nicer}), aiming to capture the X-ray spectrum at a range of flux levels. We indicate the observing dates and flux levels of these observations in Figure \ref{fig:xrt_lightcurve}.

\subsection{\emph{XMM-Newton}}\label{sec:data_xmm}

\paragraph*{Observations:} Prior to the discovery of its changing-look behavior, Mrk 590 was observed by the \emph{XMM-Newton} space telescope \citep{Turner2001,Struder2001} in 2002 (PI: Mason) and 2004 (PI: Santos-Lleo). Thereafter, it was observed during XMM-Newton Programs 86547 (PI: Miniutti) and 87084 (PI: Vestergaard), and as part of joint observations during NuSTAR Program 8233 (PI: Lawther). All observations were performed using the thin optical blocking filter. Figure \ref{fig:xrt_lightcurve} illustrates the timing of the observations since 2020, relative to the \emph{Swift} XRT lightcurve. The individual observation IDs and exposure times are listed in Table \ref{tab:observationlog}. 

\paragraph*{Data processing:} \emph{XMM-Newton} MOS and \emph{pn} data are provided by the Science Archive\footnote{URL: \url{http://nxsa.esac.esa.int/nxsa-web/}}. We reprocess all XMM-Newton data locally using the Science Analysis Software, version 20, by running the \emph{`xmmextractor'} task with its standard parameter settings. This reprocessing ensures that appropriate response matrix files for the source detector location are generated. The observation starting 11th August 2021 was split over two orbits for scheduling reasons; for each instrument (MOS1, MOS2, \emph{pn}) we stack the resulting two spectra using the task \emph{`epicspeccombine'}, and treat them as a single observation. Due to the faintness of the source during our observations, we do not find the high-resolution RGS spectra scientifically useful. We do not need the Optical Monitor data for the analyses described here.

\subsection{\emph{NuSTAR} }\label{sec:data_nustar}

\paragraph*{Observations:}

The \emph{NuSTAR} observatory \citep{Harrison2013} consists of two identical Wolter X-ray telescopes. The two focal-plane detectors, FPMA and FPMB, are sensitive to energies 3-79 keV. \emph{NuSTAR} observed Mrk 590 during February 2016 as part of the \emph{NuSTAR} Extragalactic Surveys \citep{Harrison2016}. Thereafter, Mrk 590 was observed via a DDT request during 2016, and as part of the following programs: joint \emph{Swift/NuSTAR} Program 1417159 (PI: Vestergaard), \emph{NuSTAR} Cycle 5 Program 5252 (PI: Vestergaard),  \emph{NuSTAR} Cycle 6 Program 6238 (PI: Vestergaard), \emph{NuSTAR} Cycle 7 Program 7610 (PI: Koss) and \emph{NuSTAR} Cycle 8 Program 8233 (PI: Lawther). The individual observation IDs and exposure times are listed in Table \ref{tab:observationlog}. Figure \ref{fig:xrt_lightcurve} indicates the timing of these \emph{NuSTAR} observations relative to the X-ray variability recorded by \emph{Swift}.

\paragraph*{Data processing:}

We process the \emph{NuSTAR} observations using the standard pipeline processing (\textsc{HEASoft} v. 6.32.1, \textsc{NuSTARDAS} v. 1.9.7). We extract the source and background spectra for each \emph{NuSTAR} detector (FPMA and FPMB) using the \emph{'nuproducts'} task, which also generates appropriate Auxiliary Response Files for the observations. We use a circular source extraction region of radius 60'' centered on the source PSF centroid. For the background extraction, we use a circular region of radius 67'' offset and non-overlapping with the source region, but positioned on the same detector quadrant. We ensure that neither the source nor background extraction regions include the additional soft X-ray source detected in XRT imaging \citepalias{Lawther2023}. We use Algorithm 2 of the \emph{'nucalcsaa'} task, with an 'optimized' South Atlantic Anomaly (SAA) cut and including the parameter \emph{'tentacle=yes'}, to filter out the anomalous countrates before and after the SAA passage. This preserves $>95$\% of the on-source exposure time.

\subsection{NICER}\label{sec:data_nicer}

\begin{figure*}
    \centering
    \includegraphics[trim={30 0 0 0}, scale=0.445]{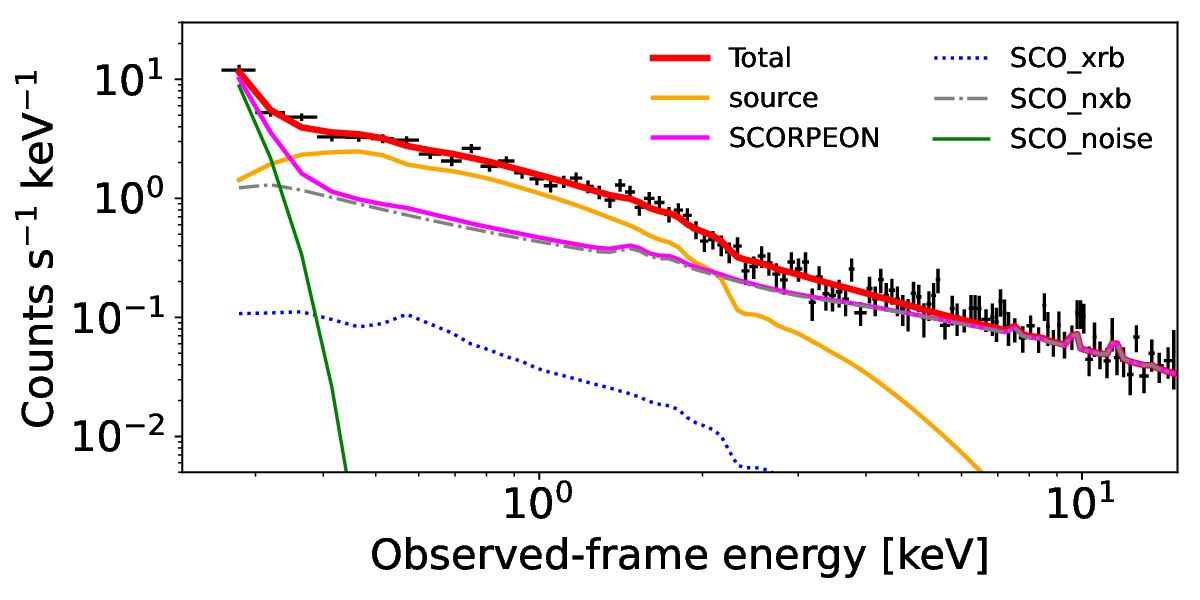}
    \includegraphics[trim={10 0 40 0}, scale=0.445]{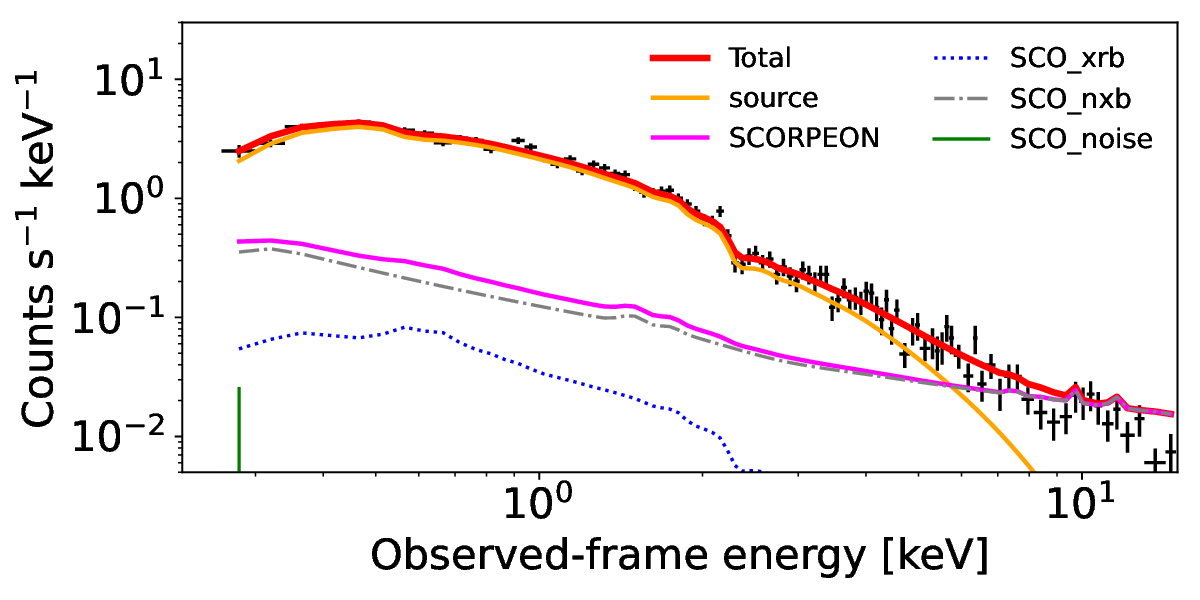}
    \caption{Examples of NICER spectra with unacceptable (left panel; observation ID 5667010101) and acceptable (right panel; observation ID 5667010401) background levels. To assess the background contribution in each NICER observation, we simultaneously fit a power-law source model and a SCORPEON background model; the latter consists of three components that are varied separately during the model fit. In the left panel, the source model (yellow curve) is fainter than the SCORPEON detector noise peak component (green curve) in the soft X-rays below $\sim$0.35 keV. Above $\sim2.5$ keV, the source is fainter than the non-X-ray background component (black dot-dash curve). We exclude observation ID 5667010101, and ten other observations with similarly high background levels, from our analyses.}\label{fig:nicer_background}
\end{figure*}

\paragraph*{Observations:} The Neutron star Interior Composition Explorer \citep[NICER,][]{Gendreau2016} observed Mrk 590 for 27 visits between June 2022 -- Feb 2023, for typical exposure times of 1.5--2.5 ks per visit (GO Program 5667, PI: Lawther).  We retrieve all observations with durations $>1$ ks from the HEAsarc NICER catalog\footnote{NICER Master Catalog: \url{https://heasarc.gsfc.nasa.gov/W3Browse/nicer/nicermastr.html} .}, resulting in 23 datasets. NICER is a non-imaging telescope, with a variable instrumental background which depends strongly on spacecraft orbital characteristics. The background features of most concern for our study include \emph{1)} a low-energy noise peak which can extend beyond $\sim300$ eV for observations with high `optical loading' (\emph{i.e.}, UV-optical stray light causing spurious X-ray events; Figure \ref{fig:nicer_background}), and \emph{2)} electron precipitation events that can exceed the source signal at $\apprge2$ keV. We use the SCORPEON background model\footnote{SCORPEON model documentation: \url{https://heasarc.gsfc.nasa.gov/docs/nicer/analysis_threads/scorpeon-overview/}} to evaluate the strength of these background features. Based thereon, we exclude 11 observations from further analysis. The 12 remaining observations are listed in Table \ref{tab:observationlog}. We present examples of acceptable and excluded \emph{NICER} observations in Figure \ref{fig:nicer_background}.

\paragraph*{Data processing:} We process the \emph{NICER} spectra using \textsc{HEASoft} v.6.32, which includes updated per-detector filtering for noisy time intervals. We use the \emph{`nicerl2'} tool to extract event files with appropriate good-time intervals. We apply the standard processing settings apart from excluding detectors FPM 14 and 34, which are known to respond strongly to optical loading \citep[e.g.,][]{Remillard2022}; after excluding these detectors, we indeed found fewer observations to be affected by the `noise peak' background feature. We extract the spectra, and the corresponding response matrices and Auxiliary Response Files required to analyze each observation, using \emph{`nicerl3-spect'}. 

%% file: sections/sec_xray_analysis.tex
\section{X-ray analysis}\label{sec:analysis}

Here, we present an analysis of the \emph{NuSTAR}, \emph{XMM-Newton} and \emph{NICER} data for Mrk 590. We first describe our \textsc{Xspec} modeling setup, and present the observed spectra (\S \ref{sec:method_xspec}). Next, we study the strength and variability of individual emission components, based on a phenomenological model (\ref{sec:analysis_bbody}). Finally, we perform in-depth modeling of combined data sets in low- and high-flux states, to probe the emission physics and reflection geometry (\S \ref{sec:analysis_joint}).

\subsection{Modeling methodology and data presentation}\label{sec:method_xspec}

We model the X-ray data using the \emph{pyxspec} (v2.1.1) implementation of the \textsc{XSPEC} (v12.13.0) X-ray analysis package \citep{Arnaud1996}. We optimize the Cash statistic with Poissonian background \citep[W-stat, \emph{e.g.},][]{Humphrey2009}, as implemented in \textsc{XSPEC}. We include energies between 0.25--12 keV for \emph{XMM-Newton} data, 3--79 keV for \emph{NuSTAR}, and 0.22--15 keV for \emph{NICER}. For \emph{XMM-Newton} and \emph{NuSTAR}, we bin each individual spectrum to have at least five \emph{background} counts per energy bin, using the \textsc{HEAsoft} task `\emph{ftgrouppha}'. This binning minimizes a bias when minimizing W-stat at faint background levels, as demonstrated computationally by Giacomo Vianello,\footnote{URL: \url{https://giacomov.github.io/Bias-in-profile-poisson-likelihood/}} and discussed by \citet{Buchner2023}. We include multiplicative flux scaling terms $C_{\mathrm{inst}}$ for \emph{NuSTAR} FPMB relative to FPMA, and for the \emph{XMM-Newton} MOS1 and MOS2 detectors relative to \emph{pn}. For data obtained simultaneously with different detectors on the same telescope, these inter-calibration offsets are in all cases $<$5\%. Throughout this work, quoted model parameter uncertainties represent 90\% confidence intervals as obtained using the XSPEC \emph{`error'} task.

\paragraph*{Background modeling for NICER:} As a non-imaging instrument, \emph{NICER} data do not allow for extraction of separate background spectra. To address this, we include the SCORPEON background model \citep{Markwardt2024} in each of our \emph{NICER} model fits. SCORPEON consists of six different physical components, governed by 26 parameters in total. The \emph{nicerl3-spect} tool suggests constraints for these parameters, based on geomagnetic and orbital conditions, which we adopt. We fit between five and six SCORPEON parameters simultaneously with the source model; the remainder are held constant at default values. As recommended\footnote{SCORPEON modeling best practices: \url{https://heasarc.gsfc.nasa.gov/docs/nicer/analysis_threads/scorpeon-xspec/} .}, we model the entire \emph{NICER} spectrum between 0.22 -- 15 keV, even though the detector is only sensitive to astronomical flux between $\sim0.3$ -- 10 keV.

\paragraph*{Presentation of individual observations:} To illustrate their salient features and overall data quality, we display each individual \emph{XMM-Newton} and \emph{NuSTAR} observation in Appendix \ref{AppendixA}, as a data/model ratio against a power-law modified by Galactic absorption (defined in \textsc{XSPEC} as \textsc{tbabs$\times$powerlaw}). Here, we exclude the spectral regions below 2 keV, 5.5--7 keV, and 12.5--30 keV, to avoid fitting to the expected soft excess and reflection features. We use the appropriate Galactic absorption column density for Mrk 590, $n_{\mathrm{H}}=2.77\times10^{20}$ cm$^{-2}$ \citep{Bekhti2016}. We note the robust detection of soft X-ray excess below $\sim2$ keV in \emph{all} our \emph{XMM-Newton} observations. Most of our spectra also show excess emission near 6.4 keV. This feature appears rather weak in the August 2021 \emph{XMM-Newton} data, and is not immediately apparent in all \emph{NuSTAR} spectra, probably due to inadequate statistics in individual visits (\emph{e.g.}, February 2016; 20 ks observed).

For presentation purposes, throughout this work we only display data from a single detector per telescope, choosing \emph{XMM-Newton pn} and \emph{NuSTAR} FPMA. Furthermore, we bin the displayed spectra such that each energy bin is detected at at least the $3\sigma$ level above the background. We emphasize that these choices only affect the figures, and not the underlying spectral modeling, which is performed using minimally binned data and including all X-ray detectors.

\subsection{Modeling of individual observations}\label{sec:analysis_bbody}

\begin{figure}
    \centering
    \includegraphics[scale=0.56, trim={16 0 0 0}]{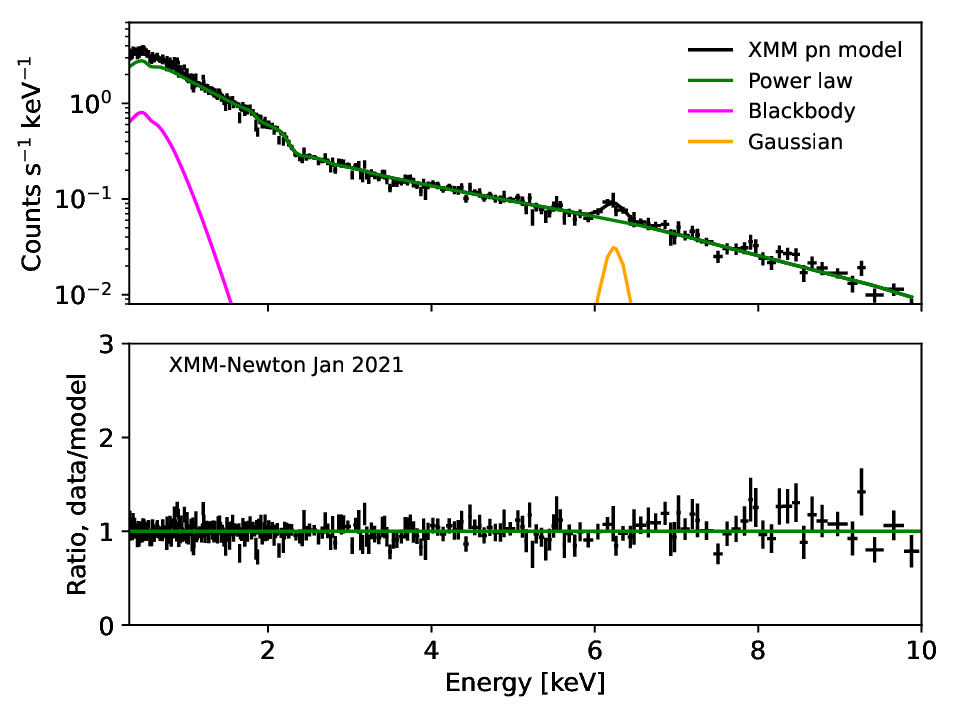}
    \caption{Top panel: Count-rate spectra and best-fit folded model components for our phenomenological model (\S \ref{sec:analysis_bbody}), consisting of a power law continuum (green curve), blackbody component (magenta curve) and Gaussian emission line (orange curve), for the January 2021 \emph{XMM-Newton} data. Bottom panel: the data/model counts ratio indicates that this model captures, to first order, the excess emission below 2 keV and near 6.4 keV. \label{fig:bbody}}
\end{figure}

\begin{figure}
    \centering
    \includegraphics[trim={16 0 0 0}, scale=0.55]{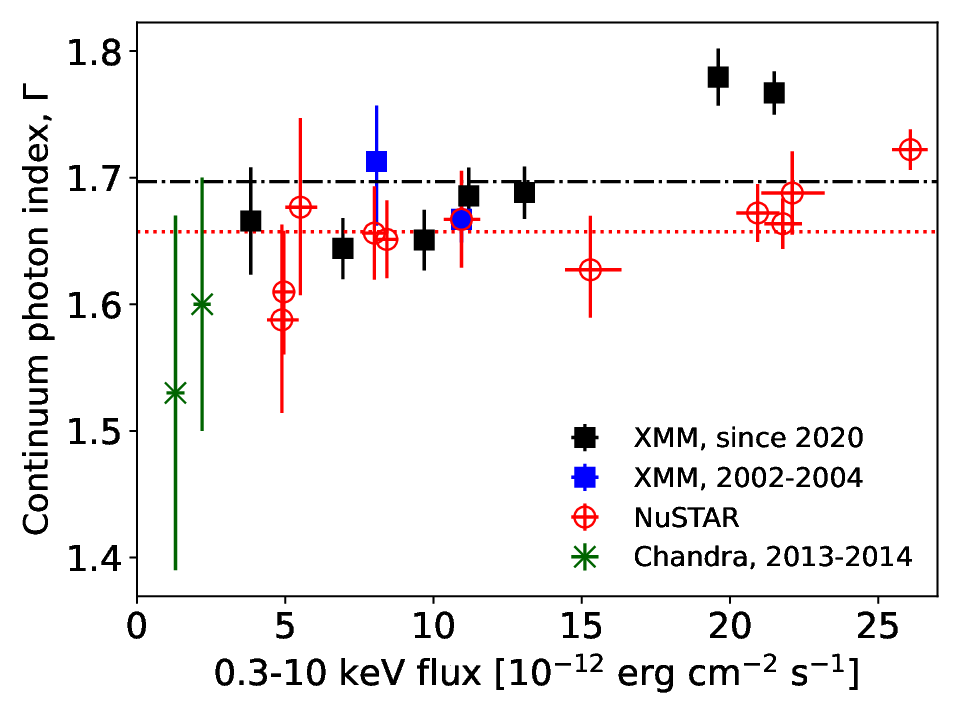}
    \caption{The 0.3-10 keV continuum photon index, measured using our phenomenological models (\S \ref{sec:analysis_bbody}), as a function of 0.3--10 keV flux, for each individual observation. The red dotted and black dash-dot lines illustrate the uncertainty-weighted mean photon indices for \emph{NuSTAR}, $\overline{\Gamma}=1.66$, and for \emph{XMM-Newton}, $\overline{\Gamma}=1.70$, respectively. For comparison, we include published results for two \emph{Chandra} observations obtained shortly after the `turn-off' event \citep{Denney2014,Mathur2018}.}
    \label{fig:flux_gamma}
\end{figure}

To study the variability of the continuum, soft excess and Fe K line emission features in our individual observations, we now define a simple phenomenological model. We represent the soft excess with a single-temperature blackbody component, and include a single Gaussian emission line with its centroid fixed to rest-frame 6.4 keV. The resulting model is \textsc{tbabs$\times$(powerlaw+zbbody+zgauss)}. Its free parameters are: the blackbody temperature $kT_{\mathrm{BB}}$, the Gaussian width $\sigma_{\mathrm{line}}$, the continuum photon index $\Gamma$, and the separate normalizations of the three additive components. For the \emph{NuSTAR} observations, which do not sample the soft X-rays, we simplify the model to \textsc{tbabs$\times$(powerlaw+zgauss)}. We present an example of our phenomenological model in Figure \ref{fig:bbody}, and list the best-fit parameters for each observation in Table \ref{tab:bbody_gauss}. Below, we explore the derived properties of the continuum, soft excess, and 6.4 keV line.

\begin{figure*}
    \centering
    \includegraphics[trim={15 0 0 0}, scale=0.56]{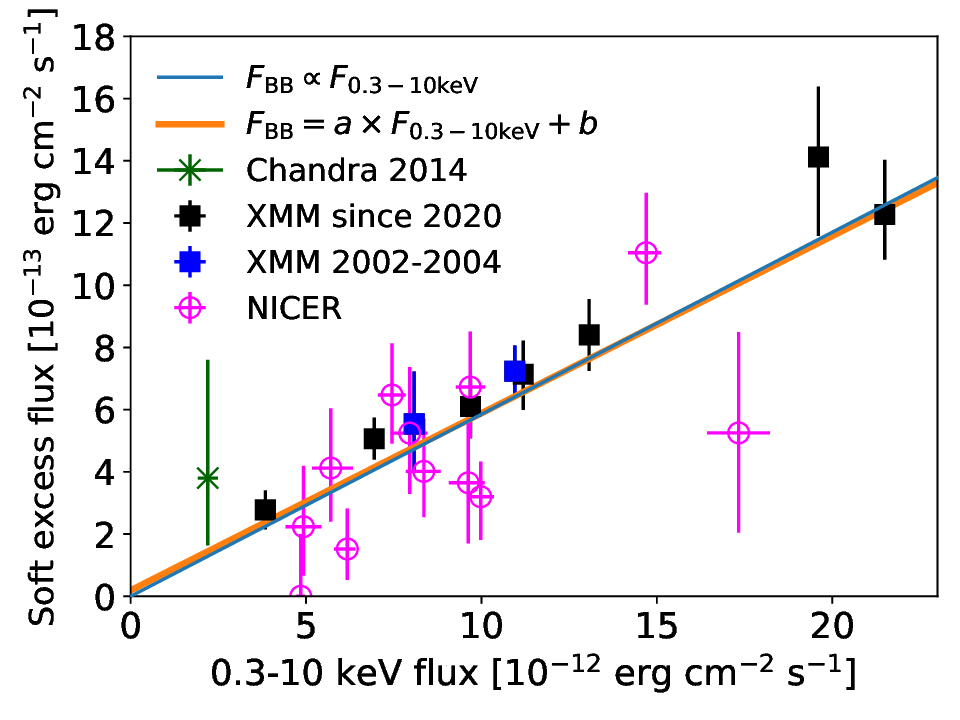}
    \includegraphics[trim={10 0 15 0}, scale=0.56]{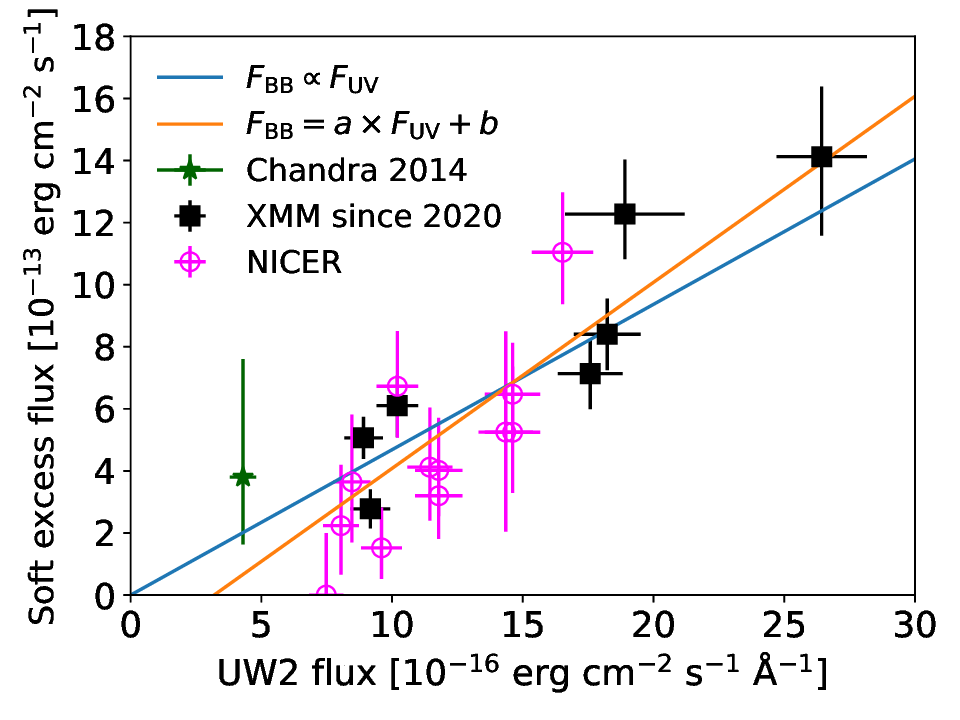}
    \caption{The flux $F_{\mathrm{BB}}$ of the blackbody component, integrated over 0.3--10 keV, versus that of the power law continuum component $F_{\mathrm{0.3-10}}$ \emph{(left panel)}, and versus the host galaxy-subtracted \emph{Swift} UVOT \emph{UW2} flux density \emph{(right panel)}. Both $F_{\mathrm{BB}}$ and  $F_{\mathrm{0.3-10}}$ are derived from our phenomenological model (\S \ref{sec:analysis_bbody}). The blue lines depict the best-fit proportionality, including \emph{NICER} (magenta circles), \emph{XMM-Newton} (black squares for recent data; blue squares for spectra taken in 2002--2004), and \emph{Chandra} \citep{Mathur2018} measurements (green asterisk). The best-fit linear relationships (orange lines) are broadly consistent with a proportionality in both cases. The early \emph{XMM-Newton} measurements are not included for the UV flux--$F_{\mathrm{BB}}$ relationship, as no UV photometry is available.}
    \label{fig:flux_bbody}
\end{figure*}

\paragraph*{Primary continuum:} For comparison purposes, we report the continuum flux between 0.3--10 keV, $F_{\mathrm{0.3-10}}$, throughout this work; we extrapolate the best-fit power law component below 3 keV for \emph{NuSTAR}. $F_{\mathrm{0.3-10}}$ varies by a factor $\sim5.5$ for the observations presented in this work, spanning a similar dynamic range for \emph{XMM-Newton} and \emph{NuSTAR}. For the continuum slope, we find an uncertainty-weighted average $\overline{\Gamma}=1.70\pm0.01$, with a standard deviation of 0.04, for the nine individual \emph{XMM-Newton} spectra (Figure \ref{fig:flux_gamma}). For \emph{NuSTAR} (11 spectra) we measure $\overline{\Gamma}=1.66\pm0.01$, with a standard deviation of 0.05. \emph{Chandra} observed Mrk 590 twice shortly after the `turn-off' event \citep{Denney2014,Mathur2018}; their reported $\Gamma$ values are rather uncertain, and fully consistent with our findings at $F_{\mathrm{0.3-10}}\apprle1.5\times10^{-11}$ erg cm$^{-2}$ s$^{-1}$. We see a possible trend of increasing $\Gamma$ (\emph{i.e.}, spectral softening) with increasing flux. However, this is primarily driven by the two highest-flux \emph{XMM-Newton} observations. A Kendall rank correlation test yields a $p$-value of 0.03 for the null hypothesis of no correlation between continuum flux and $\Gamma$. Given the rather small sample size, and that $\Gamma$ appears less variable for the \emph{NuSTAR} spectra, we do not regard this as conclusive evidence that the underlying continuum slope changes. 

For the \emph{NICER} observations, as the detector sensitivity reduces steeply above $\sim$3 keV, we found that $\Gamma$ is rather poorly constrained, and typically significantly softer than that measured for near-contemporaneous  \emph{XMM-Newton} or \emph{NuSTAR} visits. For this reason, we impose $\Gamma=1.68$ (the average value measured by \emph{XMM-Newton}) when modeling the \emph{NICER} spectra. This allows us to more efficiently use these data to constrain the soft excess, as described below.

\paragraph*{Soft excess:} Our models consistently converge on a non-zero blackbody flux $F_{\mathrm{BB}}$ for the \emph{XMM-Newton} data. We find that $F_{\mathrm{BB}}$ displays a tight proportionality to $F_{\mathrm{0.3-10}}$ for the seven \emph{XMM-Newton} observations (Figure \ref{fig:flux_bbody}, left panel). The \emph{NICER} data appear to follow the same trend, albeit with more scatter. For the \emph{NICER} observation dated 2023-01-08, at a rather faint $F_{\mathrm{0.3-10}}$ level, we find $F_{\mathrm{BB}}$ consistent with zero; all other \emph{NICER} spectra require a non-zero blackbody flux. To extend this investigation to lower $F_{\mathrm{0.3-10}}$, we include the results presented by \citet{Mathur2018} for their 2014 \emph{Chandra} observation. These authors applied a similar phenomenological model to that used here\footnote{Their approach differed only in the inclusion of an additional, ionized Fe K line at 6.7 keV. While a narrow feature at 6.7 keV is visible in some of our \emph{XMM-Newton} spectra, we do not include it in our phenomenological model, due to the varying signal to noise ratios for our individual observations. Its omission is unlikely to affect the measured flux of the continuum or soft excess components.}. Including the \emph{XMM-Newton}, \emph{NICER} and \emph{Chandra} measurements, a Kendall rank correlation test yields $p=3\times10^{-5}$ for 22 data points, indicating a statistically significant dependence. A linear least-squares fit  yields: 
\begin{equation*}
F_{\mathrm{BB}}=0.07(\pm0.01)F_{\mathrm{0.3-10}}+0.02(\pm0.09) \mathrm{\,erg\,s}^{-1}\mathrm{cm}^{-2}.
\end{equation*}

\noindent The intercept term is consistent with zero, indicating a proportionality. We note that \citet{Denney2014} report no evidence of deviations from a pure power law model, for their 2013 \emph{Chandra} observation taken at $F_{\mathrm{0.3-10}}=1.1\times10^{-12}$ erg cm$^{-2}$ s$^{-1}$. We re-analyzed those data and are unable to obtain useful constraints on a blackbody soft excess component, given the rather short (30 ks) exposure time and faint source flux. We therefore exclude the 2013 data from this analysis. We conclude that $F_{\mathrm{BB}}$ is proportional to $F_{\mathrm{0.3-10}}$ down to at least $F_{\mathrm{0.3-10}}\sim2\times10^{-12}$ erg cm$^{-2}$ s$^{-1}$.

We see a similar trend of increasing $F_{\mathrm{BB}}$ with UV flux, albeit with a larger scatter (Figure \ref{fig:flux_bbody}, right panel). Here, we compare with the nearest in time \emph{Swift} \emph{UW2} observation. We find $p=0.02$ for the null hypothesis, now considering only 18 data points as we lack UV photometry in 2002--2004. To quantify the $F_{\mathrm{UV}}$--$F_{\mathrm{BB}}$ relationship, we first subtract a constant host galaxy contribution of $2.1\times10^{-16}$ erg cm$^{-2}$ s$^{-1}$ \AA$^{-1}$ from each UVOT measurement. We base this estimate on a flux variability gradient analysis (\emph{Lawther et al., in prep.}), following the method of \citet{Gianniotis2022}. A linear fit to the host-subtracted data yields:

\begin{equation*}
    10^{-2}F_{\mathrm{BB}} = 0.60(\pm0.1)F_{\mathrm{UV}}\times1\text{\AA}-1.9(\pm1.4) \mathrm{\,erg\,}\mathrm{s}^{-1}\mathrm{cm}^{-2}.
\end{equation*}

\noindent This relationship displays a non-zero intersect, but only at the $\sim1.3\sigma$ level. As the exact value of this intersect is sensitive to the assumed host galaxy contribution, we find it premature to claim that the soft excess component `turns off' at the faintest UV flux levels, based on available data. 

\begin{figure}
    \centering
    \includegraphics[trim={15 0 0 0}, scale=0.55]{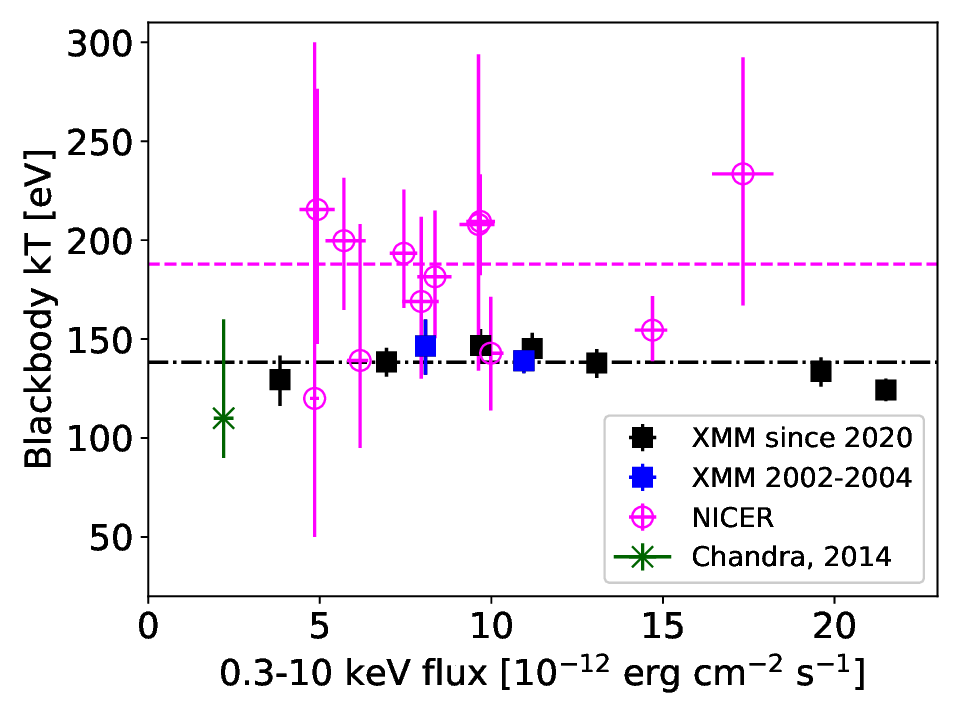}
    \caption{The temperature $kT$ of the blackbody component, for the phenomenological model described in \S \ref{sec:analysis_bbody}. The \emph{XMM-Newton} data points (squares) show negligible scatter around the average value $\overline{kT}=138\pm1$ eV (black dash-dot line). The \emph{NICER} $kT$ values (magenta circles) are poorly constrained, at a higher average temperature $\overline{kT}=187\pm3$ eV (magenta dashed line). For the \emph{Chandra} observation we adopt the blackbody temperature presented by \citet{Mathur2018}.}
    \label{fig:kT}
\end{figure}

For our \emph{XMM-Newton} spectra, the blackbody temperature $kT_{\mathrm{BB}}\sim138$ eV is roughly constant, with no dependence on $F_{\mathrm{0.3-10}}$ (Figure \ref{fig:kT}). We note that the \emph{NICER} spectra yield a systematically higher blackbody temperature, $kT_{\mathrm{BB}}\sim187$ eV, with much larger uncertainties. This is to some extent due to the lower signal to noise ratio in the short \emph{NICER} observations. We also suspect that the `noise ringer' feature in the \emph{NICER} background (Figure \ref{fig:nicer_background}, green curve), modeled with SCORPEON, may be somewhat degenerate with the soft excess component, potentially biasing the $kT_{\mathrm{BB}}$ measurements towards higher energies.

\begin{figure*}
    \centering
    \includegraphics[trim={17 0 0 0}, scale=0.56]{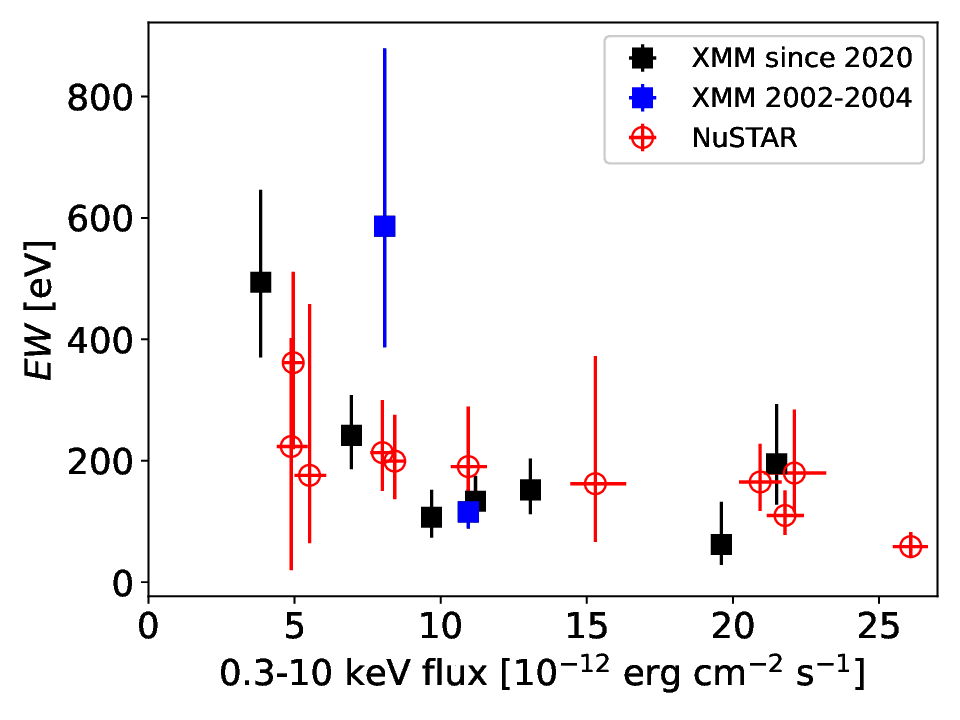}
    \includegraphics[trim={15 0 17 0}, scale=0.56]{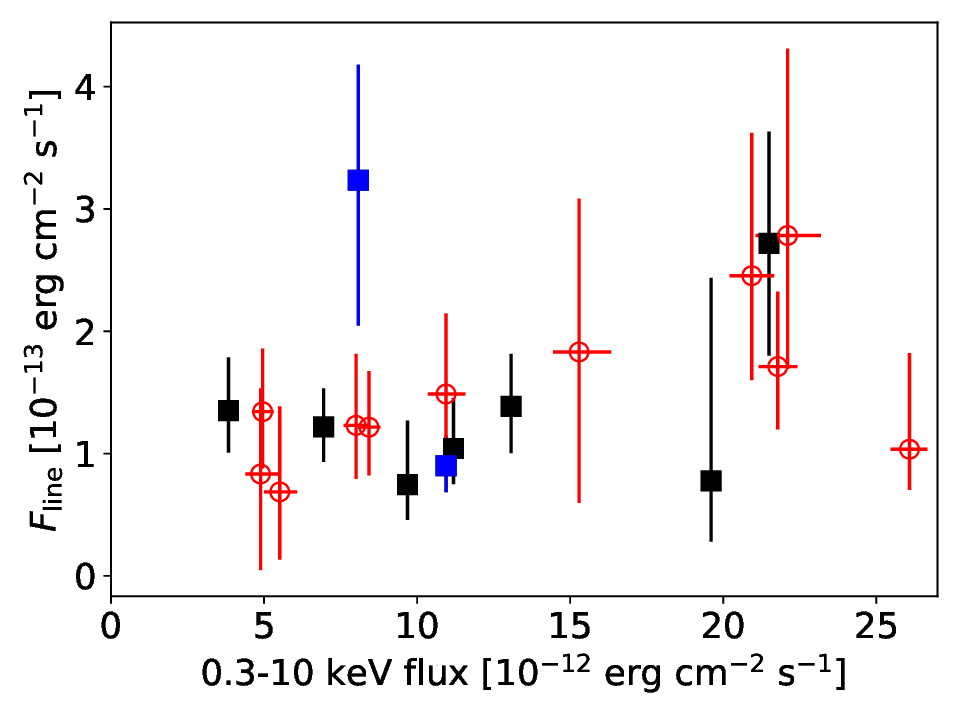}
    \caption{The equivalent width (left panel) and integrated line flux (right panel) of a single Gaussian emission line with centroid fixed to 6.4 keV, versus the X-ray continuum flux. The linewidth is allowed to vary freely during the model fit.}
    \label{fig:line_equivalent_widths}
\end{figure*}

\paragraph*{6.4 keV emission line:} For the 6.4 keV Gaussian component we find a non-zero equivalent width ($EW$) in all \emph{XMM-Newton} observations, and all but one \emph{NuSTAR} observation. The February 2016 \emph{NuSTAR} observation displays $EW\sim0$, likely due to a low signal to noise ratio. The individual \emph{NuSTAR} spectra typically provide poor constraints on $EW$, except at the highest flux levels. The line fluxes are roughly constant as a function of continuum flux, leading to a modest increase in line equivalent width at the lowest continuum flux levels observed (Figure \ref{fig:line_equivalent_widths}). The latter trend, sometimes referred to as an `X-ray Baldwin effect', is well-known from statistical samples of AGN \citep{Iwasawa1993,Nandra1997,Ricci2014}. The line width of the Gaussian component is at best marginally resolved. In most cases we obtain an upper limit, of order $\sigma_{\mathrm{line}}<160$ eV for \emph{XMM-Newton} spectra (Table \ref{tab:bbody_gauss}). This is consistent with Keplerian rotation around the central black hole at a distance of order $800$ gravitational radii, which corresponds to roughly three light-days. Thus, we cannot exclude that the main X-ray reprocessing is co-spatial with the primary UV reprocessing region, also located around three light-days from the X-ray source \citepalias{Lawther2023}. However, we caution that the distance inferred via the Fe K linewidth is only a lower limit.

Several of our \emph{XMM-Newton} observations (\emph{e.g.,} January 2022, Figure \ref{fig:xmm_windowfits}) show additional narrow excess features at energies above 6.4 keV, perhaps due to Fe K$\beta$ (7.06 keV) and/or ionized He-like K$\alpha$ (6.7 keV). We do not attempt to model these for individual observations, as they are in many cases only marginally detected. 

\subsection{Modeling of combined 0.3--79 keV data sets}\label{sec:analysis_joint}

\begin{figure*}
    \includegraphics[scale=0.565,trim={17 0 0 160}, clip]{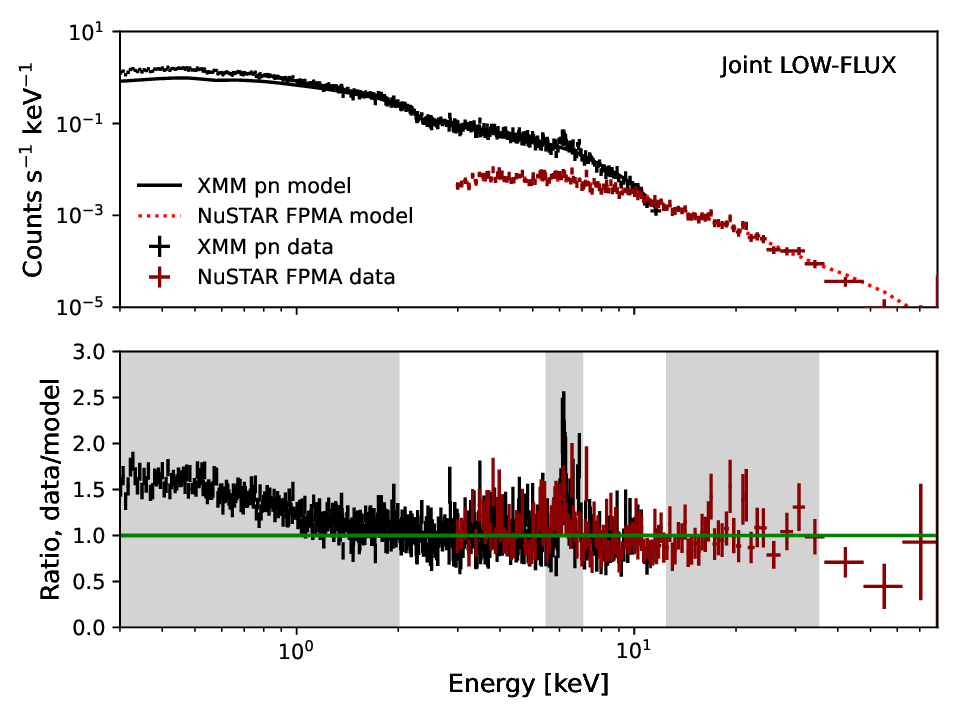}
    \includegraphics[scale=0.565,trim={15 0 10 160}, clip]{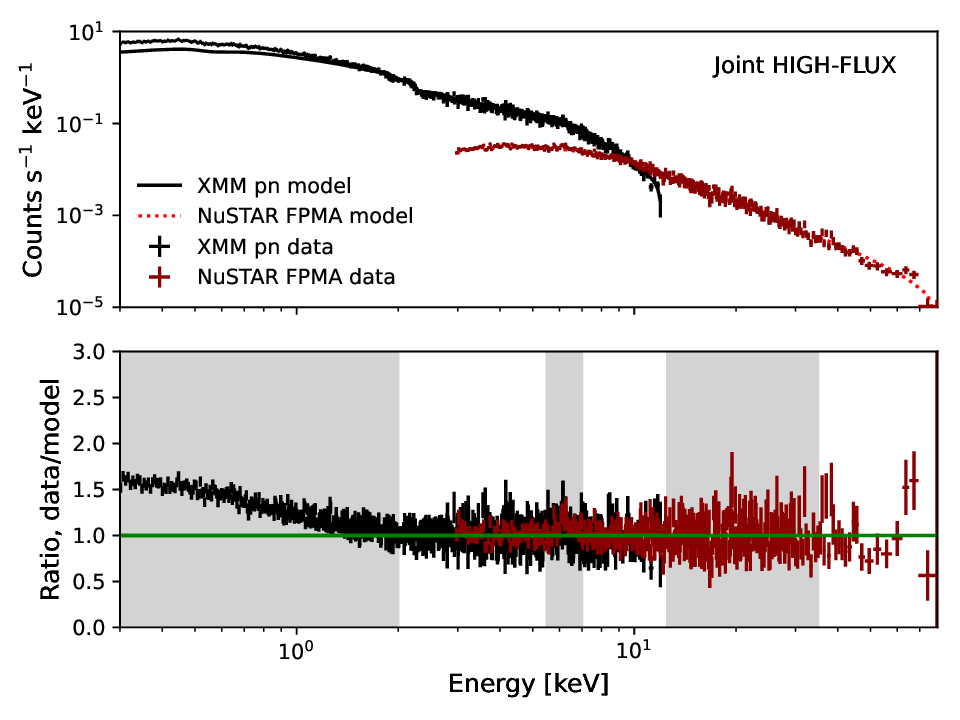}
    \includegraphics[scale=0.565,trim={9 0 10 0}]{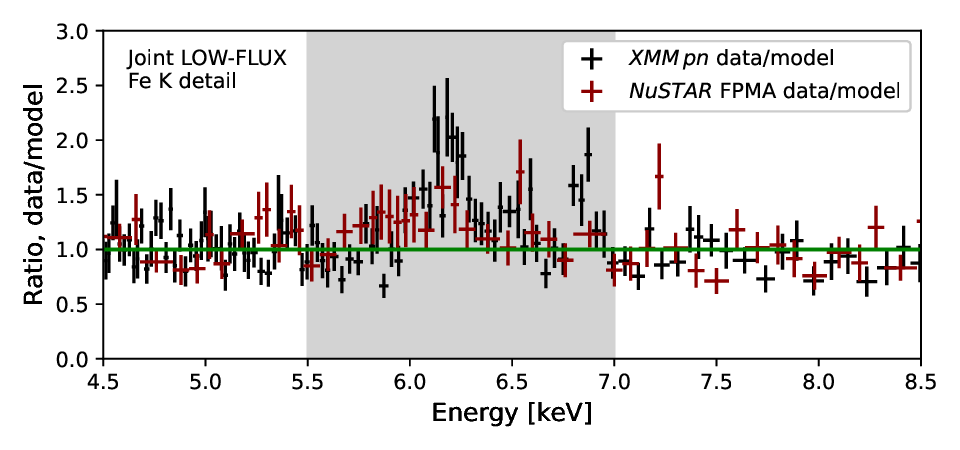}
    \includegraphics[scale=0.565,trim={9 0 20 0}]{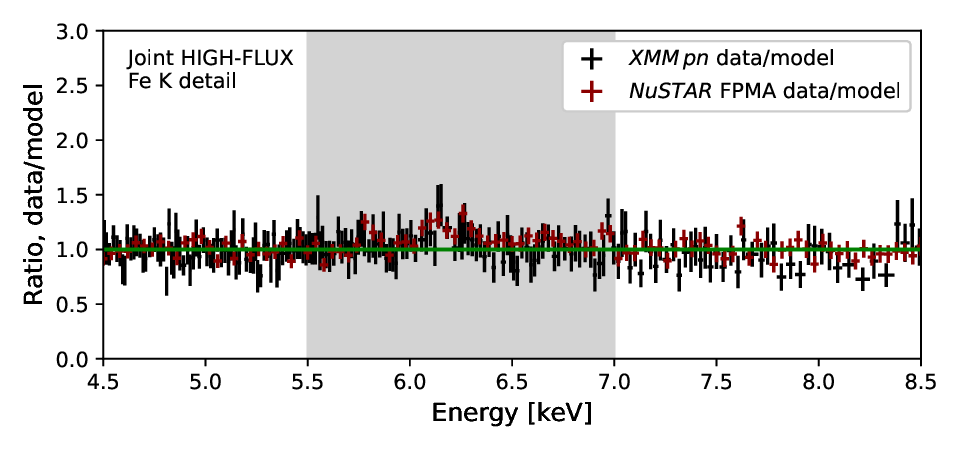}
    \caption{Joint data sets modeled against a power law with Galactic absorption, excluding the gray shaded regions from the model fit (following \S \ref{sec:method_xspec}). The low-flux (LF) data is shown in the left panels, while the right panels show the HF data set. The top panels show the data/model ratios for the entire 0.3--79 keV range. A detailed view of the Fe K emission complex is shown in the bottom panels.  \label{fig:jointdata}}
\end{figure*}

Here,  we construct combined data sets using both \emph{XMM-Newton} and \emph{NuSTAR}, to exploit the full 0.3--79 keV energy range and improve the photon counting statistics (\S \ref{sec:analysis_joint_datasets}). This facilitates an in-depth model comparison (\S \ref{sec:modelcomparison}) to determine the most likely emission mechanisms (\S \ref{sec:bestmodels_def}).
\subsubsection{Construction of combined data sets}\label{sec:analysis_joint_datasets}

According to our analysis of the individual spectra (\S \ref{sec:analysis_bbody}), the overall continuum flux level appears to be the main driver of X-ray spectral shape changes. For this reason, we elect to combine data taken at similar flux levels. We select all observations with $F_{\mathrm{0.3-10}}<7\times10^{-12}$ erg cm$^{-2}$ s$^{-1}$ to construct a `low -flux' joint data set (hereafter, LF). Similarly, we include all observations with $F_{\mathrm{0.3-10}}>18\times10^{-12}$ erg cm$^{-2}$ s$^{-1}$ in our `high-flux' data set (HF). These flux cutoffs are fairly arbitrary: our aim is to separate the brightest and faintest states captured in our observations, while achieving sufficiently long exposure times to distinguish between competing emission models. We list the observations included in each data set in \mbox{Table \ref{tab:observationlog}.}

We stack all spectra from a given instrument, in each data set, using the \textsc{HEASoft} task `\emph{addspec}' for \emph{NuSTAR} instruments, and the \textsc{SAS} task `\emph{epicspeccombine}' for \emph{XMM-Newton}. We include cross-calibration constants $C_{\mathrm{inst}}$ in our modeling, for each instrument, relative to EPIC \emph{pn} (so, $C_{\mathrm{PN}}\equiv1)$. The flux offsets reach $\pm18$\% between \emph{XMM-Newton} and \emph{NuSTAR} instruments. This is partially due to our use of non-contemporaneous \emph{XMM-Newton} and \emph{NuSTAR} observations; however, flux offsets between \emph{XMM-Newton} and \emph{NuSTAR} reach $\sim10$\% even for contemporaneous data \citep{Tsujimoto2011}. To test whether the stacking biases our results, we define two additional data sets using near-contemporaneous \emph{XMM-Newton} and \emph{NuSTAR} observations. The first of these occurred in January 2021 (hereafter, J21), at an intermediate flux state, $F_{\mathrm{0.3-10}}\sim1.2\times10^{-11}$ erg s$^{-1}$ cm$^{-2}$. The second occurred in February 2023 (hereafter, F23), at $F_{\mathrm{0.3-10}}\sim6\times10^{-12}$ erg s$^{-1}$ cm$^{-2}$.

To demonstrate the overall data quality and the prominent features in these data sets, we present power law model fits to the LF and HF spectra in Figure \ref{fig:jointdata}. Both data sets display an obvious soft excess below $\sim2$ keV, and some Fe K emission; the latter appears stronger (relative to the continuum) for the LF data. We also note hints of spectral curvature at $\sim$30--60 keV, although the statistics are rather poor at those energies, even for the combined \emph{NuSTAR} data.

\subsubsection{Bayesian model comparison}\label{sec:modelcomparison}

To determine which emission models can describe - and are warranted by - our joint data sets, we apply a Bayesian model comparison approach \citep[\emph{e.g.,}][]{Kass1995}. This approach compares the evidences $Z$ for a series of models, integrated over their respective parameter spaces. Advantages of the Bayes factor approach for model selection in astronomy are discussed by, \emph{e.g.}, \citet{Mukherjee2006, Trotta2007}. Notably, it is valid for both non-nested and nested model comparison, and, unlike null hypothesis testing, can indicate a preference for either model. It penalizes models that are \emph{non-predictive} (\emph{i.e.}, flexible) over the parameter space defined by the prior probabilities.

We describe this model selection procedure in detail, and define each model tested, in Appendix \ref{AppendixB}. Briefly, for each data set (LF, HF, J21 and F23), we  use the Bayesian X-ray Analysis package \citep{Buchner2014, Buchner2016} to calculate the evidences $Z$ for a series of models. This results in a unique `best' model with evidence $Z_\mathrm{best}$ for each data set, and a range of `acceptable' models satisfying a limiting Bayes factor, $\log(Z_\mathrm{best})-\log(Z_\mathrm{model})<3$, relative to the best model. We justify our choice of limiting Bayes factor in Appendix \ref{AppendixB}, and tabulate the Bayes factors \mbox{$\log(Z_\mathrm{best})-\log(Z_\mathrm{model})$} for each model in Table \ref{tab:bayesfactors}.

The key results of our model comparison are: \emph{1)}  \emph{inverse Compton scattering in a warm medium is formally required} to produce the observed soft excess, for all four data sets. Models not including warm-Comptonized emission are disfavored by very large Bayes factors. We note that we tested against the state-of-the-art relativistic reflection model \textsc{relxillLpCp} \citep{Dauser2022}, which allows for higher density disks, as favored for black hole masses of $\sim10^7$ M$_{\astrosun}$ and below \citep{Jiang2018, Mallick2022} and is required to produce very strong soft excess via reflection \citep{Madathil2024}. Nevertheless, models including warm-Comptonized emission are better able to reproduce the observed spectra. \emph{2)} Distant reflection is required, due to the dominant narrow Fe K line (\emph{e.g.}, Figure \ref{fig:jointdata}); models including \emph{only} disk reflection features are disfavored by very large Bayes factors, as they produce insufficient narrow emission at 6.4 keV. \emph{3)} A contribution from disk reflection is formally acceptable, but not required. Models with both distant and relativistic reflection produce a marginally higher Bayesian evidence for both LF and HF, while the simpler distant reflection-only models are marginally preferred for the J21 and F23 data sets, perhaps due to poorer statistics.

For the primary X-ray continuum, we find that an exponential high-energy cutoff is preferred (relative to a power law with no cutoff) for all data sets. Surprisingly, replacing the cutoff power law component with a more realistic hot-Comptonized emission model (\textsc{nthcomp}) is disfavored, even though this mechanism is typically assumed for the coronal emission (\S \ref{sec:introduction}). We return to this puzzle in \S \ref{sec:discussion_turnover}. Finally, we find that both neutral and ionized intrinsic absorption is in all cases strongly disfavored; the X-ray emission is thus largely unabsorbed at any observed flux level.

\subsubsection{Definitions of our preferred models}\label{sec:bestmodels_def}

\begin{figure*}
    \centering
    \includegraphics[scale=0.48,trim={15 0 0 0}]{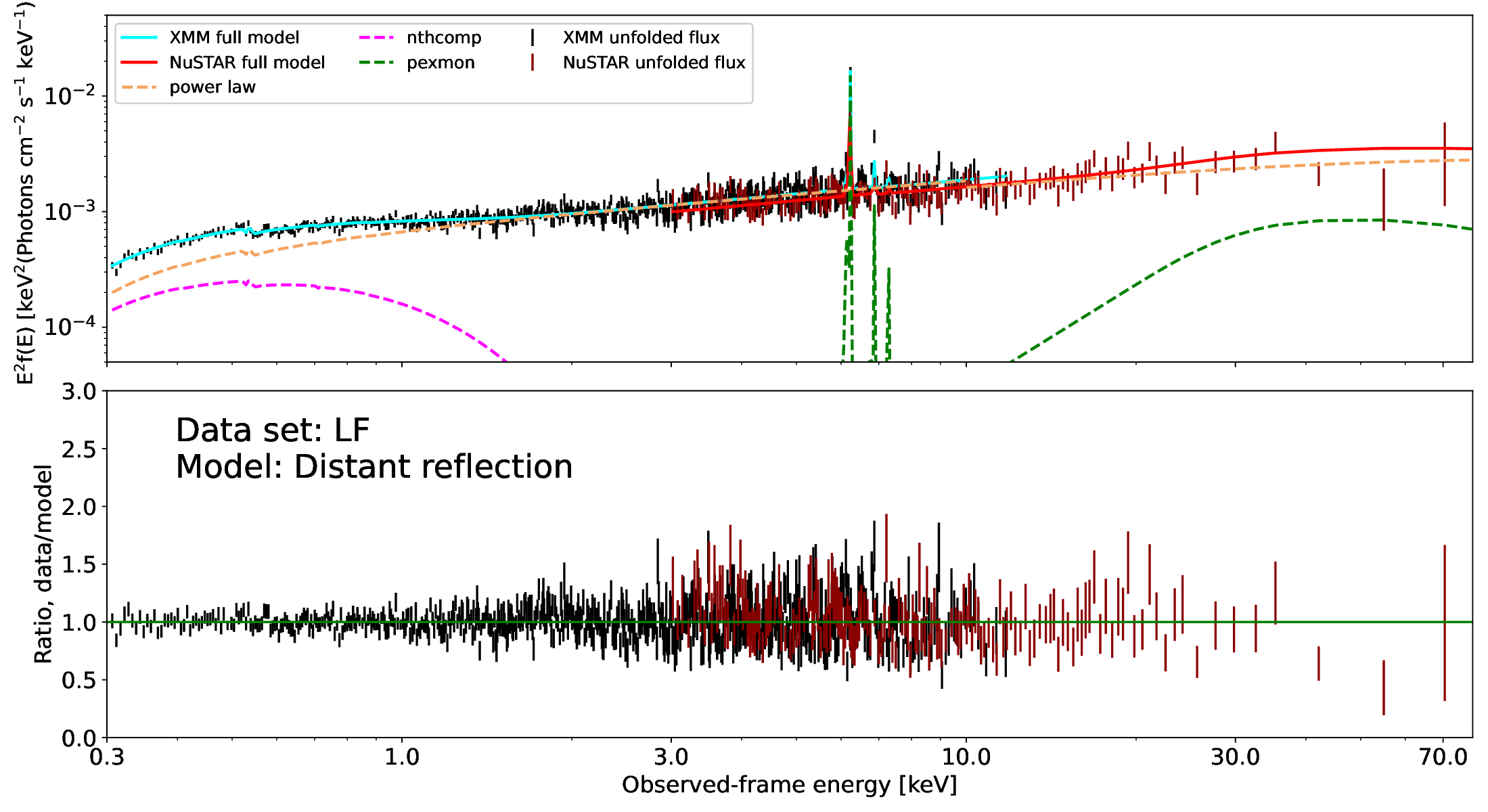}
    \caption{Best-fit distant reflector-only model for the LF data. The top panel shows the unfolded flux spectrum for \emph{XMM-Newton} (black error bars) and \emph{NuSTAR} (red error bars), along with the total model and its individual additive components; the bottom panels show the data to model ratio. The upper panel $y$-axis displays $E^2f(E)$, where $E$ is photon energy in units of keV, and $f(E)$ is the photon flux density; it is analogous to $\nu F_\nu$ as commonly used to present AGN UV--optical spectral energy distributions. The best-fit reflection fraction is $R\sim0.4$, where $R=1$ is expected for a slab reflector extending to large radii. The warm-Comptonized emission (magenta dashed curve) contributes substantially to the total model below $\sim1$ keV. \label{fig:LF_distantref}}
\end{figure*}

\begin{figure*}
    \centering
    \includegraphics[scale=0.48,trim={15 0 0 0}]{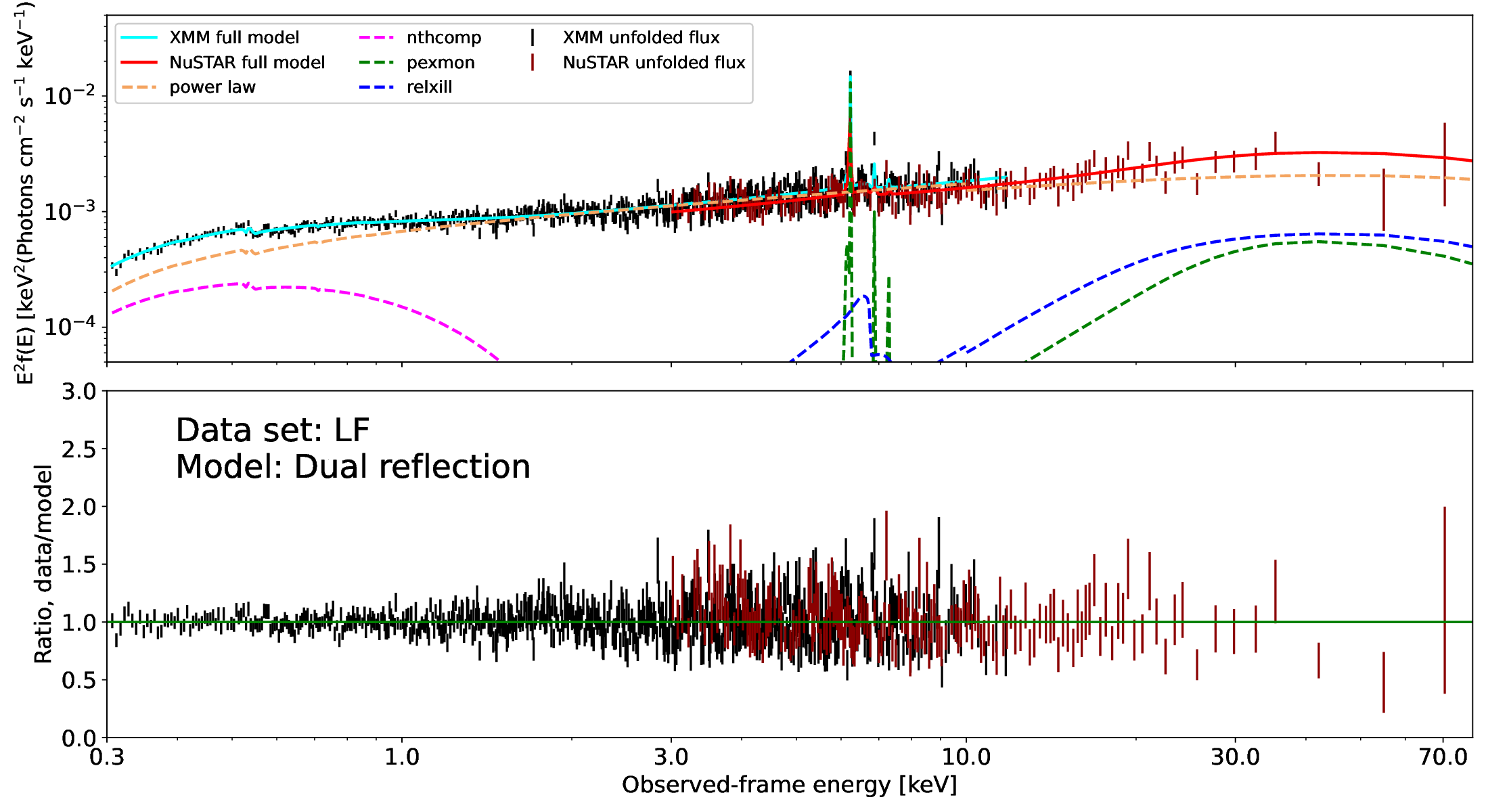}
    \caption{Best-fit model including an additional, relativistic reflection component (blue curve), for the LF data. Due to the faintness of the relativistic component, the full model (solid cyan and red curves) is near-indistinguishable from the distant reflection-only case. \label{fig:LF_dualref}}
\end{figure*}

\begin{figure*}
    \includegraphics[scale=0.48,trim={15 0 0 0}]{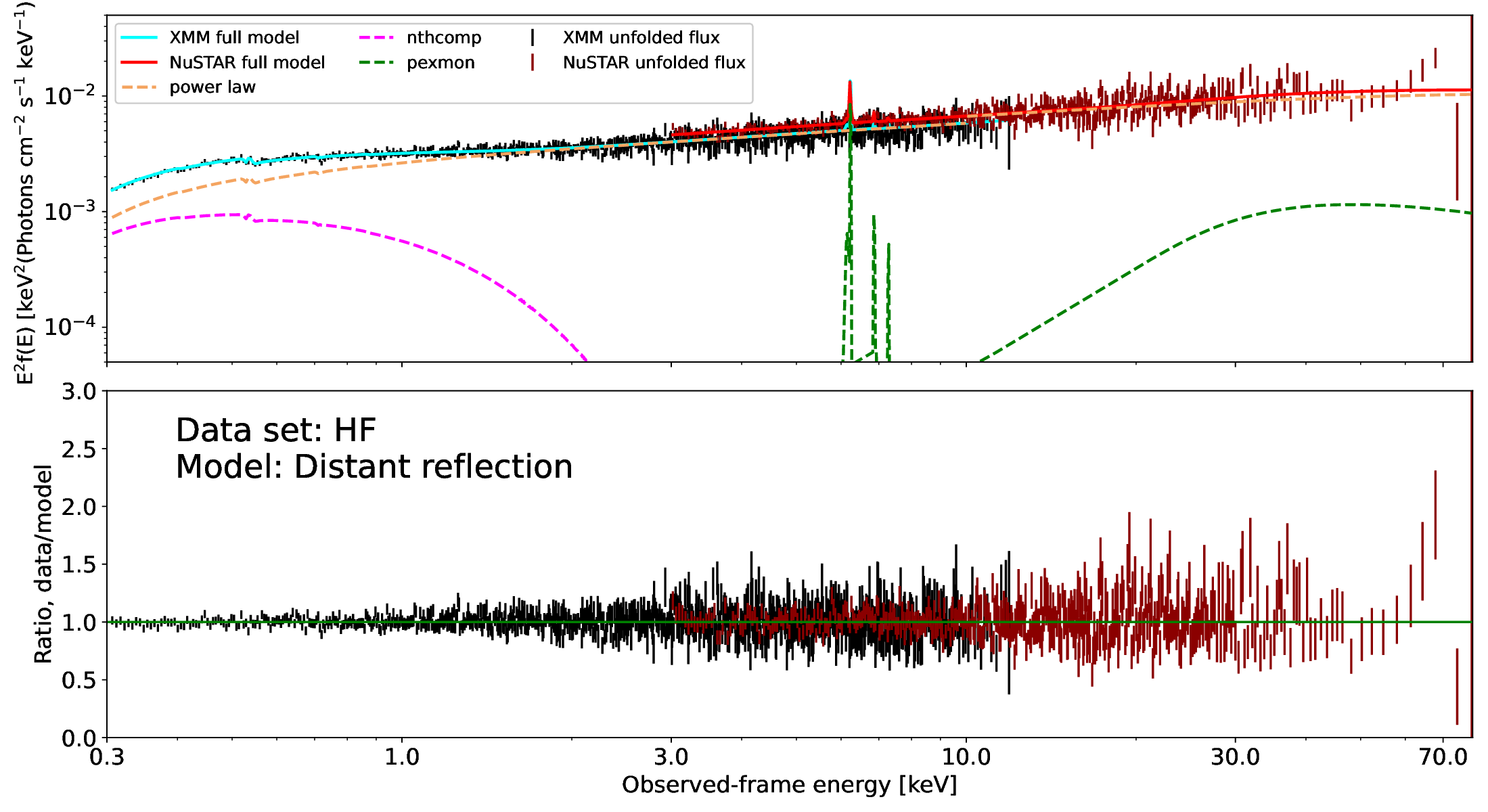}
    \caption{Best-fit distant reflector-only model for the HF data. Here, any reflection is exceedingly weak ($R\sim0.1$); this is consistent with a distant reflector that responds only slowly to continuum flares. \label{fig:HF_distantref}}
\end{figure*}

\begin{figure*}
    \includegraphics[scale=0.48,trim={15 0 0 0}]{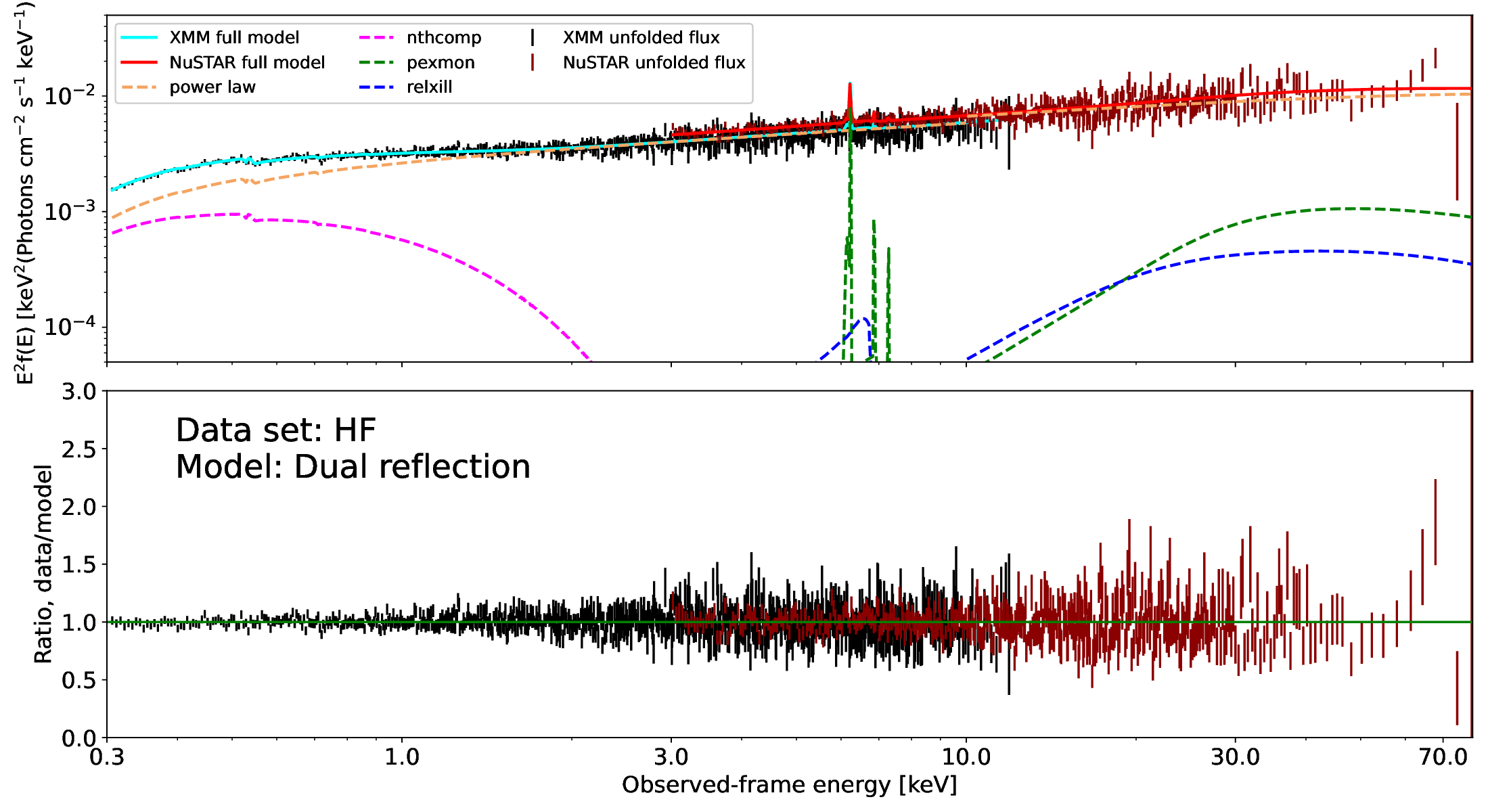}
    \caption{Best-fit model including an additional, relativistic reflection component, for the HF data. While the data allow for some soft X-ray contribution from relativistic reflection (blue dashed curve), it is clear that this cannot account for the observed soft excess flux without severely overestimating the broad iron emission. \label{fig:HF_dualref}}
\end{figure*}

Here, we define the two models that are formally acceptable for all four 0.3--79 keV data sets. These are labeled as Models C2 and G in Appendix \ref{AppendixB}.

\paragraph*{Warm-Comptonized emission and distant reflection:} This model consists of a power law continuum with an exponential cutoff, warm-Comptonized emission, and a non-rotating reflection component. In \textsc{Xspec} modeling parlance, this model is defined as \textsc{const$\times$tbabs(zcutoffpl+nthcomp+pexmon)}. Here, the \textsc{const} component represents a multiplicative scaling between different instruments, while \textsc{tbabs} represents Galactic absorption. The \textsc{nthcomp} component \citep{Zdziarski1996, Zycki1999} models inverse Compton scattering, assuming a seed photon temperature of 10 eV (\emph{i.e.}, UV seed photons), and a single electron temperature $kT_e$. The \textsc{pexmon} component \citep{Nandra1997} represents reflection in a slab of stationary material. It is parameterized by a reflection strength $R$, an inclination angle $i$, a metallicity $Z$ and an iron abundance $A_{\mathrm{Fe}}$, where the latter is treated separately due to the importance of iron emission in observed AGN spectra. We tie the incident-continuum photon index and cutoff energy of the \textsc{pexmon} component to those of the primary continuum component.

\paragraph*{Warm-Comptonized emission with dual reflection regions:} This model includes an additional relativistic reflection component. It is defined as \textsc{const$\times$ tbabs(zcutoffpl+nthcomp+pexmon+relxillLp)}. The component \textsc{relxillLp} \citep{Dauser2014, Garcia2018} represents reprocessing of a point-source X-ray continuum in a `thin disk' extending to the innermost circular stable orbit, with appropriate rotational blurring applied. We assume that the two reflection regions represent inner and outer regions of the same bulk accretion flow. Thus, we tie the \textsc{relxillLp} inclination angle $i_{\mathrm{rel}}$ and iron abundance $A_{\mathrm{Fe,rel}}$ to those of the \textsc{pexmon} component. For consistency, we tie the incident photon indices and high-energy cutoffs of both reflection components to that of the primary continuum. As the \textsc{relxillLp} component represents reflection in the inner accretion flow, the reprocessing material may be highly ionized. This is parameterized in the model as $\xi_{\mathrm{rel}}$, the ionization parameter at the inner edge of the disk; the ionization throughout the disk is then scaled according to the thin-disk prescription. The height of the continuum source is set to 10 gravitational radii above the disk, while its outflow velocity is set to zero. The relativistic reflection is rather faint in both LF and HF, as discussed below; thus, the values of these parameters do not strongly affect the total model spectrum, and are not constrained by the data.

\subsubsection{Properties of our preferred models}\label{sec:bestmodels}

To explore the behavior of these two models, we perform \emph{C-stat} minimization to each data set. We present the best-fit model parameters in Table \ref{tab:preferred_models}. We present the model spectra in Figures \ref{fig:LF_distantref} and \ref{fig:LF_dualref} (LF),  and Figures \ref{fig:HF_distantref} and \ref{fig:HF_dualref} (HF). We display these preferred models as $E^2f(E)$ spectra, unfolded from the instrumental response \citep[\emph{e.g.},][]{Gunderson2025},  to illustrate the energy output at a given photon energy. In the following, we comment on notable features of the best-fit models, and on significant differences between the LF and HF states. \\

\noindent \textbf{Continuum:} The measured primary continuum properties are insensitive to whether relativistic reflection is included. We find that the continuum is significantly harder ($\Gamma\approx1.56\pm0.02$) for the LF data, relative to the HF ($\Gamma=1.67\pm0.02$). This is broadly consistent with the trend found for our individual XMM observations (\S \ref{sec:analysis_bbody}). The high-energy cutoff is at least a \mbox{factor $\sim2$} higher in the HF data ($E_{\mathrm{cut}}>250$ keV) compared to LF ($E_{\mathrm{cut}}\approx150$ keV).\\ 

\noindent \textbf{Warm-Comptonized emission:} The soft excess component is a factor $\sim4$ brighter in the HF data, relative to LF; this is consistent with the flux dependence found for individual observations (\S \ref{sec:analysis_bbody}). The photon index $\Gamma_{\mathrm{warm}}$ parametrizes the underlying optical depth and scattering geometry of the warm region. For these models fitted to X-ray data only, we find $\Gamma_{\mathrm{warm}}\approx1.9$ for the LF data set, increasing to $\Gamma_{\mathrm{warm}}\approx2.2$ for the HF data. However, if we also consider the UV emission, we find $\Gamma_{\mathrm{warm}}\sim2.3$ in both cases, assuming a single warm Comptonizing region; we discuss this further in \S \ref{sec:discussion_SED}. The inferred temperature for the warm Comptonizing region is $kT_{\mathrm{e}}\sim0.2$--0.3 keV, corresponding to roughly $\sim3\times10^6$ K, for all data sets. 

\noindent \textbf{Reflection regions:} We are only able to obtain useful constraints on the reflection inclination angle $i$ and iron abundance $A_{\mathrm{Fe}}$ for the LF data set; we obtain $i=38^\circ\pm8^\circ$ and $A_{\mathrm{Fe}}=6.9^{+9.0}_{-3.4}$. As neither of these parameters are likely to change on timescales of a few years, we impose these values for the other data sets. Regarding $A_{\mathrm{Fe}}$, we note that the best-fit value is poorly constrained but significantly super-Solar. High iron abundances are commonly found when modeling reflection in AGN \citep{Fabian2009, Reynolds2012, Kara2015, Garcia2015}. They may be due to high particle densities in the accretion disks \citep{Garcia2018}, or due to additional iron reflection in a Compton-thin broad line region \citep{Patrick2012}. In our models, $A_{\mathrm{Fe}}$ is most usefully regarded as a degree of freedom governing the relative strengths of the Fe K line and Compton hump. These are otherwise set by the assumption of a Compton-thick, low-density reflection medium; we cannot infer the real iron abundance independently of that assumption. The black hole spin, which affects the relativistic reflection profile, is unconstrained for all data sets; we set it to $a_*\equiv0$. We confirm that setting the maximal value of $a_*=0.998$ does not meaningfully alter the total model.

Considering the distant reflection-only model, we obtain reflection factors of $R=0.38^{+0.09}_{-0.12}$ for LF and $R=0.14\pm0.02$ for HF. This is weaker than the $R=1$ expected for a Compton-thick slab extending to large radii, indicating that the reflection geometry may be truncated, patchy, and/or Compton-thin. Given that our HF observations capture Mrk 590 during short, sharp X-ray outbursts, the lower reflection fraction for HF likely indicates a delayed response of the distant reflection region. For the dual reflection model, the \emph{distant} reflection factors fall only negligibly, to $R=0.34\pm0.05$ and $R=0.12\pm0.02$, respectively; this supports that the reflection spectrum is dominated by distant reprocessing. The relativistic reflection component contributes only weakly to the overall emission spectrum (Figures \ref{fig:LF_dualref} and \ref{fig:HF_dualref}, blue curves). It displays a smooth, near-featureless profile. This necessitates a high ionization of the (putative) disk reflection surface; we find $\log(\xi)\sim3$ for both the LF and HF data sets. In summary, these model fits support the conclusions of our model comparison study (Appendix \ref{AppendixB}): weak, highly ionized relativistic reflection may be present, but is not required to explain these spectra.

%% file: sections/sec_discussion.tex
\section{Discussion}\label{sec:discussion}

\subsection{A variable yet persistent soft X-ray excess}

We detect a soft X-ray excess in Mrk 590 at all observed flux levels. Its emission strength is highly correlated with both the X-ray and the UV continuum variability (\S \ref{sec:analysis_bbody}). While we only have two \emph{XMM-Newton} observations obtained prior to the initial `turn-off' event, the soft excess level in those data is fully consistent with the overall trend (\emph{e.g.}, \mbox{Figure \ref{fig:flux_bbody})}. This is at odds with results for Mrk 590 presented by \citet{Ghosh2022}. These authors claim, based on rather short \emph{Swift} XRT observations, that the soft excess disappeared during 2016, and that it is uncorrelated with the UV emission. In particular, for \emph{Swift} XRT data taken just seven days after our January 2021 \emph{XMM-Newton} observation, they find a soft excess flux $F_{\mathrm{BB}}<0.8\times10^{-13}$ erg cm$^{-2}$ s$^{-1}$. This upper limit is a factor $\sim8$ lower than our measured $F_{\mathrm{BB}}$ for January 2021. We investigate this discrepancy in Appendix C, and find that it is due to \emph{a)} the larger modeling uncertainties when using short \emph{Swift} XRT observations instead of deep \emph{XMM-Newton} data, and \emph{b)} a possible systematic lack of soft X-ray counts in the XRT spectra. We therefore believe that our soft excess detections with \emph{XMM-Newton} (and supported by the \emph{NICER} data) are robust.

A positive correlation between the soft excess and the UV continuum flux (\emph{i.e.}, the disk emission) supports inverse Comptonization of disk seed photons as a production mechanism for the soft excess \citep{Mehdipour2015, Mehdipour2023}. Conversely, a correlation between the soft excess and the X-ray continuum flux, without any $F_{\mathrm{BB}}$--$F_{\mathrm{UV}}$ correlation, might support an origin in relativistic reflection \citep{Barua2023}. As the soft excess strength in Mrk 590 correlates with \emph{both} the X-ray and the UV continuum levels, we cannot make similar arguments based on variability data alone. However, our model comparison (\S \ref{sec:analysis_joint}) strongly favors models with warm-Comptonized emission over those without. Thus, we argue that most of the observed soft excess in Mrk 590 is warm-Comptonized emission. 

The location of the warm Comptonizing region is not constrained by these analyses. Its source of UV seed photons is often assumed to be produced in the inner accretion disk. However, in Mrk 590, the strongest UV response appears to be located $\sim3$ light-days from the X-ray source \citepalias{Lawther2023}. If the UV source and soft excess regions are co-located, the tight correlations between the UV and X-ray continua and the soft excess flux can be explained by X-ray irradiation producing additional UV seed photons. In that case, we would expect a similar $\sim3$-day delay between X-ray continuum and soft excess fluctuations. Alternatively, the warm Comptonizing region may occur physically closer to the X-ray source than the primary UV response; we discuss one possible physical scenario in \S \ref{sec:discussion_accretionflow}. Dedicated X-ray timing experiments are needed to test these scenarios.

\subsection{Warm-Comptonized emission in the UV}\label{sec:discussion_SED}

A strong contribution from warm-Comptonized emission will unavoidably affect the shape of the observed UV--optical spectrum. In particular, compared to an initial `thin disk' seed photon distribution, it flattens the spectral energy distribution by reducing the UV peak luminosity and shifting the emission peak towards higher energies. While we defer a full study of the broad-band spectral energy distribution to future work (\emph{Lawther et al., in prep.}), it is important to test whether the UV emission is consistent with our X-ray modeling. In particular, while multiple mechanisms may in principle contribute to the UV--optical emission, our model would be ruled out if it \emph{over-predicts} the UV luminosity. We now investigate this using available \emph{Swift} UVOT data.

\noindent\paragraph*{UVOT data and host galaxy subtraction:} We extract sky background-subtracted photometry for individual UVOT observations as detailed in \citetalias{Lawther2023}. These data contain a substantial contribution from host galaxy starlight, for all epochs presented in this work. We construct an approximate model of the host galaxy emission as follows. \citet{Bentz2009} present a 2D decomposition of the stellar and AGN emission for Mrk 590, based on high-resolution \emph{Hubble} Space Telescope imaging. They represent the \emph{V}-band stellar emission as three separate Sersic components: two compact `bulge' components, plus an extended disc. We calculate the relative contribution of these three components within the inner 3'' (to match our UVOT extraction aperture). We then co-add the `bulge' and `Sa' spectral templates presented by \citet{Kinney1996}, scaled in the \emph{V}-band according to the derived Sersic profile ratios. We apply \emph{Synphot} \citep{synphot2018}  to determine the host galaxy colors in the UVOT filters.  Finally we apply the probabilistic flux variation gradient technique \citep{Gianniotis2022}, using all \emph{Swift} variability data obtained since 2017, to estimate the host galaxy flux levels.\\

\begin{figure*}
    \includegraphics[scale=0.395, trim={5 0 10 0}, clip]{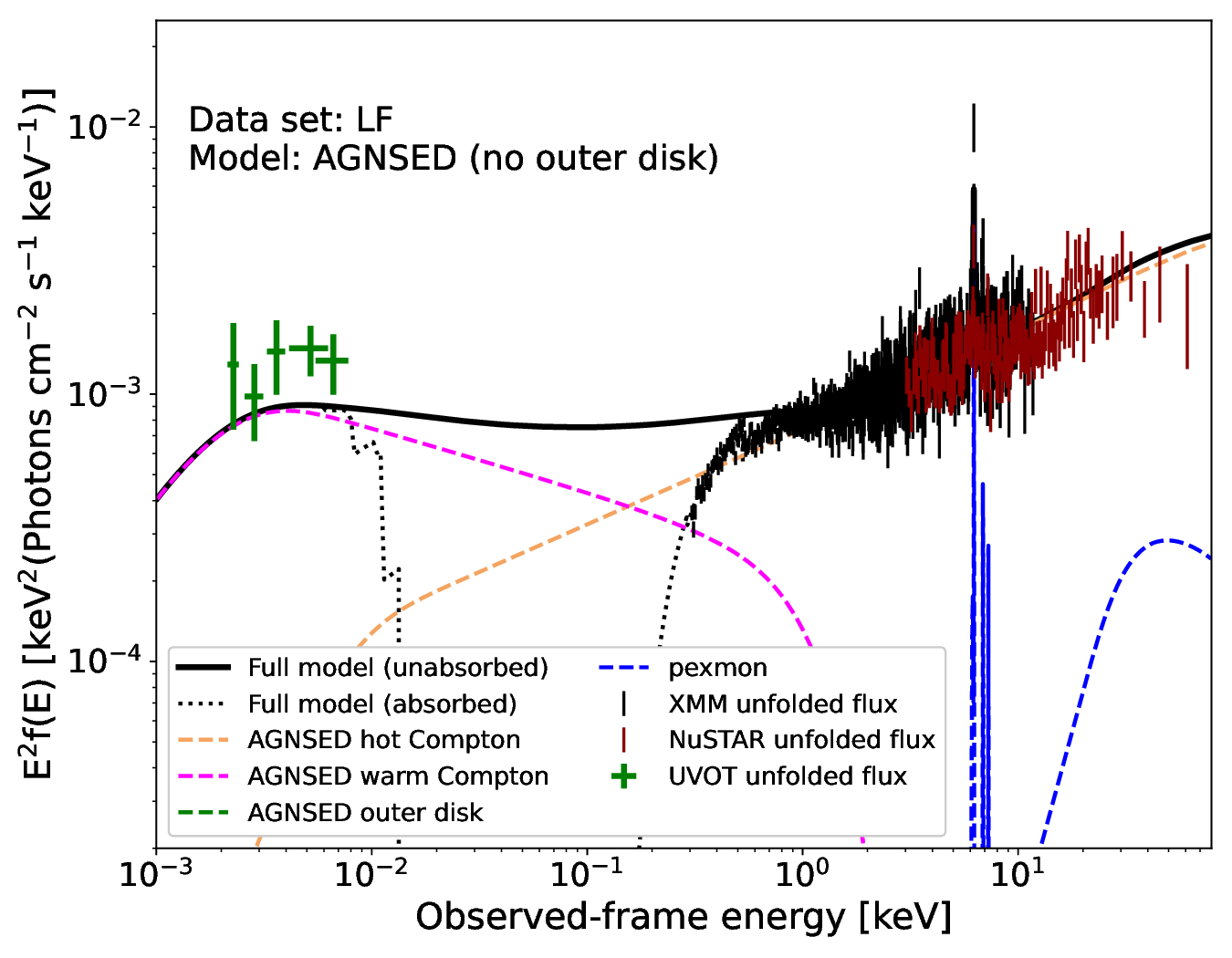}
    \includegraphics[scale=0.395, trim={30 0 10 0}, clip]{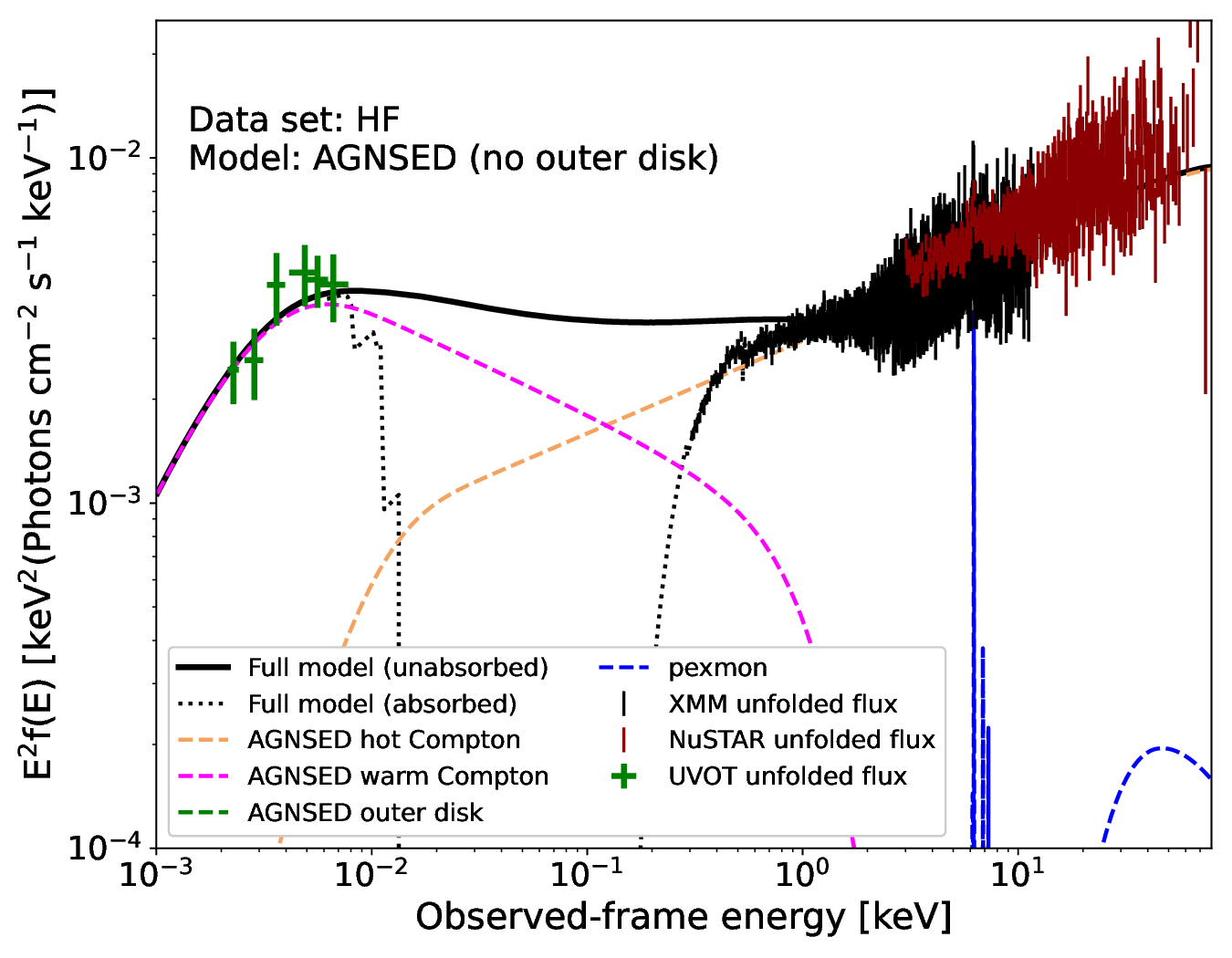}
    \includegraphics[scale=0.395, trim={5 0 10 0}, clip]{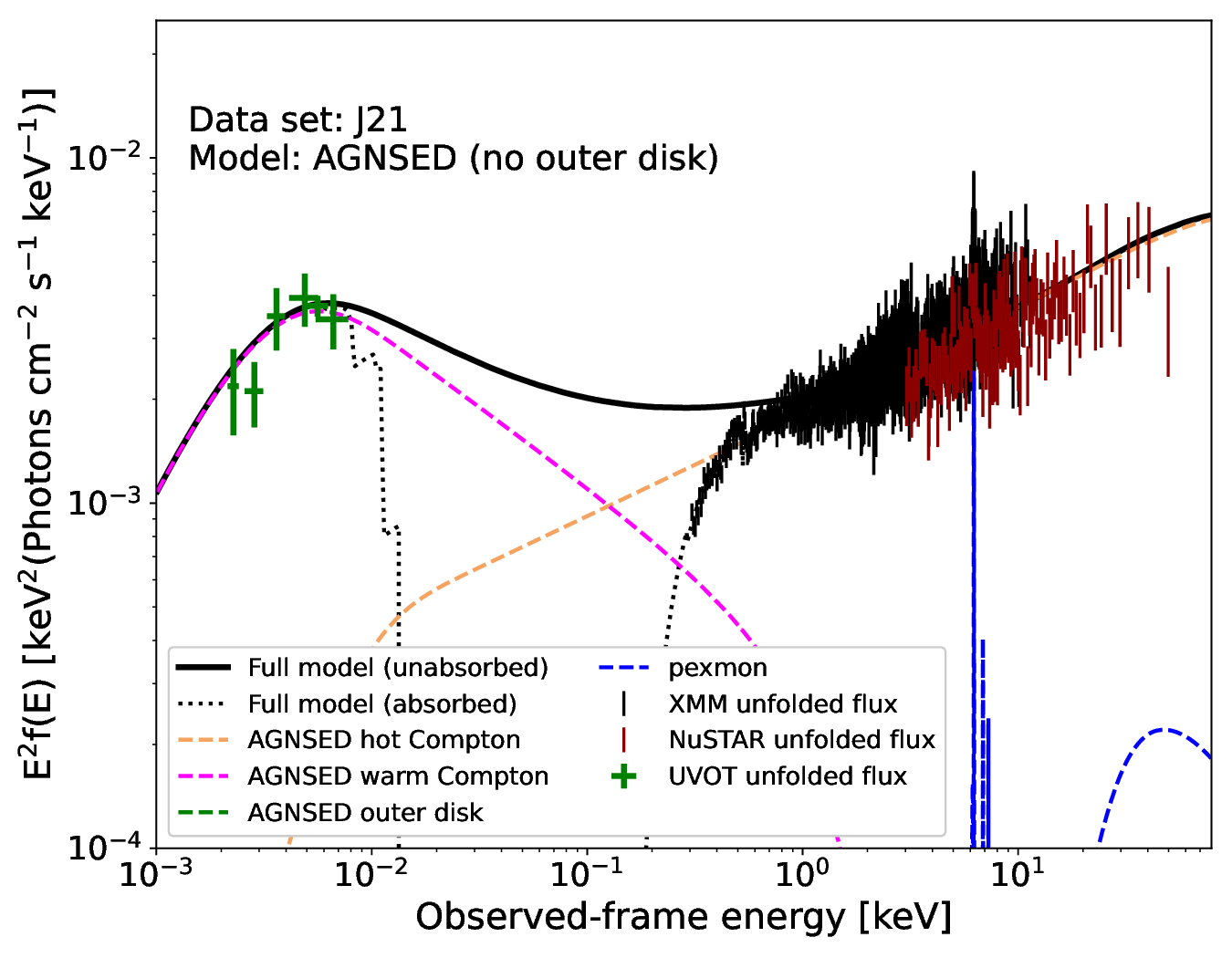}
    \includegraphics[scale=0.395, trim={30 0 10 0}, clip]{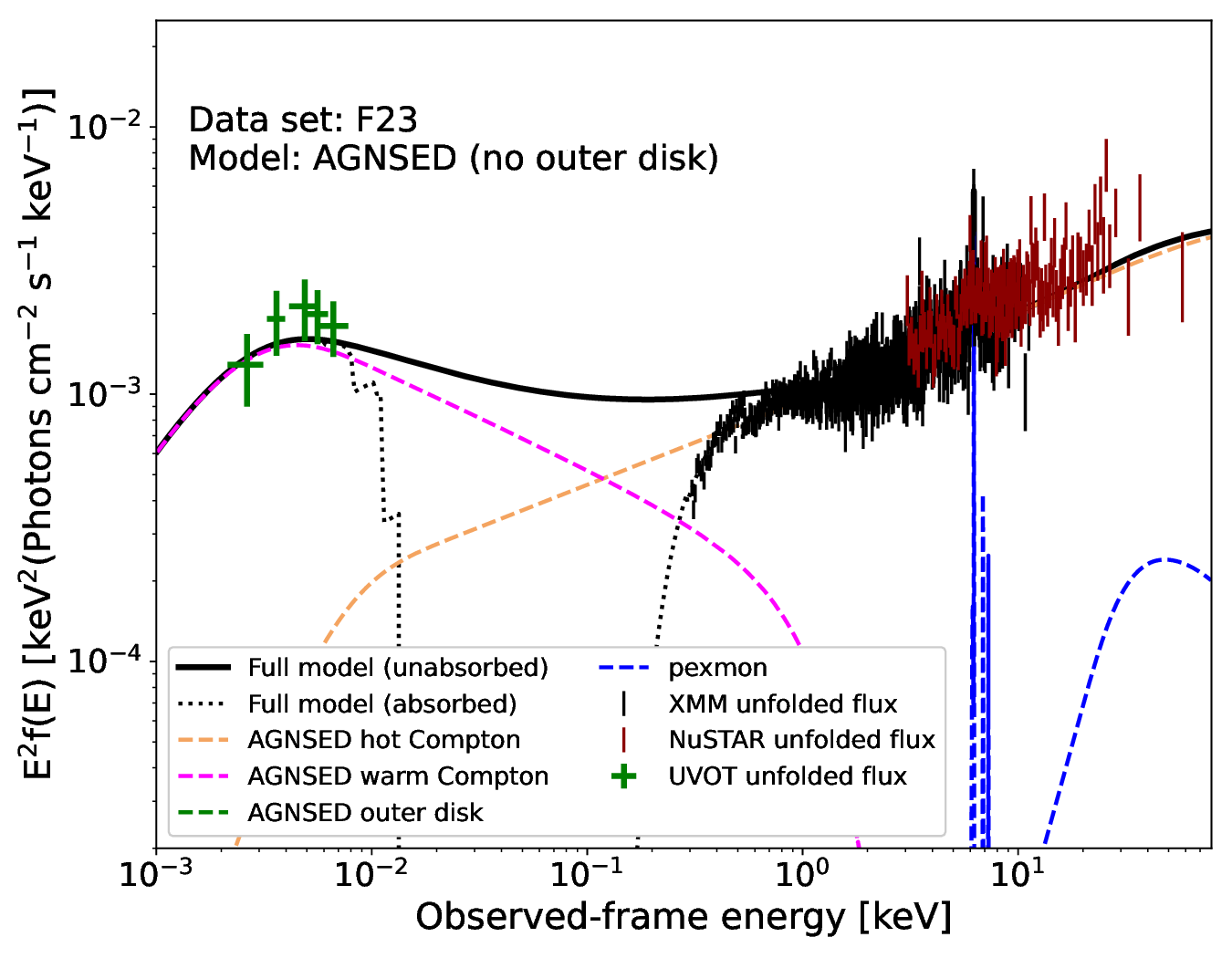}
    \caption{\textsc{agnsed} models to the optical--UV--X-ray spectral energy distributions for the LF, HF, J21 and F23 data sets. The UVOT data points represent average flux levels for observations taken within the appropriate X-ray flux regime; see main text (\S \ref{sec:discussion_SED}) for details. Here, we impose $R_{\mathrm{warm}}=R_{\mathrm{out}}$, such that no outer `thin-disk' emission is produced. We note that the model curves in these figures (black  curves) are normalized to the \emph{XMM-Newton pn} spectra. As there are substantial flux offsets between \emph{XMM-Newton} and \emph{NuSTAR} in these combined data sets (\S \ref{sec:analysis_joint_datasets}), the \emph{NuSTAR} spectra appear offset from the model curves in this unfolded presentation; these offsets are accounted for in the underlying modeling. We present alternative configurations of the \textsc{agnsed} model (\emph{i.e.}, including a cool disk; excluding warm Comptonization; limiting the size of the hot corona; imposing a lower temperature for the hot corona) in Appendix \ref{AppendixD}. } \label{fig:agnsed}
\end{figure*}

\noindent Short \emph{Swift} observations were carried out during the \emph{XMM} observations. However, for the LF data set, the contemporaneous UVOT observations are very faint; after subtracting our host galaxy model, we find negative residuals in the \emph{V} and \emph{B} bands. This suggests that we are slightly overestimating the host galaxy emission in \emph{V} and \emph{B}, and/or that these particular UVOT observations suffer flux calibration issues. To avoid biasing our UV modeling for the LF data set due to these negative residuals, we turn to non-contemporaneous data taken at similar X-ray flux levels. We select all observations for which the \emph{Swift} XRT flux satisfies $3.8\times10^{-12}<F_{\mathrm{0.3-10}}< 6.8\times10^{-12}$ erg cm$^{-2}$ $s^{-1}$ for the LF dataset, and $18.8\times10^{-12}<F_{\mathrm{0.3-10}} < 24.7\times10^{-12}$ erg cm$^{-2}$ $s^{-1}$ for HF. This corresponds to the X-ray flux ranges spanned by each data set (\S \ref{sec:analysis_joint_datasets}). In each UVOT filter, we adopt the sample average flux of the flux-matched set, and use the 1$\sigma$ sample scatter in measured fluxes as the uncertainty. This provides an indication of the typical UV--optical fluxes emitted for each X-ray flux regime (Figure \ref{fig:agnsed}, green crosses). As this `X-ray flux-matched' approach is non-standard, we also present the contemporaneous UVOT data for LF for comparison purposes.

\noindent\paragraph*{UV--optical--X-ray modeling:} We replace both the power-law continuum and the warm-Comptonized emission components in our distant reflection model (\S \ref{sec:bestmodels_def}) with a single \textsc{agnsed} \citep{Kubota2018} component. The resulting model is defined in \textsc{Xspec} as \textsc{const$\times$tbabs(agnsed+pexmon)}. We allow the constant offsets $C_{\mathrm{inst}}$ to differ for each X-ray detector, as before, but require $C_{\mathrm{UVOT}}\equiv1$ as we account for observed UV--optical flux variability in the uncertainties. The \textsc{agnsed} model assumes that accretion energy is dissipated radially according to the `thin-disk' prediction, but that the energy release within a radius $R_{\mathrm{hot}}$ occurs as hot-Comptonized emission to produce the X-ray continuum. Thus, $R_{\mathrm{hot}}$ is effectively the size of the hot corona. For radii $R_{\mathrm{hot}}<R<R_{\mathrm{warm}}$, the energy is released as warm-Comptonized emission, while for $R_{\mathrm{warm}}<R<R_{\mathrm{out}}$, `thin-disk' thermal emission is observed without any reprocessing. Thus, \textsc{agnsed} calculates the emission from an inner corona, an intermediate warm-Comptonizing region, and an outer thin disk, in an energetically consistent way. The \textsc{pexmon} component accounts for the distant reflection features, as before. We do not include any relativistic reflection component here, as it is not formally required by the X-ray data (\S \ref{sec:bestmodels}). For this preliminary analysis of the optical--UV emission, we hold several \textsc{agnsed} parameters constant at their default values, as we only aim to test whether warm Comptonization is in broad agreement with the observed spectral energy distribution shape. In particular, we adopt a disk inclination angle $i=38^\circ$ as derived from our X-ray modeling, and assume zero black hole spin. We assume a black hole mass of $3.7\times10^{7}$ $M_{\astrosun}$ \citep{Peterson2004}. We use the \textsc{pyXspec} Monte Carlo Markov Chain implementation to explore the parameter space, with a chain length of 300,000, and optimize the model starting at the posterior median to obtain a final fit. As the UVOT photometry is obtained in imaging mode, we adopt $\chi^2$ statistics for this process, rebinning our X-ray data to a minimum of 25 counts per bin.

\begin{figure}
    \centering
    \includegraphics[scale=0.395, trim={10 0 10 0}, clip]{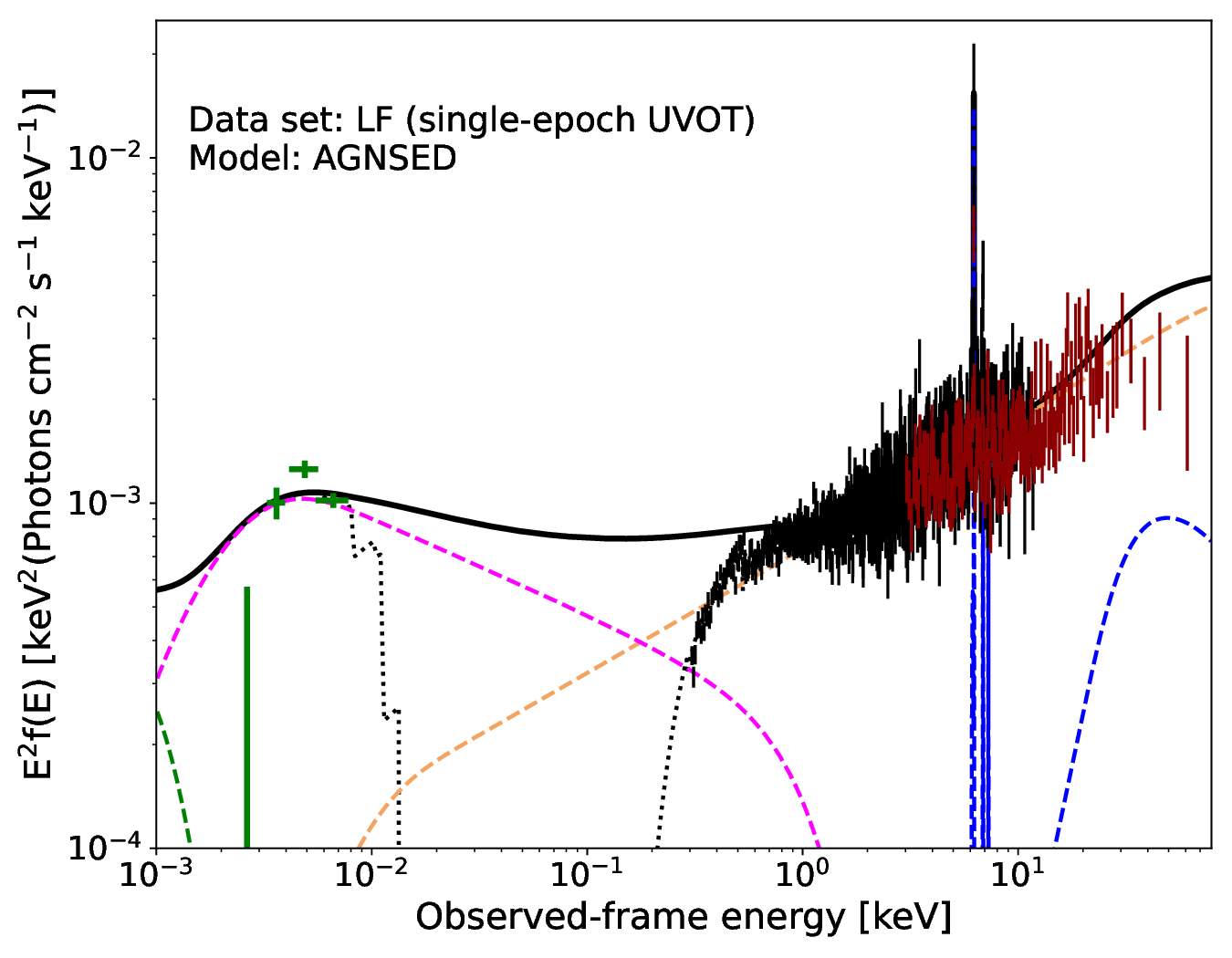}
    \caption{An alternative spectral energy distribution for the LF data set, using a 2 ks contemporaneous UVOT observation. Unfortunately, our host galaxy model over-subtracts the \emph{B} and \emph{V}-band flux: here, we show $3\sigma$ statistical upper limits, which are inconsistent with the best-fit warm Comptonization component. However, the \emph{U} and far-UV bands are well-modeled by warm Comptonization. An improved host galaxy model is required to fully test the warm-Comptonization scenario at the lowest observed flux levels; we will address this in future work. }
    \label{fig:agnsed_LF_singleuvot}
\end{figure}

\noindent\paragraph*{SED modeling results:} The LF data set has an accretion rate of around 0.9\% of the Eddington rate, while HF is accreting at around 3\% Eddington, according to the \textsc{agnsed} model. We find $R_{\mathrm{hot}}\sim100$ gravitational radii for all four data sets (LF, HF, J21 and F23). In terms of the assumed \textsc{agnsed} geometry, this indicates that the inner accretion flow is hot and optically thin; the disk is heavily truncated. In a model-agnostic sense, it simply means that Mrk 590 has a relatively bright primary X-ray continuum compared to the UV, \emph{e.g.}, considering the ensemble UV versus X-ray luminosity relationship for AGN \citep{Lusso2016}. Intriguingly, \citet{Krishnan2024} report a similarly large $R_{\mathrm{hot}}$ in the low-flux state of a flaring Seyfert AGN, accreting at $\sim4$\% of the Eddington rate. They find a compact corona ($R_{\mathrm{hot}}\sim20$ $r_g$) for the same source in its high-flux state  ($\sim10$\% Eddington). Compact $R_{\mathrm{hot}}$ are also reported for Mrk 110 \citep{Porquet2024} and ESO 141-G55 \citep{Porquet2024b}, both accreting at roughly 10\% of the Eddington rate. As Mrk 590 achieves only 3\% Eddington, even in the HF data, our findings are consistent with these results. It appears that the onset of inner-disk formation may occur at $\apprge5$--10\% Eddington; additional X-ray and UV data at higher flux levels are needed to test this explicitly for Mrk 590.\\

\noindent `Pure' warm-Comptonized emission (\emph{i.e.}, $R_{\mathrm{warm}}\equiv R_{\mathrm{out}}$) provides a better match to the UVOT data than models dominated by a cool outer disk. Because the disk is truncated to power the bright hard X-ray emission, the inner edge of the disk is at $\sim100$ $r_g$, and is thus cooler than a `standard' disk extending to the innermost stable orbit. Without warm-Comptonized emission, this predicts a spectral turnover in the optical, contrary to our observations. Warm-Comptonized emission peaks in the extreme-UV, which is broadly consistent with the rather flat UVOT spectral energy distributions (Figure \ref{fig:agnsed}). Thus, we suspect that Mrk 590 may lack a standard, unobscured `thin-disk' component during our observations. We tabulate the best-fitting model parameters, for \textsc{agnsed} with no cool outer disk, in Table \ref{tab:agnsed}. We demonstrate that alternative geometries, \emph{e.g.}, requiring a compact hot region with $R_{\mathrm{hot}}=10$, or replacing the warm Comptonizing region with a standard `thin disk', do not substantially improve the modeling (Appendix \ref{AppendixD}). While a hybrid model including both warm-Comptonized and thin-disk emission can roughly match the UV flux levels (Appendix \ref{AppendixD}), the best-fit solutions are dominated by the warm-Comptonized emission at the energies probed by UVOT. 

\noindent The overall UV--optical fluxes are matched to first order by the \textsc{agnsed} model for the HF, J21 and F23 data. Meanwhile, for the LF data set, the model underestimates the observed UV flux (Figure \ref{fig:agnsed}, top left). While this does not exclude the warm-Comptonization scenario, it would mean that an additional source of UV emission is required in the low-flux state. It is unclear whether this represents the onset of a real physical change at the lowest flux levels, or is due to our use of non-contemporaneous UVOT data to construct the low-state SED. For comparison purposes, we present \textsc{agnsed} modeling of LF based on contemporaneous UVOT data in Figure \ref{fig:agnsed_LF_singleuvot}. While the \emph{V}- and \emph{B}-band upper limits (likely affected by host galaxy over-subtraction) cannot be reconciled with the \textsc{agnsed} model, the UV bands are broadly consistent with it. An improved host galaxy model is needed to fully test the warm-Comptonization scenario for the LF data, as the SED modeling is more sensitive to over-subtraction issues at low AGN flux levels. We note that the F23 data set is successfully modeled by \textsc{agnsed} with no outer disk (Figure \ref{fig:agnsed}, bottom right). This suggests that warm Comptonization is the main UV--optical emission mechanism at X-ray flux levels down to at least $F_{0.3-10}\sim6\times10^{-12}$ erg cm$^{-2}$ s$^{-1}$. \\

\noindent The \textsc{agnsed} modeling yields softer photon indices for the warm-Comptonized emission ($\Gamma_{\mathrm{warm}}\sim2.3$), compared to our X-ray analysis ($\Gamma_{\mathrm{warm}}\sim1.9$). This distinction is important, as $\Gamma_{\mathrm{warm}}\apprle2$ likely requires both a non-dissipative underlying disk, and a patchy, low covering fraction atmosphere \citep{Petrucci2018}. However, we cannot fully exclude $\Gamma_{\mathrm{warm}}<2$ based on our UVOT data, given the substantial uncertainties (Table \ref{tab:agnsed}). We will present additional spectroscopic observations, to robustly test this and alternative scenarios, in upcoming work (\emph{Lawther et al., in prep.}).

\subsection{Is the X-ray reflector Compton-thin?}\label{sec:discussion_turnover}

\begin{figure}
    \centering
    \includegraphics[scale=0.39]{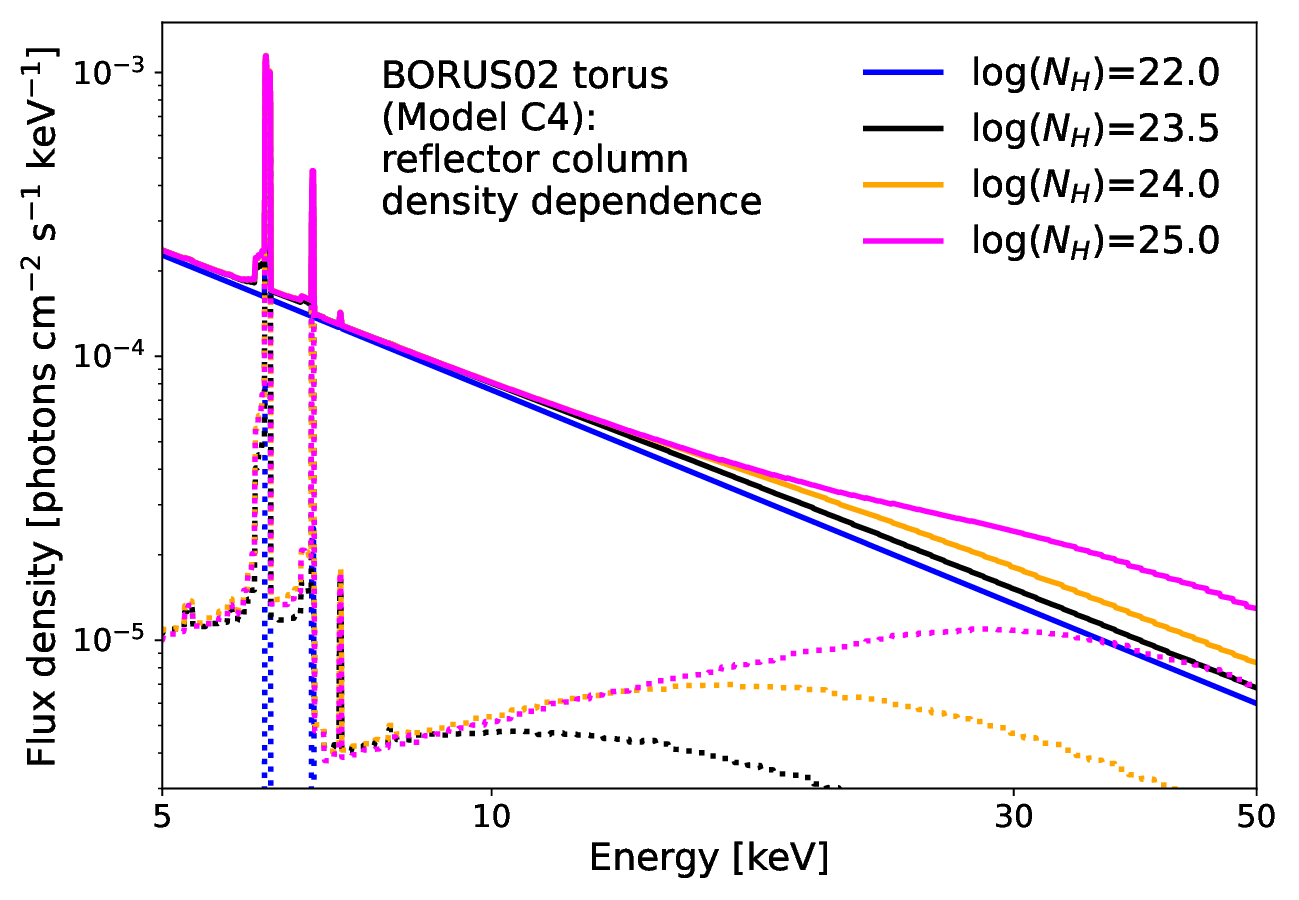}
    \caption{Dependence of a model including variable-density torus reflection component (\textsc{borus02}), on the column density of the reflecting material. The solid lines illustrate the full model; dotted lines correspond to the reflection component only. For $\log(N_{\mathrm{H}})\leq23$, the hard X-ray Compton reflection hump is negligible. This is one possible explanation for the rather weak Compton reflection feature seen for Mrk 590, where models with a Compton-thick reflection component overestimate the hard X-ray flux (\emph{e.g.}, Figure \ref{fig:appendixC_modelC_LF}).}
    \label{fig:borus_comparison}
\end{figure}

An exponential cutoff in the primary continuum is warranted for all four joint data sets according to our model comparison (\ref{sec:analysis_joint}). However, we find that replacing the cutoff power law with a \textsc{nthcomp} component is strongly disfavored in terms of Bayes factors (Appendix \ref{AppendixB_modeldefs}). This is unexpected if the observed continuum is indeed due to a hot Comptonizing `corona' region, as typically assumed. Also, the cutoff power law does not fully capture the spectral curvature at 30--50 keV for the LF data (Figure \ref{fig:LF_distantref}). An alternative explanation is that the assumption of Compton-thick reflection might be incorrect. In that case, our preferred models would over-predict the Compton hump (for a given reflection factor $R$). This could then be ameliorated to first order by inclusion of a high-energy cutoff, even if the true continuum is power law-like over the observed energy range. To test this, we replace the \textsc{pexmon} component with a \textsc{borus02} component \citep{Balokovic2018}, which approximates a torus geometry, and includes the column density of the reflection region as an additional parameter. We include this variant in our Bayesian comparison (Appendix \ref{AppendixB}); it is acceptable  for three data sets (LF, HF and J21). It is disfavored for F23, but this may be due to signal-to-noise issues: F23 has the least total X-ray counts. Using the \textsc{borus02} model, we find column densities of order $10^{23}$ cm$^{-2}$, which indeed produces a weaker Compton hump than does a Compton-thick reflector (Figure \ref{fig:borus_comparison}). We find this interpretation plausible, given that a cutoff power law spectrum is not physically motivated. \citet{Diaz2023} find a weak trend for low-luminosity AGN to display Compton-thin reflection, perhaps because they tend to be more gas-starved, as explored below. Alternatively, additional reflection from Compton-thin BLR clouds would produce substantial iron line emission while only contributing weakly to the Compton hump \citep{Patrick2012}. Assuming Keplerian rotation, the narrow Fe K width implies a radius of at least three light-days for the distant reflector (\S \ref{sec:analysis_bbody}). This is fully consistent with reflection in the BLR: \citet{Mandal2021} estimate a BLR size of $\sim20$ days for Mrk 590 during 2018.

\subsection{Hints of a gas-starved nucleus}\label{sec:discussion_absorption}

\citet{Denney2014} reported a negligible level of intrinsic absorption upon their discovery of the initial `turn-off' event. Our model comparison confirms that the X-ray source in Mrk 590 is largely unabsorbed, at all observed continuum flux levels, over $\sim20$ years of \emph{XMM-Newton} observations. To constrain how much intrinsic absorption might still be present, we fit a model including an intrinsic neutral absorber to the HF and LF data sets. We find an intrinsic column density of 7.9($\pm9.1$)$\times10^{19}$ cm$^{-2}$ for HF and 1.1($\pm0.9$)$\times10^{20}$ cm$^{-2}$ for LF. Thus, Mrk 590 can essentially be classified as a `bare' AGN, defined by \citet{Nandi2023} as sources with intrinsic column densities below \mbox{$10^{20}$ cm$^{-2}$}. 

While these constraints only directly apply to measurements along our line-of-sight, we note a few other indications that the central engine of Mrk 590 may be gas poor. \emph{1)} Mrk 590 displays an unusually strong (for AGN) UV response to X-ray variability \citepalias{Lawther2023}; this suggests unobscured sight-lines between the X-ray continuum and the reprocessor. \emph{2)} While we see clear evidence of X-ray reflection, the reflector may be Compton-thin (\S \ref{sec:discussion_turnover}). \emph{3)} Molecular gas is not detected in the inner $\sim100$ pc nucleus of Mrk 590, although the derived upper limit is $10^5$ $M_{\astrosun}$, in principle sufficient to power the AGN for thousands of years \citep{Koay2016}. While inconclusive, our findings are \emph{consistent} with a scenario where the nucleus is gas-poor compared to typical Seyfert AGN. \citet{Ricci2022} posit an evolutionary scenario where young AGN reside in gas-rich nuclei, displaying strong absorption along any line of sight. They then enter a 'blow-out' phase at high Eddington ratios, becoming bright, unabsorbed sources (unless the line of sight goes through the accretion flow, in which case a Type 2 AGN is observed). The later stage of the AGN lifetime is then gas-starved; the continuum luminosity dims due to reduced accretion rate, and the accretion flow itself becomes Compton-thin. Our findings so far are consistent with such a scenario.

\subsection{The inner accretion flow of Mrk 590}\label{sec:discussion_accretionflow}

\begin{figure}
    \includegraphics[scale=0.26, trim={25 0 0 0}]{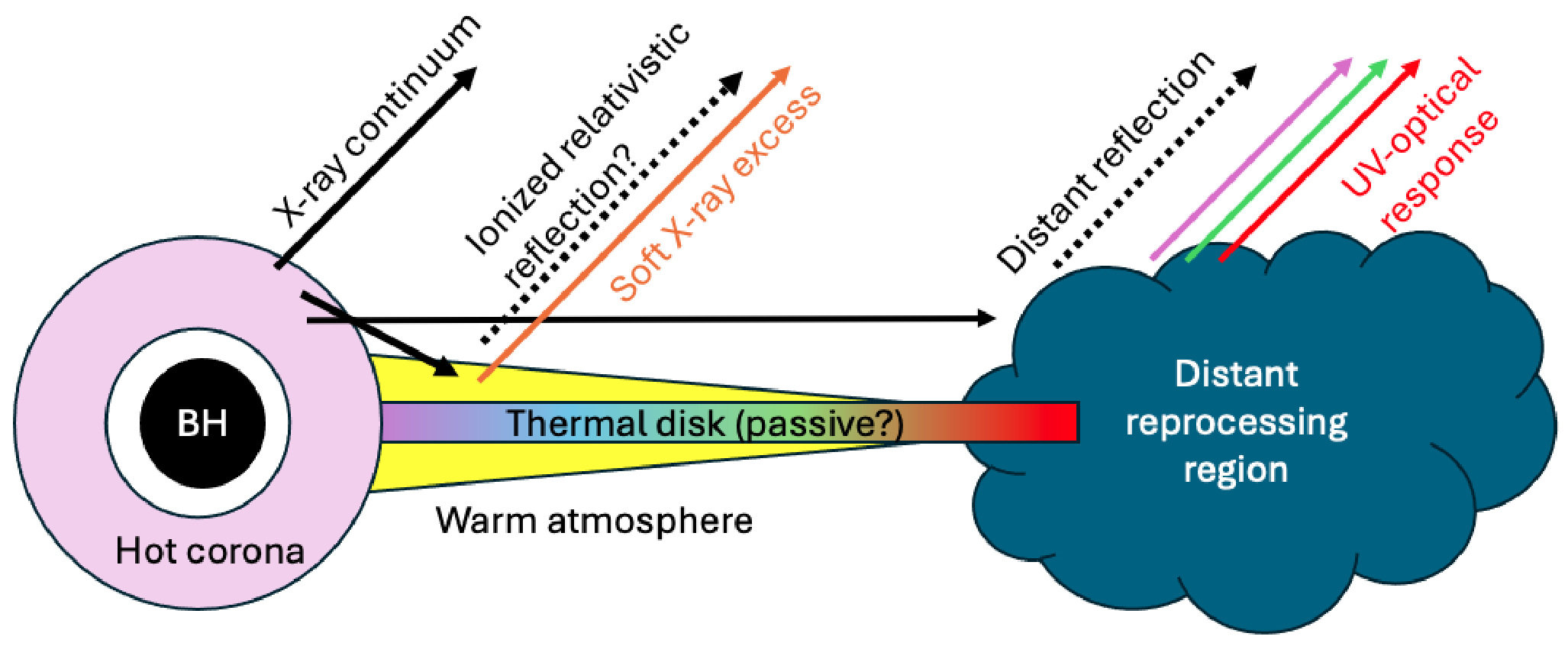}
    \caption{Possible arrangement of the emission regions in Mrk 590 during the flaring episodes since 2017, inspired by the `passive disk' scenarios invoked by \citet{Petrucci2018, Petrucci2020, Ballantyne2024, Paliti2024}. For Mrk 590, the bright X-ray emission (relative to the UV) requires that a substantial fraction of the accretion energy is released in the form of hot-Comptonized emission, here represented by a hot corona extending to $\sim100r_g$ (based on the energetic assumptions of the \textsc{agnsed} model, \S \ref{sec:discussion_SED}). The warm-Comptonized emission is then produced in an extended, optically thick disk atmosphere, which reprocesses seed photons form an underlying thermal disk. As the thermal disk is obscured by this optically thick atmosphere out to $\sim10^5r_g$ (\S \ref{sec:discussion_SED}), and/or truncated at the inner edge due to the large corona, it does not produce a prompt UV response to X-ray illumination. Instead, the UV response may be dominated by a `pure reprocessing' component emitted at larger radii. This may explain both the 3-day X-ray to UV delay \citepalias{Lawther2023}, and the unusually (for AGN) coherent UV response to X-ray variability. We note that our analyses do not constrain the shape, size or orientation of the distant reprocessor. Nor do they demand that the hot corona be spherically symmetric, as opposed to, \emph{e.g.}, a `lamp-post' geometry, perhaps related to jet processes. \label{fig:sketch_disk}}
\end{figure}

We find warm-Comptonized emission in Mrk 590 in both the low- and high-flux states (\S \ref{sec:analysis_joint}), and demonstrate that this emission component may be responsible for most or all of the UV--optical flux (\S \ref{sec:discussion_SED}). Meanwhile, any inner-disk reflection is fairly weak, and (if present) occurs in a highly ionized reflective region. Here, we explore a scenario originally suggested by \citet{Petrucci2013}, for the Type 1 AGN Mrk 509, that can potentially explain these observables. In this scenario, the thin accretion disk is covered by an optically thick `atmosphere' \citep{Haardt1993, Obrien2001, Janiuk2001}, and some of the accretion energy is dissipated directly into the atmosphere \citep{Petrucci2020, Ballantyne2024, Paliti2024}. The underlying disk is both irradiated by the atmosphere, and provides a source of seed photons. The energy balance of this disk--atmosphere system is likely accretion rate-dependent. In particular, if magnetic fields support the disk, an increasing fraction of the accretion energy is dissipated in the atmosphere at lower accretion rates \citep{Begelman2015}. In that case, a largely passive disk with a dissipative atmosphere occurs at low-intermediate accretion rates. A full transition into a `puffed-up', radiatively inefficient advective state is predicted if the accretion slows further. We speculate that Mrk 590 may be in just such an intermediate state during our observations.

For a plane-parallel `sandwich' geometry of a warm-Comptonizing atmosphere above and below a thermal disk, the photon index $\Gamma_{\mathrm{warm}}$ and electron temperature $kT_e$ serve as diagnostics of the disk--atmosphere energy balance. In particular, $\Gamma_{\mathrm{warm}}\sim2$ indicates a passive underlying disk \citep{Petrucci2020}. While we find $\Gamma_{\mathrm{warm}}\sim1.9\pm0.2$ for the LF data set, considering only the X-ray observations, we find $\Gamma_{\mathrm{warm}}\sim2.3\pm0.3$ when we include the UVOT data (\S \ref{sec:discussion_SED}). Although both measurements are consistent with  $\Gamma_{\mathrm{warm}}\sim2$ given their substantial uncertainties, the optical--UV--X-ray data sets are more sensitive to the overall shape of the warm-Comptonized emission. Thus, we cannot strongly constrain the dissipative balance of the disk.

Any relativistic reflection from a warm-Comptonizing disk atmosphere would be highly ionized, with $\log(\xi)\sim3$ \citep{Janiuk2001}, in agreement with our dual reflection region model fits (\S \ref{sec:analysis_joint}). Also, the $\sim3$-day delayed UV response to X-ray flares \citepalias{Lawther2023} may be compatible with this scenario: an optically thick atmosphere with $kT\sim0.3$ keV would presumably partially obscure the underlying disk from X-ray illumination, and in any case would alter its wavelength-dependent response. Much of the observed UV response might then occur in surrounding dense gas, such as a heavily truncated outer disk (if present), broad line-emitting region, or the inner edge of the larger-scale accretion flow. We illustrate one possible `toy model' geometry in Figure \ref{fig:sketch_disk}, but note that the nature of the putative distant reprocessor is highly speculative. We aim to address these issues in future work, harnessing both improved optical--UV wavelength coverage and high-cadence timing data.

\subsection{The diverse X-ray behaviors of CLAGN}

Mrk 590 displays no intrinsic absorption (\S \ref{sec:discussion_absorption}); its changing-look behavior must therefore be due to intrinsic variability. Here, we compare its observed X-ray behavior with other `changing-state' sources. We note that the observations presented here capture Mrk 590 in epochs with and without optical broad emission lines \citep[][\emph{Lawther et al., in prep.}]{Denney2014, Raimundo2019}, thus probing both `turned-off' and `turned-on' states. It is remarkable that the overall X-ray spectral shape does not vary strongly. The 6.4 keV emission lines do display a smaller equivalent width at high luminosities, as expected for flaring activity, given their origin in distant reprocessing material (\S \ref{sec:modelcomparison}). However, the overall spectral shape varies only mildly, as evidenced by the linear $F_{0.3-10}$--$F_{\mathrm{BB}}$ relationship and the weak dependence of $\Gamma$ on $F_{0.3-10}$ (\S \ref{sec:analysis_bbody}). To investigate the long-term evolution, we highlight measurements derived from the \emph{XMM-Newton} observations in 2002 and 2004 (blue squares in Figures \ref{fig:flux_gamma}, \ref{fig:flux_bbody}, \ref{fig:kT} and \ref{fig:line_equivalent_widths}). These measurements largely follow the overall observed trends. Thus, Mrk 590 displayed similar X-ray behavior even before the first discovered `turn-off' event. A similarly constant X-ray spectral shape is observed in the CLAGN LEDA 1154204 \citep{Saha2023} and IRAS 23226-3843 \citep{Kollatschny2023}, despite X-ray variability by factors of $\sim17$ and $\sim10$, respectively.

This is a marked contrast to the changing-look AGN 1ES 1927+654, for which the X-ray continuum disappears in low-luminosity states, with only the soft excess remaining \citep{Ricci2020}. Those authors attribute the apparent disappearance of the continuum to a rapid cooling of the hot X-ray corona. Conversely, \citet{Noda2018} report the disappearance of the soft excess component (but not the hard continuum) in the CLAGN Mrk 1018 as it enters the `turn-off' state. They interpret this as evidence for a transition of the accretion flow into an ADAF state; \citet{Veronese2024} suggest that this transition is linked to a disturbance in the inner disk, with subsequent jet production. Clearly, some individual CLAGN undergo drastic X-ray evolution, which is not seen in Mrk 590. We note that UV--optical CLAGN are essentially defined by the \emph{response} of the broad-line region to changes in the ionizing continuum. The variety of X-ray behaviors observed for different CLAGN suggests that several distinct physical mechanisms in the innermost regions can lead to UV--optical changing look events.

%% file: sections/conclusion.tex
\section{Conclusion}\label{sec:conclusion}

The CLAGN Mrk 590 displayed repeating strong X-ray flares during 2017--2024. We capture the X-ray spectrum several times during this flaring behavior, and supplement these data with archival X-ray observations (2002--2016) to build a comprehensive record of its X-ray behavior. We analyze these data both as individual `snapshots' and as combined, multi-instrument data sets. Our findings are as follows.

\begin{itemize}
    \item Soft X-ray excess is detected at all flux levels. This component displays a tight correlation with both the X-ray and the UV continuum emission from the AGN. Warm Comptonization of UV seed photons is strongly favored over inner-disk reflection as the source of the soft excess. We confirm that a warm Comptonizing region covering most of the accretion flow, out to $\sim10^5r_g$, is broadly consistent with the observed optical--UV emission.
    \item Iron K emission lines are present at both low and high continuum flux levels, although their fluxes cannot be constrained in some individual `snapshot' observations due to poor statistics. The iron emission is dominated by distant reprocessing. Some relativistic reflection may be present, but is not formally required. The relativistic reflecting medium, if present, must be highly ionized.
    \item Mrk 590 displays very little obscuration in any of our observations. It may harbor a Compton-thin outer accretion flow, as supported by the faintness of the observed high-energy reflection hump. Furthermore, the outer accretion flow is likely directly irradiated by the X-ray continuum, given the strong and delayed UV response \citepalias{Lawther2023}, which suggests a lack of absorbing material in the central engine. Considering also the intermittent accretion activity, we speculate that the nucleus may be gas starved to some degree.
    \item The X-ray spectral shape appears roughly constant as a function of continuum flux. Mrk 590 displays neither a disappearing soft excess (as observed for the CLAGN Mrk 1018) or hard continuum (as for the CLAGN 1ES 1927+654). The variety of X-ray behaviors observed among known CLAGN suggests that several distinct physical mechanisms may produce the observed UV--optical CLAGN phenomenon.
    \item The optical--UV data are broadly consistent with a warm-Comptonized emission component, with no direct emission from a thermal `thin-disk' required by the data. Due to host galaxy subtraction issues, this result is less robust at the lowest flux levels observed; however, as our X-ray model comparison strongly favors a warm-Comptonized component at all flux levels, it is likely present in all our data. We suggest a scenario where much of the accretion energy is dissipated in a warm, optically thick disk atmosphere. The seed photons for the Comptonized emission would then be produced in an underlying disk, which may be largely passive, heated radiatively by its atmosphere. This is compatible with our finding of weak inner-disk reflection from a highly ionized medium. It also explains the $\sim$3-day delayed X-ray to UV response observed during recent flaring episodes: if the inner disk is obscured by an optically thick plasma, the strongest UV response to X-ray illumination may occur in more distant reprocessing regions.
\end{itemize}

\noindent In future work we will explore these scenarios by way of \emph{i)} an in-depth study of the evolution of the optical--UV--X-ray spectral energy distribution since 2012, \emph{ii)} further timing experiments, currently in-progress, to determine the geometry of the inner regions of Mrk 590.

\paragraph*{Data availability:} The astronomical observations studied here are available from NASA's HEASArc archive; the tools used to process and analyze the data, as described in the text, are all freely available. Ancillary code used to produce figures, etc., is available from DL on request.

\paragraph*{Acknowledgements:} We thank the referee for a thoughtful and useful report that improved the quality of this work; in particular for their suggestions regarding inclusion of the UV data. We are grateful to Giovanni Minuitti and coworkers for granting us early access to the \emph{XMM-Newton} spectra observed during 2020--2021, and for guidance on our observing strategy for subsequent observations. We thank Thomas Dauser for guidance regarding the various \textsc{relxill} model variants, and Aya Kubota for feedback on \textsc{agnsed} parameter limits. Much of the analysis presented in this paper relies on the \textsc{HEASoft}, \textsc{Ftools}, and \textsc{XSPEC} software packages and related online resources. DL acknowledges financial support from NASA through Guest Observer proposals 8233/21-NUSTAR21-0025, 6238/19-NUSTAR19-0028, 5252/18-NUSTAR18-0042, and 5167/21-NICER21-0015. SIR acknowledges support from the Science and Technology Facilities Council (STFC) of the UK Research and Innovation via grant reference ST/Y002644/1. This work was supported by the Independent Research Fund Denmark via grants DFF-4002-00275 and DFF-8021-0013 and the Carlsberg Foundation via grant CF21-0649.

%% file: tables/xray_fit_tables.tex
\begin{deluxetable}{lccccc}
	\tabletypesize{\normalsize}
	\tablewidth{0pt}
	\tablecaption{X-ray observation log, and 0.3-10 keV fluxes \label{tab:observationlog}}
	\tablehead{
	\colhead{Telescope} & \colhead{Observation} & \colhead{Observation} & \colhead{Exposure} & \colhead{$F_{\mathrm{0.3-10}}$} & \colhead{Joint}\\
	\colhead{} & \colhead{date} & \colhead{ID} & \colhead{time [s]} & \colhead{[10$^{-12}$ erg cm$^{-2}$ s$^{-1}$]} & \colhead{data set}\\
	\colhead{(1)} & \colhead{(2)} & \colhead{(3)} & \colhead{(4)} & \colhead{(5)} & \colhead{(6)}
	}
	\startdata
    \emph{XMM-Newton} & 2002-01-01 & 0109130301 & 11349 & 8.6 & \\
        & 2004-07-04 & 0201020201 & 112674 & 11.3 \\
        & 2020-07-04 & 0865470201 & 27000 & 10.8 \\
        & 2021-01-03 & 0865470301 & 27000 & 12.6 & J21 \\
        & 2021-08-11 & 0870840101, 0870840401& 17998 & 18.8 & HF \\
        %& 2021-08-13 & 0870840401 & 7198 &  & \\
        & 2022-01-24 & 0870840201 & 25800 & 3.8 & LF \\
        & 2022-07-28 & 0912400101 & 27000 & 9.3 \\
        & 2023-02-06 & 0870840301 & 38000 & 6.8 & LF; F23 \\
        & 2024-01-25 & 0912400201 & 27000 & 21.5 & HF \\
        \hline
    \emph{NuSTAR} & 2016-02-05 & 60160095002 & 21205 & 4.9 & LF \\
	    & 2016-12-02 & 90201043002 & 51001 & 5.0 & LF \\
	    & 2018-10-27 & 80402610002 & 21069 & 14.8 &  \\
	    & 2019-08-31 & 80502630002 & 68123 & 21.8 & HF \\
	    & 2020-01-21 & 80502630004 & 50168 & 20.9 & HF \\
        & 2021-01-10 & 80502630006 & 41517 & 10.9 & J21\\
        & 2021-08-18 & 60761012002 & 18649 & 22.1 & HF \\
        & 2021-12-22 & 80602604002 & 53311 & 8.4 \\
        & 2023-02-06 & 80602604004 & 40977 & 8.0 & F23 \\
        & 2023-09-29 & 80802652002 & 20825 & 5.5 & LF \\
        & 2024-01-25 & 80802652004 & 62046 & 24.7 & HF \\
    \hline
        \emph{NICER} & 2022-07-16 & 5667010401 & 1743 & 9.6 \\
        & 2022-07-26 & 5667010501 & 3371 & 10.4 \\
        & 2022-08-15 & 5667010702 & 2052 & 14.7 \\
        & 2022-08-24 & 5667010801 & 1015 & 8.0 \\
        & 2022-08-25 & 5667010802 & 2043 & 7.7 \\
        & 2022-09-14 & 5667011001 & 2636 & 9.9 \\
        & 2022-09-25 & 5667011101 & 2522 & 8.7 \\
        & 2022-10-04 & 5667011201 & 1786 & 17.6 \\
        & 2022-10-24 & 5667011401 & 1917 & 6.1 \\
        & 2022-11-12 & 5667011601 & 1815 & 6.2 \\
        & 2022-12-03 & 5667011801 & 1505 & 5.2 \\
        %& 2022-12-13 & 5667011901 & 2220 & 4.5 \\
        & 2023-01-08 & 5667012101 & 3498 & 5.0 \\
        %& 2023-01-09 & 5667012102 & 1843 & 3.5 \\
        %& 2023-01-20 & 5667012402 & 1594 & 5.5 \\
        %& 2023-02-01 & 5667012501 & 1662 & 5.7 \\
	\enddata
	\tablecomments{All parameter uncertainties are quoted at the 90th percentile confidence interval. (1) Telescope name. (2) Date of observation start, YYYY-MM-DD. (3) Observation identifier in the \emph{HEASarc} archive. (4) On-source integration time. This is measured before event screening. (5) Integrated 0.3--10 keV model flux, determined using the phenomenological model (\S \ref{sec:analysis_bbody}). As \emph{NuSTAR} is not sensitive below 3 keV, the best-fit model is extrapolated, to facilitate comparison with the other telescopes. (6) Data set(s) that this observation is included in for the joint analysis described in \S \ref{sec:analysis_joint}.}
\end{deluxetable}

\begin{deluxetable}{lccccccc}
	\tabletypesize{\small}
	\tablewidth{0pt}
	\tablecaption{Phenomenological modeling of soft excess and Iron emission in individual spectra\label{tab:bbody_gauss}}
	\tablehead{
	\colhead{Telescope} & \colhead{Observation} & \colhead{$\Gamma$} & 
 $F_{\mathrm{BB}}$ & \colhead{$kT_{\mathrm{BB}}$} & \colhead{$\sigma_{\mathrm{line}}$} & \colhead{$EW$} & 
	\colhead{$F_{\mathrm{line}}$} \\ & \colhead{date} & & \colhead{[$10^{-13}$erg cm$^{-2}$s$^{-1}$]} & \colhead{[keV]} & \colhead{[eV]} & \colhead{[eV]} & \colhead{[$10^{-13}$ erg cm$^{-2}$s$^{-1}$]} \\
	\colhead{(1)} & \colhead{(2)} & \colhead{(3)} & \colhead{(4)} & \colhead{(5)} & \colhead{(6)} & \colhead{(7)} & \colhead{(8)} 
	}
	\startdata
    \emph{XMM-Newton} & 2002-01-01 & $1.71\pm0.03$ & $5.5^{+1.7}_{-1.6}$ & $147^{+13}_{-14}$ & --- & $586_{-202}^{+298}$ & $3.2_{-1.2}^{+0.9}$\\
    & 2004-07-04 & $1.67\pm0.02$ & $7.2\pm0.8$ & 139$_{-6}^{+5}$ & $<68$ & 116$_{-32}^{+34}$ & $0.9\pm0.2$ \\
        & 2020-07-04 & $1.69\pm0.02$ & $7.1\pm1.1$ & 147$\pm8$ & $<115$ & 140$^{+47}_{-44}$ & $1.0_{-0.3}^{+0.4}$ \\
        & 2021-01-03 & $1.69\pm0.02$ & $8.4\pm1.2$ & 140$_{-7}^{+6}$ & $<141$ & 152$\pm46$ & $1.4_{-0.4}^{+0.4}$\\
        & 2021-08-11 & $1.78\pm0.02$ & $14.1_{-2.5}^{+2.3}$ & 133$\pm7$ & $<75$ & 62$_{-48}^{+44}$ & $0.8_{-0.5}^{+1.7}$\\
        & 2022-01-24 & $1.67\pm0.04$ & $2.8\pm0.6$ & 129$_{-13}^{+12} $ & 187$_{-64}^{+115}$ & 496$_{-141}^{+148}$ & $1.4_{-0.3}^{+0.4}$ \\
        & 2022-07-28 & $1.65\pm0.02$ & $6.1\pm1.0$ & 147$\pm8$ & $<203$ & 111$_{-51}^{+50}$ & $ 0.7_{-0.3}^{+0.5}$\\
        & 2023-02-06 & $1.64\pm0.02$ & $5.1\pm0.7$ & 141$\pm7$ & $129_{-48}^{+59}$ & 241$_{-64}^{+61}$ & $1.2\pm0.3$\\
        & 2024-01-25 & $1.77\pm0.02$ & $12.2_{-1.5}^{+1.8}$ & $124\pm6$ & --- & $195_{-71}^{+106}$ & $2.7\pm0.9$\\
    \hline
    \emph{NuSTAR} & 2016-02-05 & $ 1.59_{-0.07}^{+0.08}$ & --- & --- & $<756$ & 216$_{-199}^{+182}$ & $0.8_{-0.8 }^{+0.7}$ \\
	    & 2016-12-02 & $1.61\pm0.05$ & --- & --- & 398$_{-192}^{+205}$ & 357$_{-138}^{+154}$ & $1.3_{-0.5 }^{+0.5 }$\\
	    & 2018-10-27 & $1.61\pm0.04$ & --- & --- & $<434$ & $300_{-237}^{+273}$ & $ 3.4_{-2.1 }^{+1.8 }$\\
	    & 2019-08-31 & $1.66\pm0.02$ & --- & --- & $<201$ & 101$_{-32}^{+35}$ & $ 1.7_{-0.5}^{+0.6}$\\
	    & 2020-01-21 & $1.67\pm0.02$ & --- & --- & 377$_{-200}^{+186}$ & 194$_{-68}^{+61}$ & $ 2.5_{-0.9}^{+1.2}$\\
        & 2021-01-10 & $1.67\pm0.04$ & --- & --- & $<360$ & 190$_{-78}^{+82}$ & $ 1.5_{-0.6}^{+0.7}$ \\
        & 2021-08-18 & $1.69\pm0.03$ & --- & --- & $<488$ & 165$\pm64$ & $2.8_{-1.1}^{+1.5}$ \\
        & 2021-12-22 & $1.65\pm0.03$ & --- & --- & $<437$ & 186$_{-63}^{+64}$ & $1.2_{-0.4}^{+0.5}$ \\
        & 2023-02-06 & $1.66\pm0.04$ & --- & --- & $<444$ & 215$_{-82}^{+142}$ & $1.2_{-0.4}^{+0.6}$ \\
        & 2023-09-29 & $1.68\pm0.07$ & --- & --- &$<416$ & $176_{-113}^{+316}$ & $0.7_{-0.6}^{+0.7}$ \\
        & 2024-01-25 & $1.72\pm0.02$ & --- & --- & --- & $79_{-39}^{+5}$ & $1.0_{-0.3}^{+0.8}$\\
        
    \hline
    \emph{NICER} & 2022-07-16 & --- & $3.6_{-1.9}^{+2.2}$ & $208_{-74}^{+86}$ & --- & --- & --- \\
        & 2022-07-26 & --- & $6.7_{-1.7}^{+1.8}$ & $209_{-27}^{+24}$ & --- & --- & --- \\
        & 2022-08-15 & --- & $11.0_{-1.7}^{+1.9}$ & $155_{-16}^{+17}$ & --- & --- & --- \\
        & 2022-08-24 & --- & $5.2_{-2.0}^{+2.1}$ & $169_{-39}^{+43}$ & --- & --- & --- \\
        & 2022-08-25 & --- & $6.5_{-1.6}^{+1.7}$ & $193_{-28}^{+32}$ & --- & --- & --- \\
        & 2022-09-14 & --- & $3.2_{-1.4}^{+1.1}$ & $143_{-29}^{+28}$ & --- & --- & --- \\
        & 2022-09-25 & --- & $4.0_{-1.5}^{+1.7}$ & $182_{-31}^{+33}$ & --- & --- & --- \\
        & 2022-10-04 & --- & $5.3_{-3.2}^{+3.2}$ & $234_{-67}^{+59}$ & --- & --- & --- \\
        & 2022-10-24 & --- & $4.1_{-1.7}^{+1.9}$ & $200_{-35}^{+32}$ & --- & --- & --- \\
        & 2022-11-12 & --- & $1.5_{-1.0}^{+1.3}$ & $139_{-44}^{+69}$ & --- & --- & --- \\
        & 2022-12-03 & --- & $2.2_{-1.6}^{+2.0}$ & $216_{-68}^{+61}$ & --- & --- & --- \\
        %& 2022-12-13 & --- & \\
        & 2023-01-08 & --- & $<2.1$ & --- & --- & --- & --- \\
        %& 2023-01-09 & --- & \\
        %& 2023-01-20 & --- & \\
        %& 2023-02-01 & --- & \\
	\enddata
	\tablecomments{Here, we list best-fit parameters for the phenomenological model \textsc{tbabs*(powerlaw+zgauss+zbbody)} (\S \ref{sec:analysis_bbody}). All parameter uncertainties are quoted at the 90th percentile confidence interval. (1) Telescope name. (2) Date of observation start, YYYY-MM-DD. (3) Continuum photon index. As the model is fitted to the full instrumental energy range, the spectral slope may vary systematically between instruments, even if it is not time-variant. We find that \emph{NICER} cannot robustly recover $\Gamma$ for these short individual exposures (\S \ref{sec:analysis_bbody}) (4) Integrated flux of the blackbody component; this is only included for \emph{XMM-Newton} and \emph{NICER} spectra. (5) Temperature of the blackbody component. (6) Line width of the 6.4 keV Gaussian emission component. Most of these are upper limits at the 90\% confidence interval; the nominal energy resolutions near the Fe K line are 400 eV for \emph{NuSTAR} and 150 eV for \emph{XMM-Newton pn}. Thus, the emission lines are largely unresolved. (7) Equivalent width of the Gaussian component. (8) Integrated flux of the Gaussian component.}
\end{deluxetable}

\begin{deluxetable}{cccccc}
	\tabletypesize{\small}
	\tablewidth{0pt}
	\tablecaption{Best-fitting parameters for our preferred models \label{tab:preferred_models}}
	\tablehead{
    \colhead{Row} & \colhead{Parameter} & \colhead{LF} & \colhead{HF} & \colhead{J21} & \colhead{F23} \\
    }
    \startdata
    (1) & $N_{\mathrm{H,Gal}}$ & \textcolor{Gray}{2.77$\times10^{20}$} & \textcolor{Gray}{2.77$\times10^{20}$} & \textcolor{Gray}{2.77$\times10^{20}$} & \textcolor{Gray}{2.77$\times10^{20}$}  \\
    (2) & $Z$ & \textcolor{Gray}{1} & \textcolor{Gray}{1} & \textcolor{Gray}{1} & \textcolor{Gray}{1} \\ 
    (3) & $i$ & 38$^\circ\pm7^\circ$ & \textcolor{Gray}{38$^\circ$} & \textcolor{Gray}{38$^\circ$} & \textcolor{Gray}{38$^\circ$} \\ 
    (4) &$A_{\mathrm{Fe}}$ &  6.9$_{-3.3}^{+9.0}$ & \textcolor{Gray}{6.9} & \textcolor{Gray}{6.9} & \textcolor{Gray}{6.9}\\ 
    (5) &$z$ & \textcolor{Gray}{0.026385}  & \textcolor{Gray}{0.026385} & \textcolor{Gray}{0.026385} & \textcolor{Gray}{0.026385} \\ 
    \hline
    \multicolumn{6}{c}{\textbf{Distant reflector only,} \textsc{const}$\times$\textsc{tbabs}(\textsc{zcutoffpl+nthcomp+pexmon})} \\
     & \emph{Continuum}\\
    (6) & N$_{\mathrm{cont}}$ & 7.5$(\pm0.2)\times10^{-4}$ & 3.0$(\pm0.4)\times10^{-3}$ & 1.7$(\pm0.1)\times10^{-3}$ & 9.2$(\pm0.4)\times10^{-4}$ \\ 
    (7) & $\Gamma$ & 1.57$_{-0.03}^{+0.02}$ & 1.66$\pm0.01$ & 1.59$^{+0.04}_{-0.05}$ & 1.58$_{-0.05}^{+0.04}$\\ 
    (8) & $E_{\mathrm{cut}}$ & 142$\pm43$ & $>255$ & 85$_{-26}^{+56}$ & $84_{-26}^{+50}$ \\
     & \emph{Soft excess}\\
    (9) & N$_{\mathrm{warm}}$ & 1.7$(\pm0.2)\times10^{-4}$ & 5.9($\pm0.4)\times10^{-4}$ & 4.6($\pm0.9)\times10^{-4}$ & 2.0($\pm0.4)\times10^{-4}$ \\ 
    (10) & $kT_e$ & 0.21$\pm0.03$ & 0.28$\pm0.03$ & 0.24$^{+0.05}_{-0.04}$ & 0.20$_{-0.02}^{+0.04}$ \\ 
    (11) & $\Gamma_{\mathrm{warm}}$ & 1.90$_{-0.17}^{+0.14}$ & 2.23$\pm0.08$ & 2.01$_{-0.23}^{+0.17}$ & 1.83$\pm0.23$ \\
     & \emph{Distant refl.}\\
    (12) & $R$ & -0.38$_{-0.12}^{+0.09}$ & -0.14$\pm0.02$ & -0.25$\pm0.05$ & -0.34$\pm0.06$ \\
     & \emph{Instrumental}\\
    (13) & $C_{\mathrm{MOS1}}$ & 1.03$\pm0.01$ & 1.05$\pm0.01$ & 0.98$\pm0.01$ & 1.01$\pm0.01$ \\
    (14) & $C_{\mathrm{MOS2}}$ & 1.04$\pm0.01$ & 1.03$\pm0.01$ & 0.97$\pm0.01$ & 1.03$\pm0.01$ \\
    (15) &$C_{\mathrm{FPMA}}$ & 0.87$\pm0.01$ & 1.15$\pm0.01$ & 0.84$\pm0.01$ & 1.13$\pm0.01$ \\
    (16) &$C_{\mathrm{FPMB}}$ & 0.88$\pm0.01$ & 1.19$\pm0.01$ & 0.87$\pm0.01$ & 1.16$\pm0.01$\\
    \hline
    \multicolumn{6}{c}{\textbf{Dual reflectors,} \textsc{const}$\times$\textsc{tbabs}(\textsc{zcutoffpl+nthcomp+pexmon+relxillLp})} \\
     & \emph{Continuum}\\
     & N$_{\mathrm{cont}}$ & 7.3$(\pm0.2)\times10^{-4}$ & 2.9$(\pm0.1)\times10^{-3}$ & 1.7$(\pm0.1)\times10^{-3}$ & 9.2$(\pm0.4)\times10^{-4}$ \\ 
     & $\Gamma$ & 1.57$\pm0.03$ &1.66$\pm0.02$ & 1.59$\pm0.04$ & 1.59$_{-0.05}^{+0.04}$ \\ 
     & $E_{\mathrm{cut}}$ & 166$_{-57}^{+154}$ & $>242$ & 80$_{-23}^{+50}$ & 79$_{-24}^{+47}$ \\
     &  \emph{Soft excess}\\
     & N$_{\mathrm{warm}}$ & 1.1($\pm0.3)\times10^{-4}$ & 5.4($\pm0.6)\times10^{-4}$ & 4.1$(\pm0.9)\times10^{-4}$ & 1.9$(\pm0.5)\times10^{-4}$ \\ 
     & $kT_e$ & 0.20$\pm0.04$ &  0.26$\pm0.02$ & 0.24$_{-0.04}^{+0.06}$ & 0.21$_{-0.03}^{+0.05}$ \\ 
     & $\Gamma_{\mathrm{warm}}$ & 1.98$_{-0.25}^{+0.20}$ & 2.27$_{-0.10}^{+0.11}$ & 2.02$_{-0.22}^{+0.19}$ & 1.91$\pm0.24$ \\
     & \emph{Distant refl.}\\
     & $R$ & -0.34$\pm0.05$ & -0.12$\pm0.02$ & -0.24$\pm0.04$ & -0.30$\pm0.06$ \\
     & \emph{Rel. refl.}\\
    (17) & $N_{\mathrm{rel}}$ & 7.2$(\pm6.4)\times10^{-6}$ & 6.4$(\pm4.0)\times10^{-6}$ & 4.2$(_{-4.2}^{+10.1})\times10^{-6}$ & 1.7$\pm0.8$ \\
    (18) & $R_{\mathrm{rel}}$ & \textcolor{Gray}{1} & \textcolor{Gray}{1} & \textcolor{Gray}{1} & \textcolor{Gray}{1} \\ 
    (19) & $\log(\xi)_\mathrm{rel}$ & 2.8$\pm0.3$ & 3.2$\pm0.2$ & 2.3$_{-2.3}^{+0.4}$ & 1.7$_{-1.4}^{+1.1}$ \\ 
    (20) & $h_{\mathrm{rel}}$ & \textcolor{Gray}{10} & \textcolor{Gray}{10} & \textcolor{Gray}{10} & \textcolor{Gray}{10} \\ 
    (21) & $\beta_{\mathrm{rel}}$ & \textcolor{Gray}{0} & \textcolor{Gray}{0} & \textcolor{Gray}{0} & \textcolor{Gray}{0} \\
    %$n_\mathrm{rel}$ &  \\ 
     & \emph{Instrumental}\\
     & $C_{\mathrm{MOS1}}$ & 1.03$\pm0.01$ & 1.05$\pm0.01$ & 0.98$\pm0.01$ & 1.01$\pm0.01$ \\
     & $C_{\mathrm{MOS2}}$ & 1.02$\pm0.01$ & 1.02$\pm0.01$ & 1.00$\pm0.01$ & 1.01$\pm0.01$ \\
     & $C_{\mathrm{FPMA}}$ & 0.87$\pm0.01$ & 1.15$\pm0.01$ & 0.84$\pm0.02$ & 1.13$\pm0.02$ \\
     & $C_{\mathrm{FPMB}}$ & 0.89$\pm0.01$ & 1.19$\pm0.01$ & 0.87$\pm0.02$ & 1.16$\pm0.02$ \\
    \enddata
    \tablecomments{Best-fit model parameters for our preferred models (\S \ref{sec:bestmodels}), for the low-flux (LF), high-flux (HF), January 2021 (J21), and February 2023 (F23) data sets. The uncertainties correspond to 90\% confidence intervals. Parameters listed in \textcolor{Gray}{gray} are held constant during the model fit procedure. In particular, the reflection component iron abundance $A_{\textrm{Fe}}$ and inclination $i$ are only constrained for LF; we use the best-fit LF values for all other models. (1) Column density for Galactic absorption, units of cm${^{-2}}$. (2) Metallicity relative to Solar. (3) Inclination angle of reflection slab relative to line-of-sight; for the dual-reflector model, the same inclination is used for both reflection components. (4) Iron abundance relative to Solar. (5) Source redshift. (6) Normalization at 1 keV for cutoff power law continuum. (7) Continuum photon index. (8) Continuum cutoff energy in keV. (9) Normalization for warm Comptonization component. (10) Electron temperature of Comptonizing region in keV. (11) Photon index of warm Comptonization component. (12) Reflection fraction for distant reflection, where $R=1$ corresponds to an infinite Compton-thick slab. (13--16) Cross-calibration constants for the MOS1, MOS2, FPMA and FPMB detectors, relative to EPIC \emph{pn}. (17) Normalization of the relativistic reflection component, as defined by \citet{Dauser2022}. (18) Reflection strength of the relativistic reflection, as defined by \citet{Dauser2022}; we hold this constant and allow the normalization to vary. (19) Ionization parameter at the innermost stable circular orbit for the relativistic reflection component. (20) X-ray continuum height above the disk, in gravitational radii. (21) X-ray continuum source velocity, relative to the disk; we assume a stationary continuum source. }
\end{deluxetable}

\begin{deluxetable}{cccccc}
	\tabletypesize{\small}
	\tablewidth{0pt}
	\tablecaption{Best-fitting parameters for the optical--UV--X-ray model, AGNSED \label{tab:agnsed}}
	\tablehead{
    \colhead{Row} & \colhead{Parameter} & \colhead{LF} & \colhead{HF} & \colhead{J21} & \colhead{F23} \\
    }
    \startdata

    (1) & $N_{\mathrm{H,Gal}}$ & \textcolor{Gray}{2.77$\times10^{20}$} & \textcolor{Gray}{2.77$\times10^{20}$} & \textcolor{Gray}{2.77$\times10^{20}$} & \textcolor{Gray}{2.77$\times10^{20}$}  \\
    (2) & $Z$ & \textcolor{Gray}{1} & \textcolor{Gray}{1} & \textcolor{Gray}{1} & \textcolor{Gray}{1} \\ 
    (3) &$A_{\mathrm{Fe}}$ &  6.9$_{-3.3}^{+9.0}$ & \textcolor{Gray}{6.9} & \textcolor{Gray}{6.9} & \textcolor{Gray}{6.9}\\ 
    (4) &$z$ & \textcolor{Gray}{0.026385}  & \textcolor{Gray}{0.026385} & \textcolor{Gray}{0.026385} & \textcolor{Gray}{0.026385} \\ 
    \hline
     & \textsc{agnsed} \\
    (5) & $M_{\mathrm{BH}}$ & \textcolor{Gray}{$3.7\times10^7$} & \textcolor{Gray}{$3.7\times10^7$} & \textcolor{Gray}{$3.7\times10^7$} & \textcolor{Gray}{$3.7\times10^7$} \\
    (6) & $D$ & \textcolor{Gray}{112} & \textcolor{Gray}{112} & \textcolor{Gray}{112} & \textcolor{Gray}{112} \\
    (7) & $\log(\dot{m})$ & -2.05$\pm0.06$ & -1.57$\pm0.01$ & -1.72$\pm0.02$ & -1.99$\pm0.04$ \\
    (8) & $a_{*}$ & \textcolor{Gray}{0} & \textcolor{Gray}{0} & \textcolor{Gray}{0} & \textcolor{Gray}{0} \\
    (9) & $\cos(i)$ & \textcolor{Gray}{0.788} & \textcolor{Gray}{0.788} & \textcolor{Gray}{0.788} & \textcolor{Gray}{0.788} \\
    (10) & $kT_{\mathrm{hot}}$ & \textcolor{Gray}{300} & \textcolor{Gray}{300} & \textcolor{Gray}{300} & \textcolor{Gray}{300} \\
    (11) & $kT_{\mathrm{warm}}$ & 0.25$\pm0.02$ & 0.25$\pm0.03$ & 0.30$\pm0.03$ & 0.26$\pm0.03$ \\
    (12) & $\Gamma_{\mathrm{hot}}$ & 1.65$\pm0.06$ & 1.73$\pm0.03$ & 1.70$\pm0.03$ & 1.68$\pm0.09$ \\
    (13) & $\Gamma_{\mathrm{warm}}$ & 2.23$\pm0.32$ & 2.29$\pm0.39$ & 2.45$\pm0.16$ & 2.38$\pm0.30$ \\
    (14) & $R_{\mathrm{hot}}$ & 168$\pm35$ & 124$\pm47$ & 104$\pm8$ & 127$\pm24$ \\
    (15) & $R_{\mathrm{warm}}$ & $4.7(\pm2.8)\times10^4$ & $5.7(\pm2.6)\times10^4$ & $4.3(\pm2.9)\times10^4$ & $6.0(\pm2.8)\times10^4$ \\
    (16) & $\log R_{\mathrm{out}}$ & \textcolor{Gray}{$\equiv R_{\mathrm{warm}}$} & \textcolor{Gray}{$\equiv R_{\mathrm{warm}}$} & \textcolor{Gray}{$\equiv R_{\mathrm{warm}}$} & \textcolor{Gray}{$\equiv R_{\mathrm{warm}}$} \\ 
    (17) & $h_{\mathrm{max}}$ & \textcolor{Gray}{10} & \textcolor{Gray}{10} & \textcolor{Gray}{10} & \textcolor{Gray}{10} \\
    \hline
     & \textsc{pexmon} \\
    (18) & N$_{\mathrm{cont}}$ & $3.4(\pm1.0)\times10^{-4}$ & 4.2$(\pm4.0)\times10^{-4}$ & $(5.5\pm1.1)\times10^{-4}$ & $(3.8_{-3.8}^{+8.2})\times10^{-4}$ \\ 
    \hline
     & \emph{Instrumental}\\
    (19) & $C_{\mathrm{MOS1}}$ & 1.03$\pm0.01$ & 1.05$\pm0.01$ & 0.99$\pm0.01$ & 1.01$\pm0.01$ \\
    (20) & $C_{\mathrm{MOS2}}$ & 1.04$\pm0.01$ & 1.03$\pm0.01$ & 0.99$\pm0.01$ & 1.03$\pm0.01$ \\
    (21) &$C_{\mathrm{FPMA}}$ & 0.89$\pm0.04$ & 1.18$\pm0.06$ & 0.84$\pm0.01$ & 1.14$\pm0.01$ \\
    (22) &$C_{\mathrm{FPMB}}$ & 0.90$\pm0.03$ & 1.20$\pm0.03$ & 0.87$\pm0.01$ & 1.17$\pm0.01$\\
    (23) &$C_{\mathrm{UVOT}}$ & \textcolor{Gray}{1} & \textcolor{Gray}{1} & \textcolor{Gray}{1} & \textcolor{Gray}{1} \\
    \enddata
    \tablecomments{Best-fit parameters for our optical--UV--X-ray model based on AGNSED (\S \ref{sec:discussion_SED}), for the low-flux (LF) and high-flux (HF) data sets. The uncertainties correspond to 1$\sigma$ intervals derived from the MCMC posterior distributions. Parameters listed in \textcolor{Gray}{gray} are held constant during the model fit procedure.  (1) Column density for Galactic absorption, units of cm${^{-2}}$. (2) Metallicity relative to Solar. (3) Iron abundance relative to Solar. (4) Source redshift. (5) Black hole mass in units of Solar mass, as estimated via reverberation mapping \citep{Peterson2004}. (6) Co-moving distance in Mpc; as the redshift is small ($z=0.026385$) we set this to the luminosity distance. (7) Mass accretion rate, scaled by the Eddington accretion rate. (8) Black hole rotation parameter. (9) Inclination angle; here, we constrain the \textsc{agnsed} and \textsc{pexmon} inclinations to the best-fit value from our X-ray analysis. (10) Electron temperature in the hot Comptonizing region, units of keV. (11) Electron temperature in the warm Comptonizing region. (12) Photon index for hot Comptonized emission. (13) Photon index for warm Comptonized emission. (14) Radius of the hot Comptonization region, in units of the gravitational radius $r_g$. (15) Radius of the warm Comptonization region. (16) Outer radius; for this model we do not include an outer disk, so it is equal to $R_{\mathrm{warm}}$. (19) Height above the disk of the `lamp-post' X-ray source that irradiates the disk. We set this to 10 $r_g$, as for the disk reflection model in our X-ray analysis. (18) Incident continuum normalization for the distant reflection component. For the optical--UV--X-ray modeling, as we are mainly interested in the overall spectral energy distribution, we set the \textsc{pexmon} reflection strength to $R=-1$ and use $N_{\mathrm{cont}}$ to scale the reflection spectrum. (19--23) Instrumental scaling factors relative to \emph{XMM-Newton pn}. We set $C_{\mathrm{UVOT}}\equiv1$, as we are testing whether the overall UV flux level can be reproduced by the soft excess model; allowing for UV to X-ray flux offsets would defeat this purpose. Instead, our use of averaged UVOT photometry should minimize the influence of luminosity variability. }
\end{deluxetable}

% The \textsc{pexmon} reflection strength $R$ is defined to be negative for the model construction described here: this causes \textsc{pexmon} to only return the reflected component, allowing the incident continuum to be modeled independently. Its meaning is otherwise unchanged: $R=-1$ corresponds to a semi-infinite reflecting slab. We quote $R$ values without the negative sign elsewhere in the text. The black hole spin $a_*$ for the dual reflector model is entirely unconstrained within the parameter bounds for all data sets; we set $a_*\equiv0$ in all model fits. Most parameters of the \textsc{relxillLp} component are poorly constrained due to the faintness of the relativistic reflection component; we do not attempt to fit the X-ray source height $h_{\mathrm{rel}}$ or velocity $\beta_{\mathrm{rel}}$. 

%% file: appendixA/appendixA.tex
\begin{figure*}
    \section{Individual X-ray spectra}\label{AppendixA}
    \centering
    \includegraphics[scale=0.52]{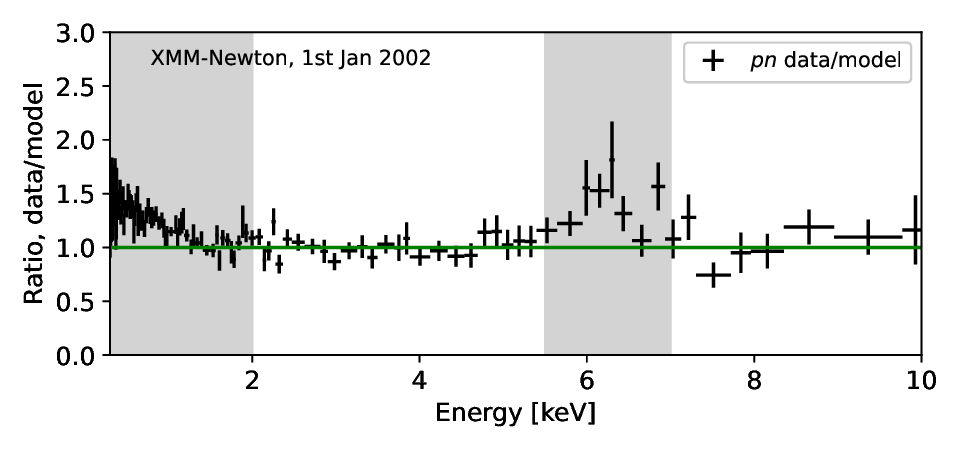}
    \includegraphics[scale=0.52]{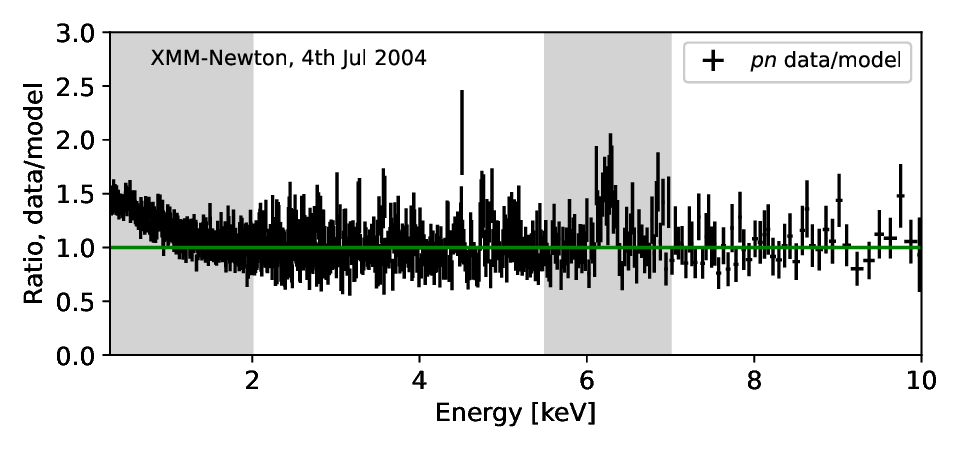}
    \includegraphics[scale=0.52]{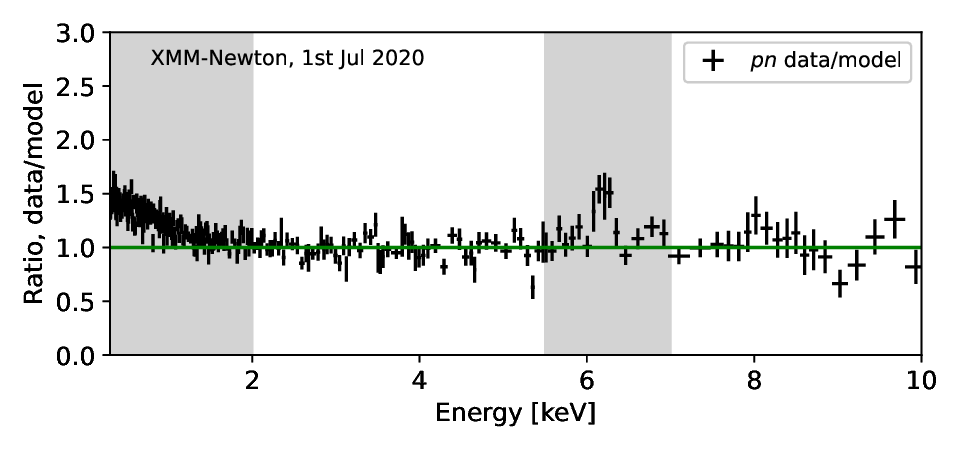}
    \includegraphics[scale=0.52]{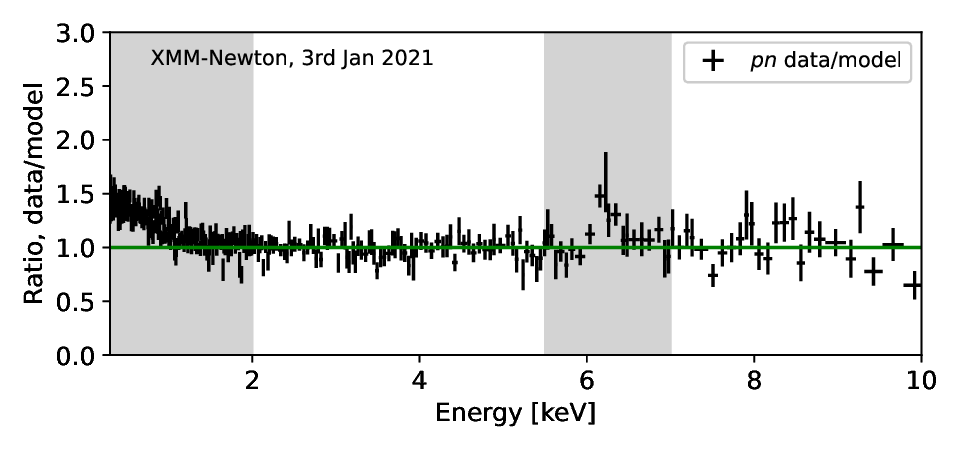}
    \includegraphics[scale=0.52]{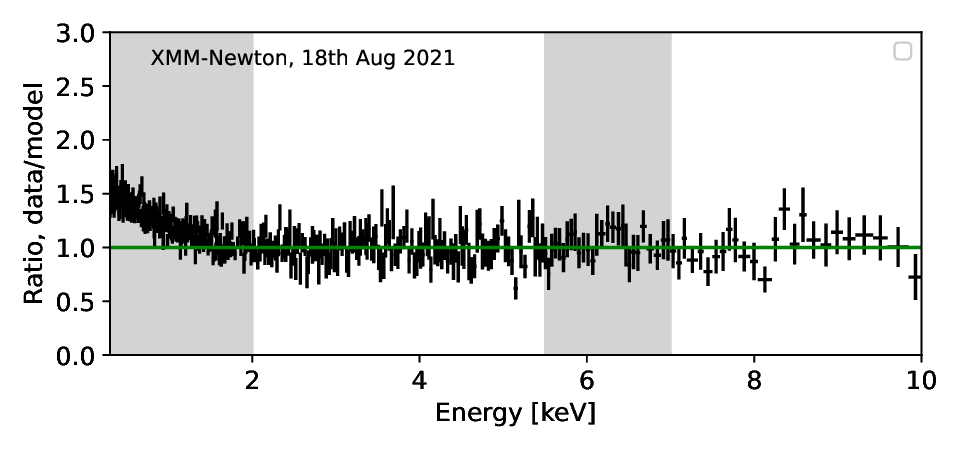}
    \includegraphics[scale=0.52]{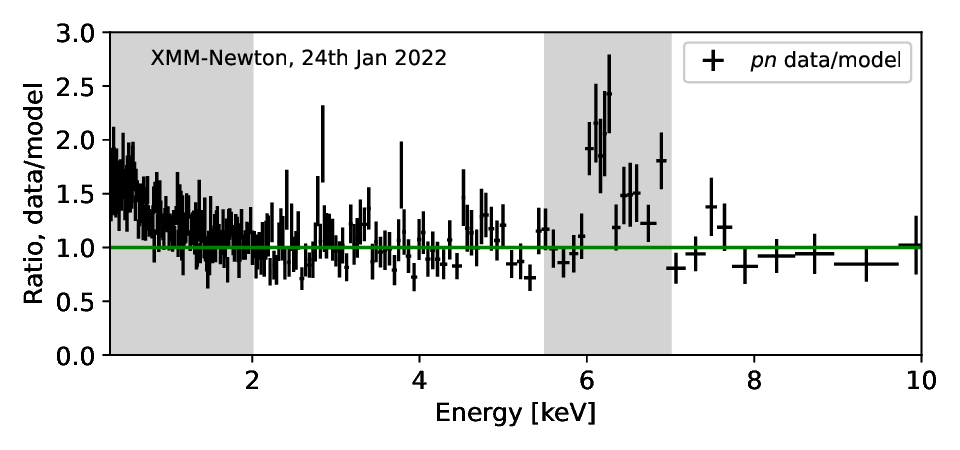}
    \includegraphics[scale=0.52]{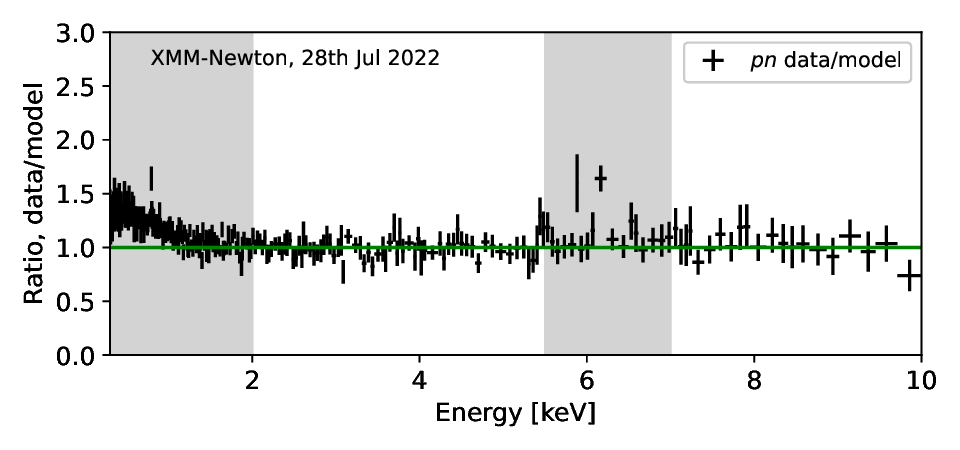}
    \includegraphics[scale=0.52]{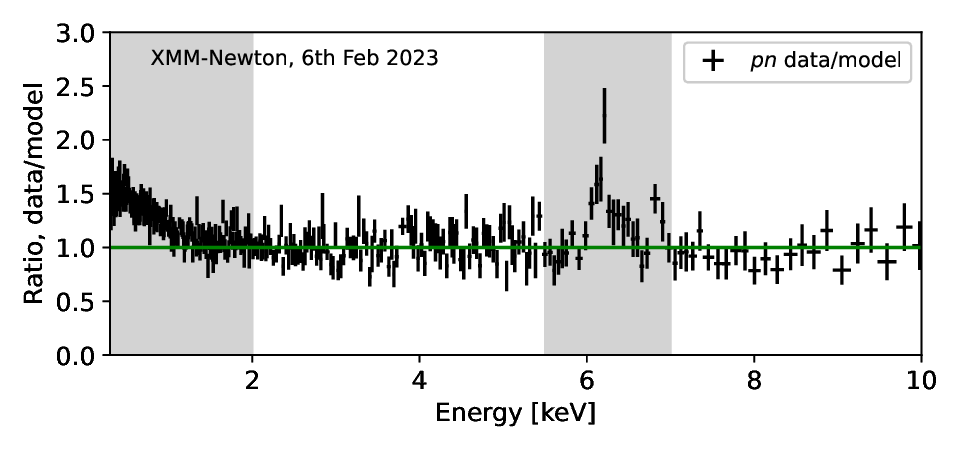}
    \begin{minipage}{.5\textwidth}\centering
    \,\,\,\,\,\,\includegraphics[scale=0.52]{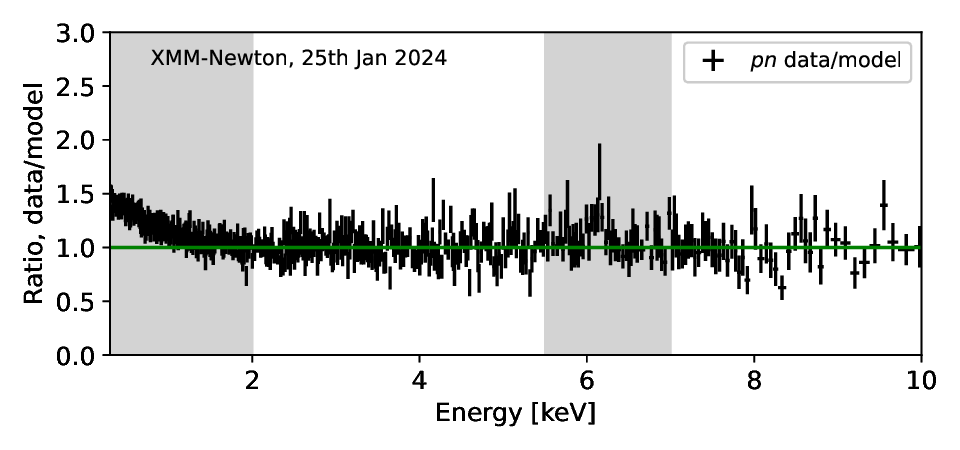}
    \end{minipage}\hfill
    \caption{Data-to-model ratios, against a power law model, for individual \emph{XMM-Newton} observations. All energies are given in the observed frame. The gray regions indicate spectral regions excluded from the model fit (\S \ref{sec:method_xspec}). A substantial soft excess above the extrapolated power law model is seen in all spectra. Excess emission near 6.4 keV is robustly detected in most cases, but appears rather weak for August 2021 and July 2022.}
    \label{fig:xmm_windowfits}
\end{figure*}

\begin{figure*}
    \centering
    \includegraphics[scale=0.49]{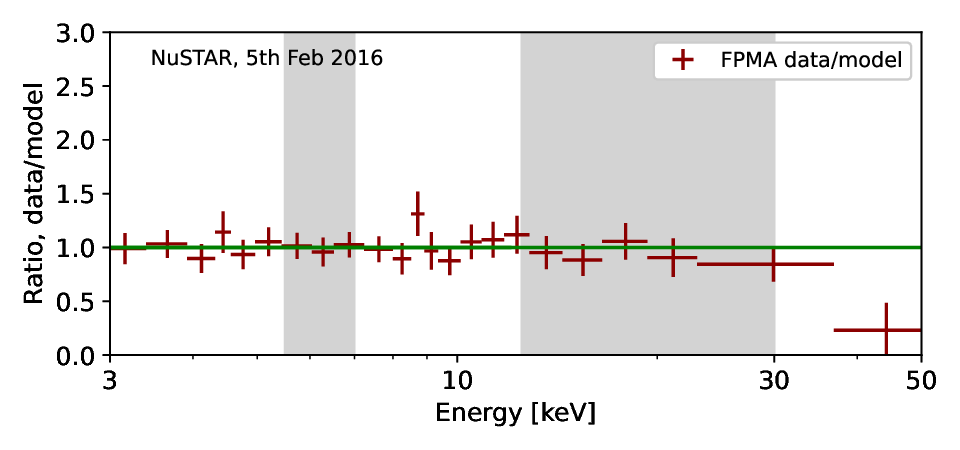}
    \includegraphics[scale=0.49]{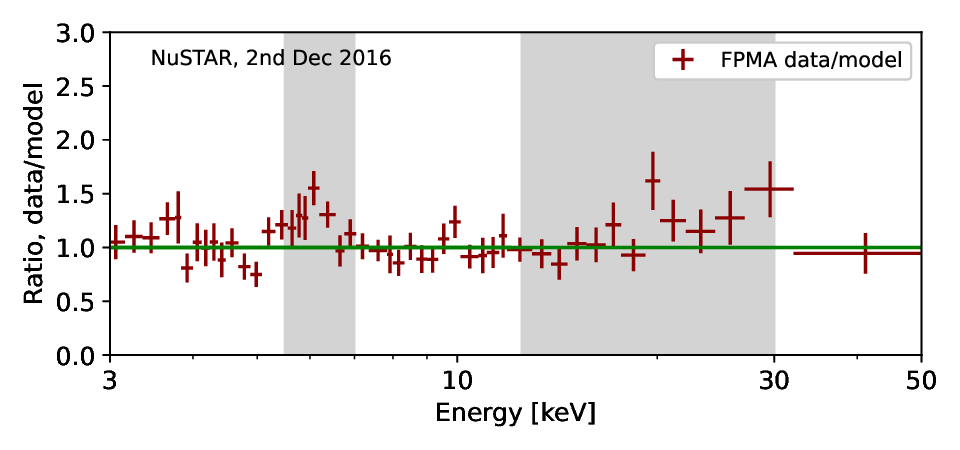}
    \includegraphics[scale=0.49]{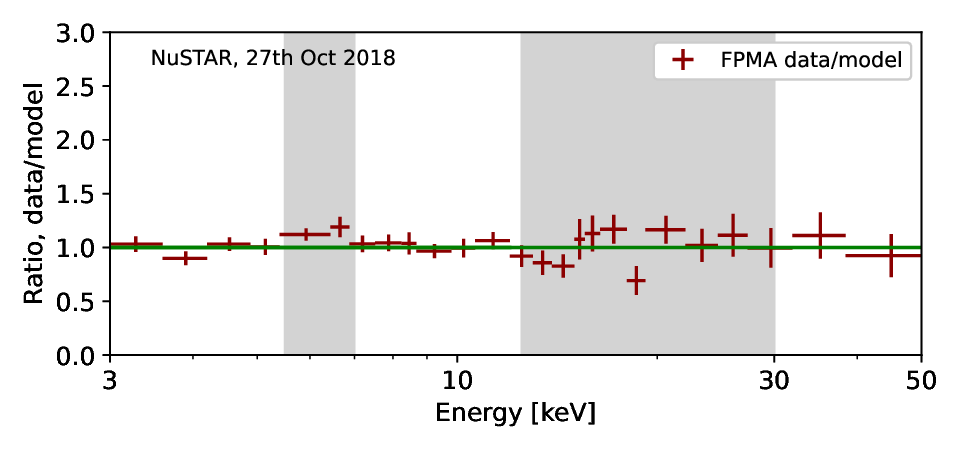}
    \includegraphics[scale=0.49]{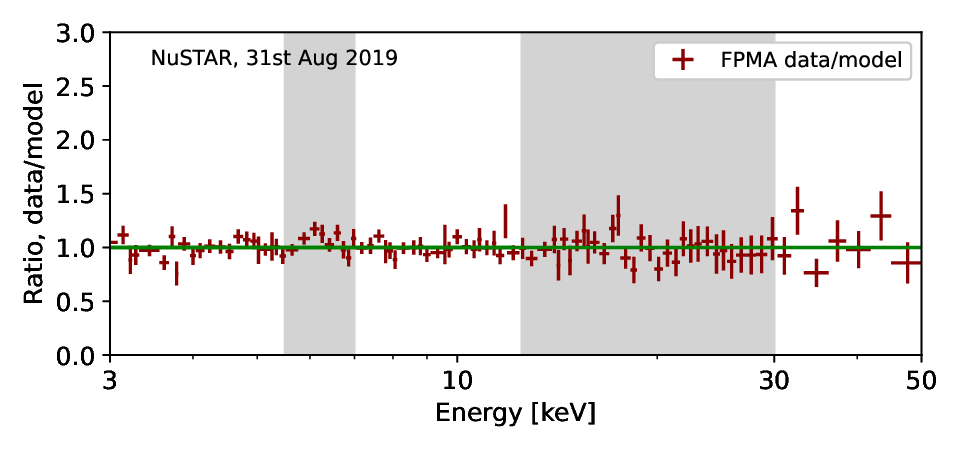}
    \includegraphics[scale=0.49]{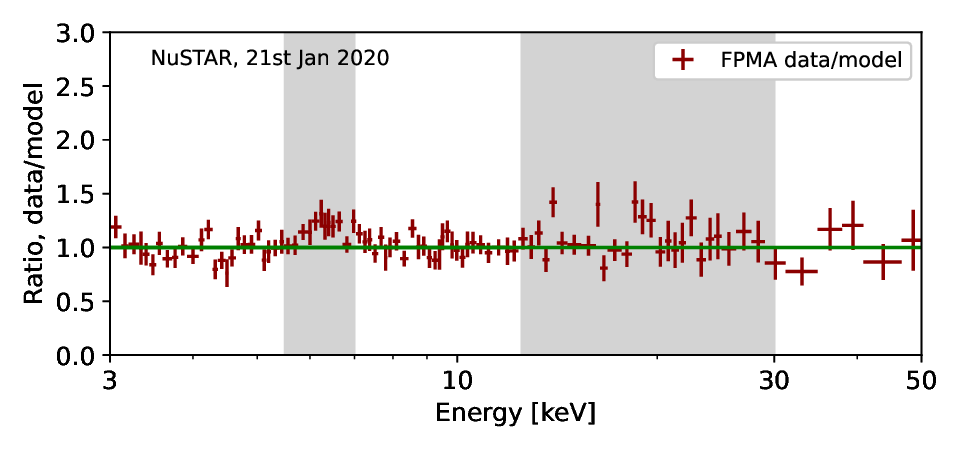}
    \includegraphics[scale=0.49]{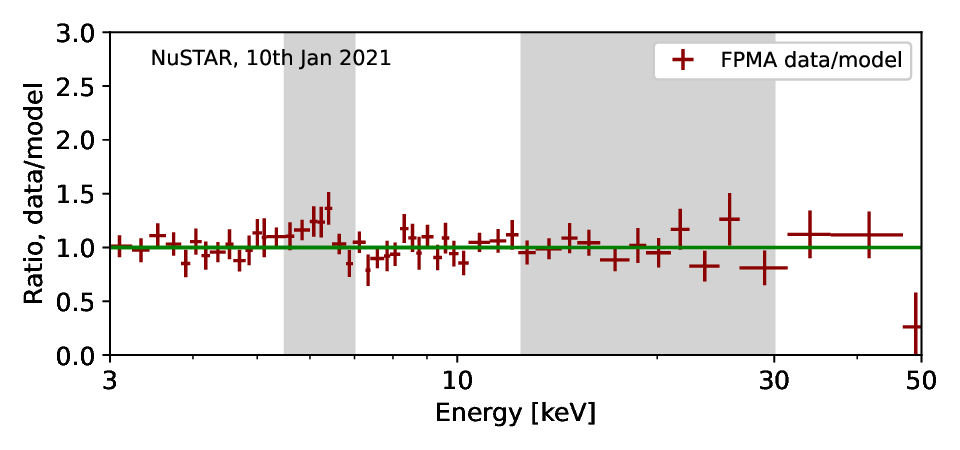}
    \includegraphics[scale=0.49]{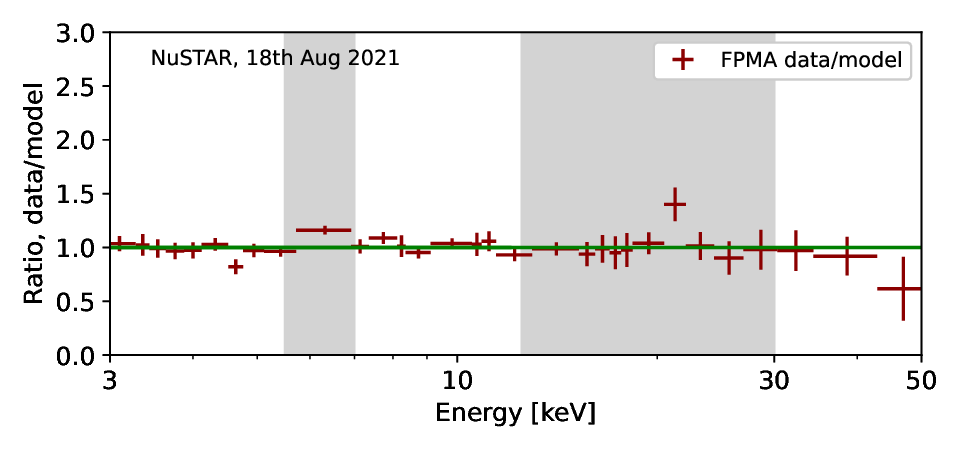}
    \includegraphics[scale=0.49]{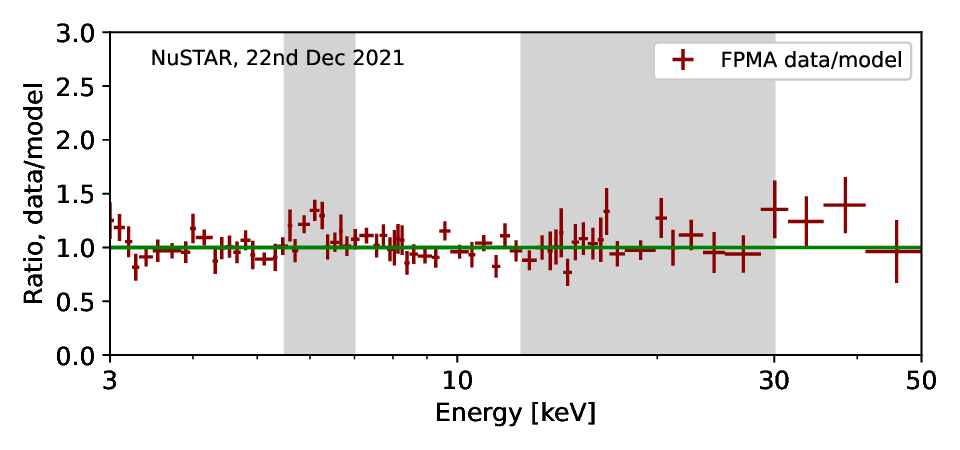}
    \includegraphics[scale=0.49]{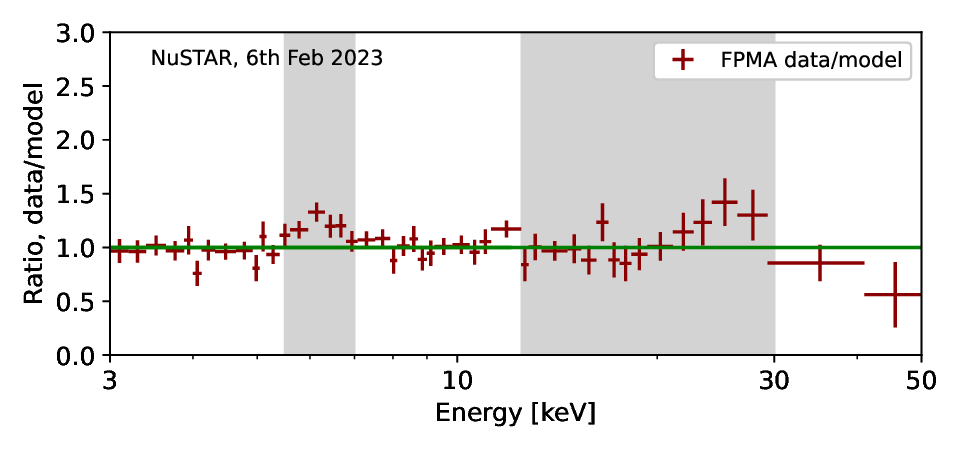}
    \includegraphics[scale=0.49]{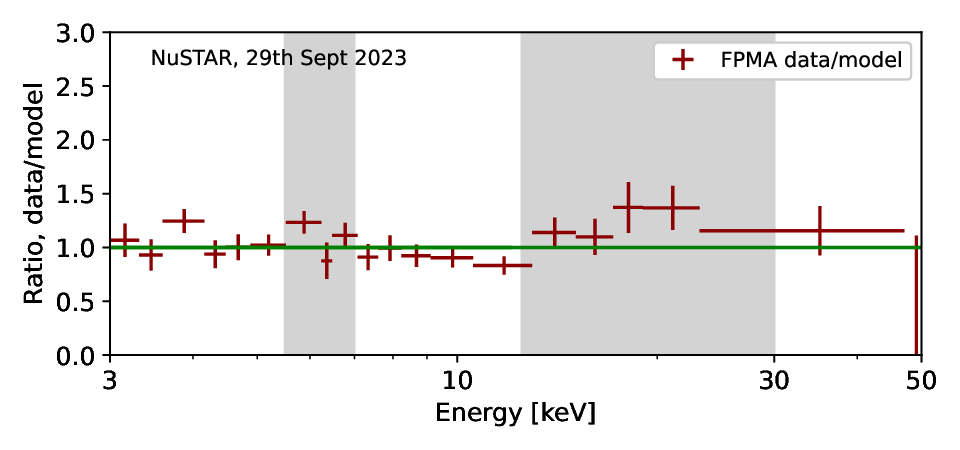}
    \begin{minipage}{.5\textwidth}\centering
    \,\,\,\,\,\,\includegraphics[scale=0.5]{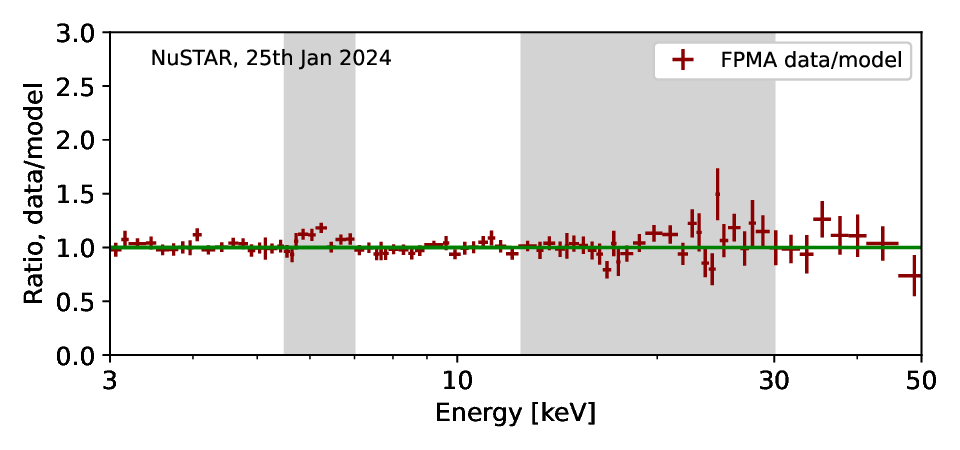}
    \end{minipage}\hfill
    \caption{Data-to-model ratios, against a power law model, for individual \emph{NuSTAR} observations. All energies are given in the observed frame. The gray regions indicate spectral regions excluded from the model fit. }
    \label{fig:nustar_windowfits}
\end{figure*}

%% file: appendixB/AppendixB.tex
\section{Bayesian model comparison}\label{AppendixB}

Here, we detail the model comparison procedure used to arrive at the most suitable models for the HF, LF, J21 and F23 joint data sets (\S \ref{sec:analysis_joint}).

\subsection{Procedure for Bayesian X-ray Analysis}

To estimate the evidence $Z$ for each model, we use the Bayesian X-ray Analysis package \citep[BXA,][]{Buchner2014,Buchner2019}, which runs the \emph{UltraNest} algorithm \citep{Buchner2016} to sample the full parameter space. BXA calculates the evidence $Z=\int\pi(\boldsymbol{\theta})\exp(-\frac{1}{2}C(\boldsymbol{\theta}))\mathrm{d}\boldsymbol{\theta}$, where $C(\boldsymbol{\theta})$ is the Cash statistic for a given parameter vector $\boldsymbol{\theta}$, and $\pi(\boldsymbol{\theta})$ is the prior weighting of those particular parameters. $C(\boldsymbol{\theta})$ is calculated via an \textsc{Xspec} model call.  We adopt log-uniform priors for normalization constants and column densities, which may vary over several orders of magnitude, and uniform priors for all other parameters. Thus, $\pi(\boldsymbol{\theta})$ is fully determined by our choice of parameter bounds. We run BXA once per data set for each model, using 400 live points, and using the standard nested sampling routine (\emph{i.e.,} the BXA parameter \emph{speed=`safe'}). 

As our data sets have between $\sim10^5$ (F23) and $\sim4\times10^5$ (HF) total photon counts, each \textsc{Xspec} model call is fairly computationally expensive. BXA runs with more than six or seven free parameters are therefore not feasible for these data, as the number of \textsc{Xspec} model calls required to map out $C(\boldsymbol{\theta})$ over the parameter space increases geometrically. We set certain model parameters to constant values, to limit the dimensionality of the BXA analysis; we discuss these choices in detail in our model definitions (Appendix \ref{AppendixB_modeldefs}). For all models, we set the cross-normalization terms $C_{\mathrm{inst}}$ to their best-fit values for Model C1. We verify via preliminary model fits that the $C_{\mathrm{inst}}$ terms for a given data set are consistent to within 2\%, irrespective of which model is used.

\subsection{Defining a threshold for acceptable models}\label{sec:deltalogz_cutoff} 

While the model with the highest evidence $Z_{\mathrm{best}}$ is in a sense the `best' model of those tested, the numerical values of the un-normalized evidences $Z_i$ for individual models are not otherwise particularly meaningful. The Bayes factors (\emph{i.e.}, evidence ratios $Z_i/Z_j$) between models $i$ and $j$ are the useful quantitative results for model comparison. As is typical in Bayesian model comparison, we work with the logarithms of the evidences, and express our results in terms of the Bayes factor between each model and the `best' (highest evidence) model tested. Thus, we express the logarithmic Bayes factor between a given model and the `best' model as $\Delta\log(Z)=\log(Z_{\mathrm{best}})-\log(Z_{\mathrm{model}})$.

A Bayes factor of $\Delta\log(Z)>2$ is traditionally interpreted as `decisive' evidence for the better model in the broader statistical literature \citep[\emph{e.g.},][]{Kass1995}. Recently, \citet{Waddell2023} investigated this interpretation empirically using a statistical sample of simulated AGN X-ray observations. These authors measure the frequency with which BXA returns higher evidence for an incorrect model, and determine a `purity' threshold for $\Delta\log(Z)$ based thereon. They find that a threshold of $\Delta\log(Z)=2.6$ is required to ensure a false positive rate below 2.5\% when testing for the presence of soft X-ray excess. However, in general the appropriate threshold will depend on the details of both the models and the observations (\emph{e.g.}, exposure times, instrumental response matrices, and wavelength-dependent sensitivities). It is prohibitively computationally expensive to perform similar simulations for our deep, multi-instrument data sets. We therefore instead adopt a conservative threshold for our model selection, requiring $\Delta\log(Z)>3$ to formally prefer one model over the other. This criterion is used to select our `preferred' models (\S \ref{sec:bestmodels_def}), which correspond to Models C2 and G as defined below. Reassuringly, these preferred models (which include warm Comptonization components) display Bayes factors of $\Delta\log(Z)>10$ compared to any models that lack warm Comptonization. This is also true for models lacking distant reflection components; in fact, those models are suppressed by even larger Bayes factors, $\Delta\log(Z)>29$. Thus, our main findings (that warm Comptonization and distant reflection are required; \S \ref{sec:modelcomparison}) are robust to any reasonable choice of $\Delta\log(Z)$ threshold. 

\subsection{Model definitions}\label{AppendixB_modeldefs}

\begin{figure*}
    \includegraphics[scale=0.65,trim={15 0 0 0}]{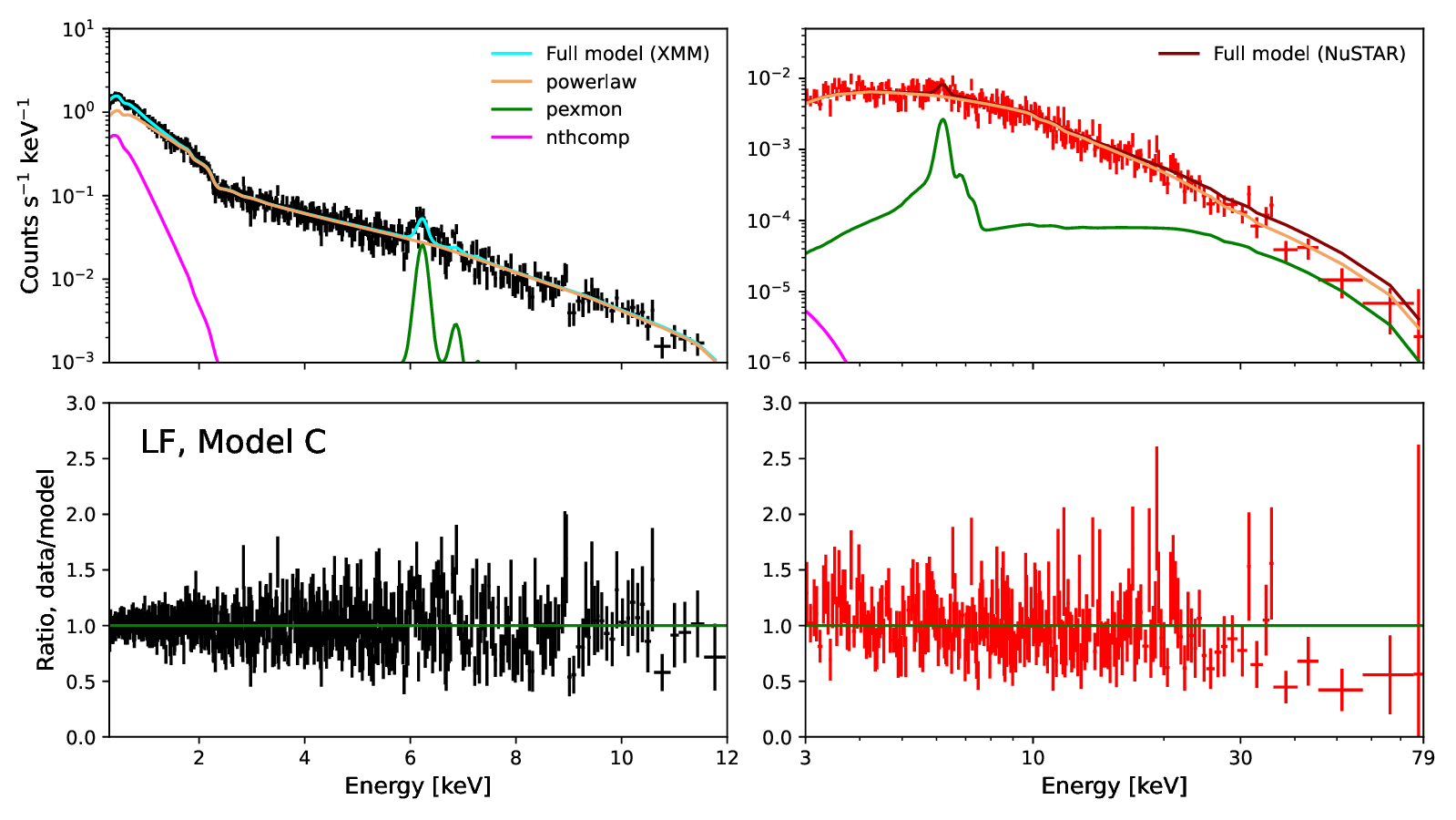}
    \caption{Model C (warm Comptonization and distant reflection) fitted to the LF data set. The upper panel shows the count rate spectrum, while the lower panel shows the data/model ratio. Photon energies are provided in the observed frame. The cyan curve (\emph{XMM-Newton pn}) and red-brown curve (\emph{NuSTAR}) indicate the composite models, including Galactic absorption and instrumental scaling factors. The orange (\textsc{powerlaw}), green (\textsc{pexmon}) and magenta (soft Comptonization \textsc{nthcomp}) curves indicate the predicted count rates for the individual additive model components. \label{fig:appendixC_modelC_LF}}
\end{figure*}

\begin{figure*}
    \includegraphics[scale=0.65,trim={15 0 0 0}]{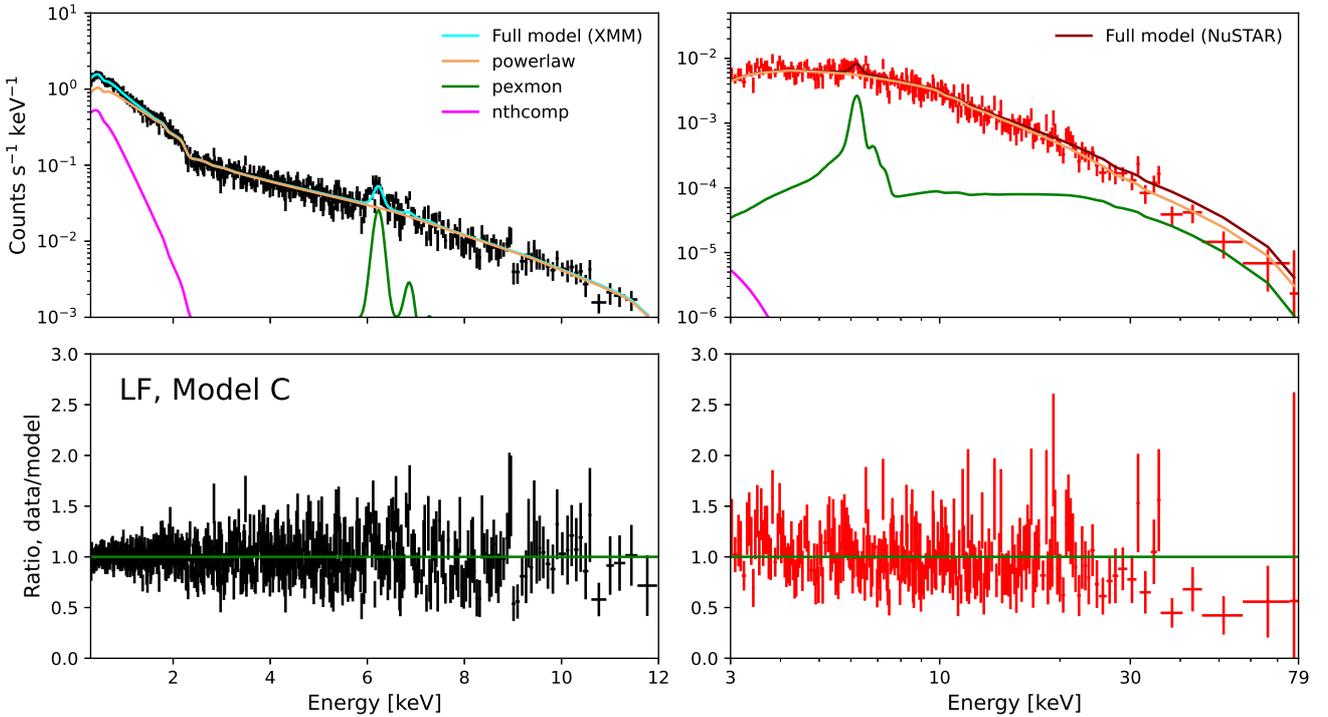}
    \caption{As Figure \ref{fig:appendixC_modelC_LF} but for the HF data set. \label{fig:appendixC_modelC_HF}}
\end{figure*}

\begin{figure*}
    \includegraphics[scale=0.65,trim={15 0 0 0}]{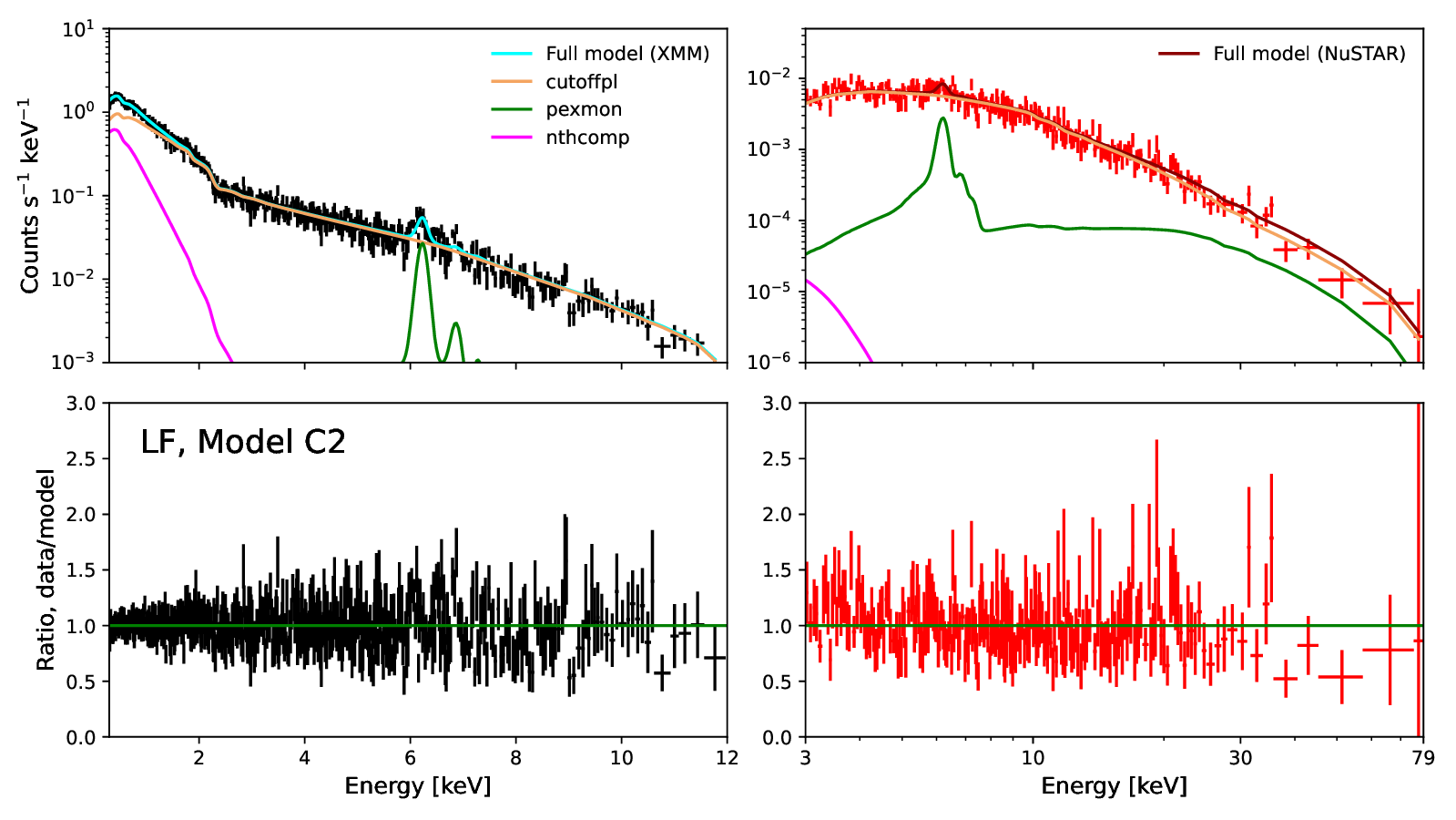}
    \caption{As Figure \ref{fig:appendixC_modelC_LF} but for Model C2, for the LF data set. \label{fig:appendixC_modelC2_LF}}
\end{figure*}

\begin{figure*}
    \includegraphics[scale=0.65,trim={15 0 0 0}]{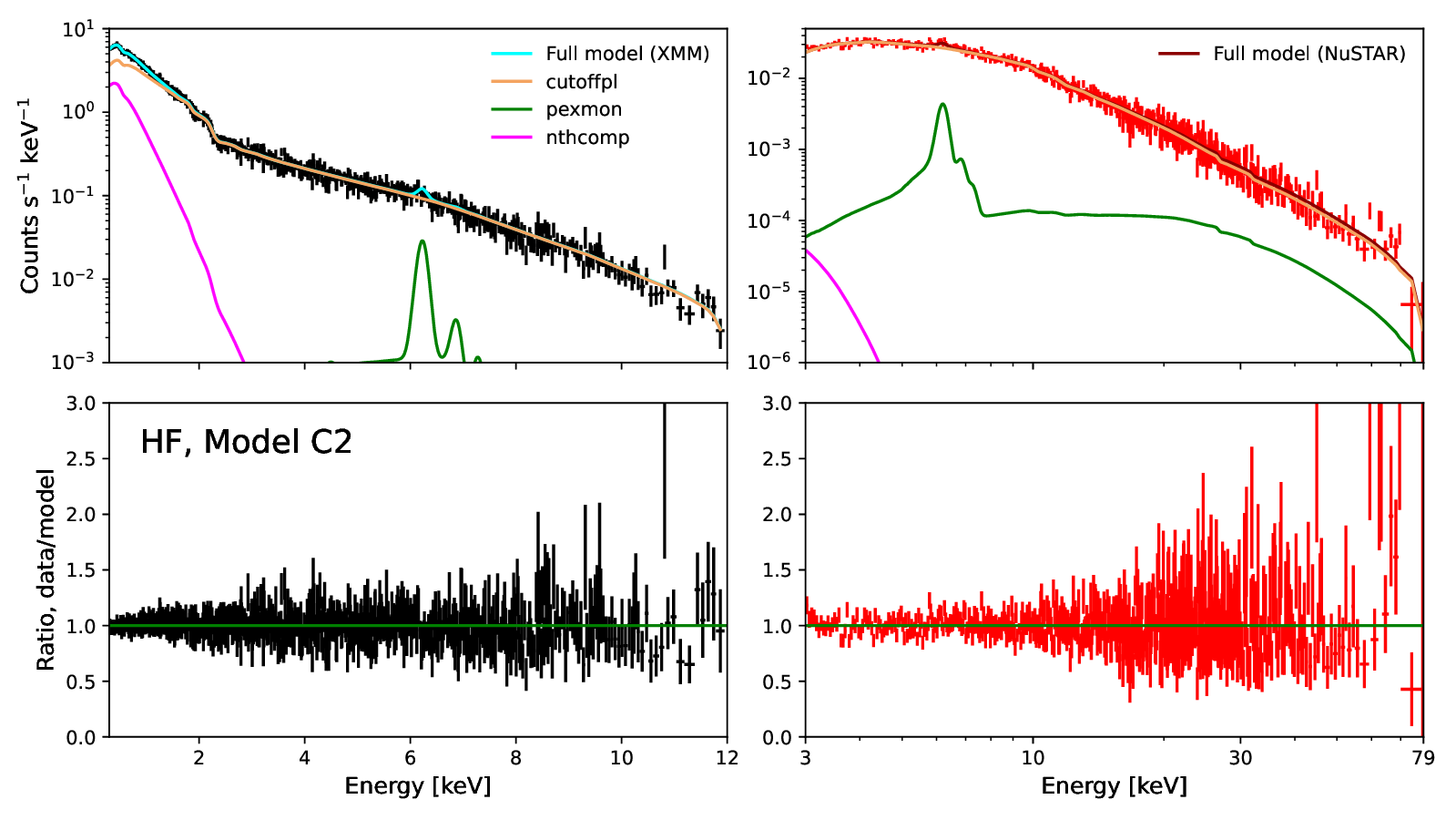}
    \caption{As Figure \ref{fig:appendixC_modelC_LF} but for Model C2, for the HF data set. \label{fig:appendixC_modelC2_HF}}
\end{figure*}

\begin{figure*}
    \includegraphics[scale=0.65,trim={15 0 0 0}]{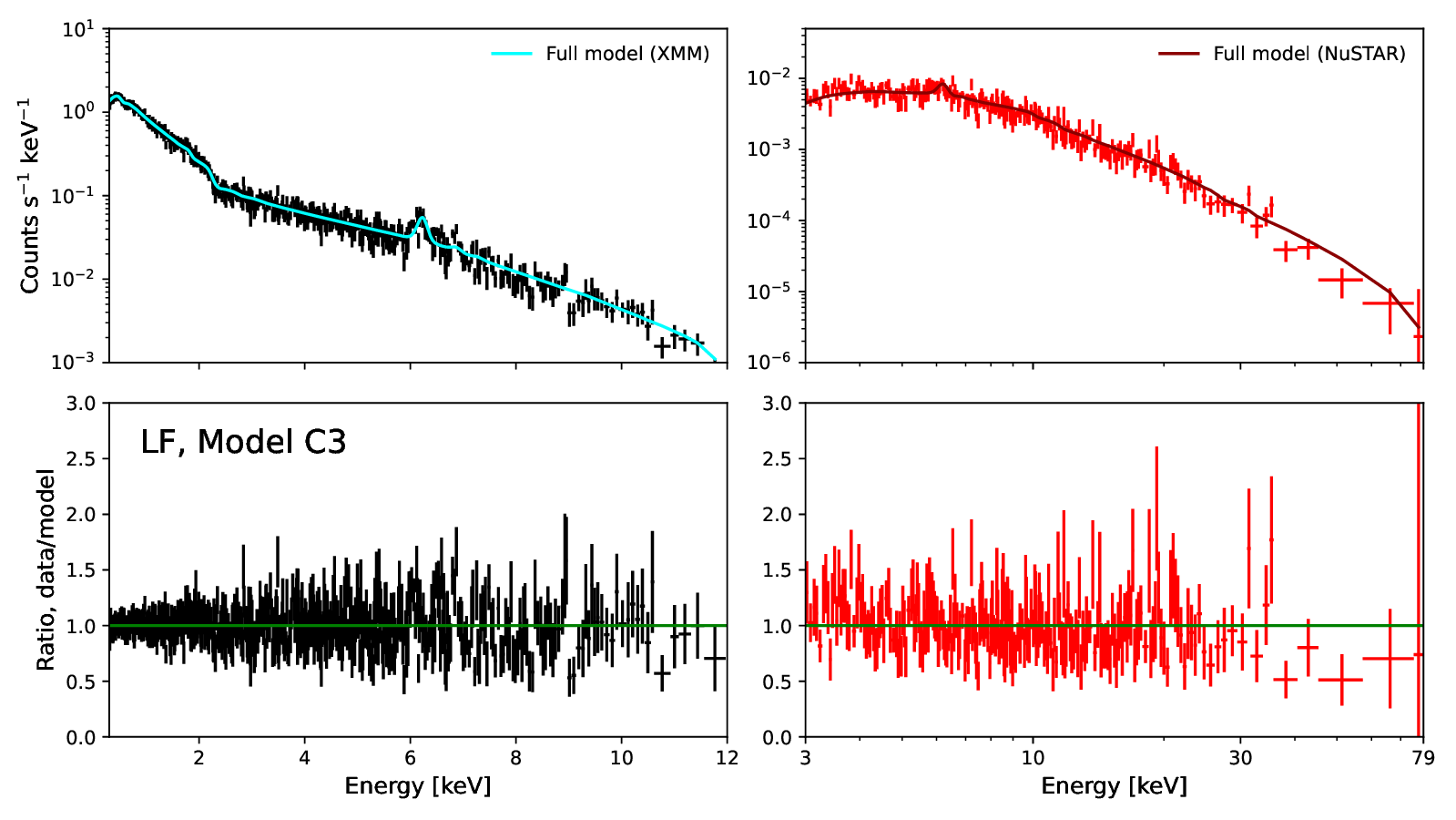}
    \caption{As Figure \ref{fig:appendixC_modelC_LF} but for Model C3, for the LF data set. \label{fig:appendixC_modelC3_LF}}
\end{figure*}

\begin{figure*}
    \includegraphics[scale=0.65,trim={15 0 0 0}]{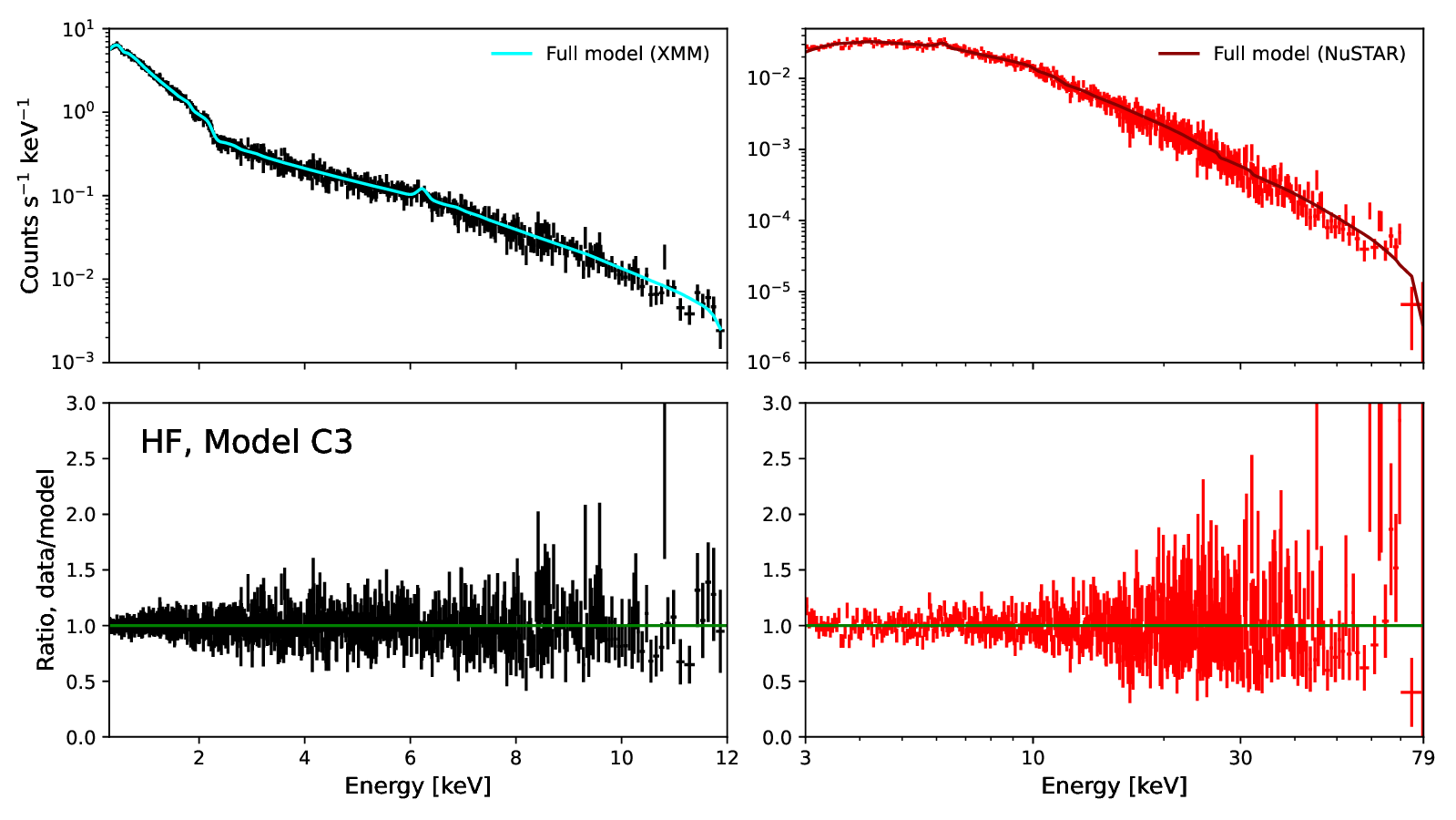}
    \caption{As Figure \ref{fig:appendixC_modelC_LF} but for Model C3, for the HF data set. \label{fig:appendixC_modelC3_HF}}
\end{figure*}

\begin{figure*}
    \includegraphics[scale=0.65,trim={15 0 0 0}]{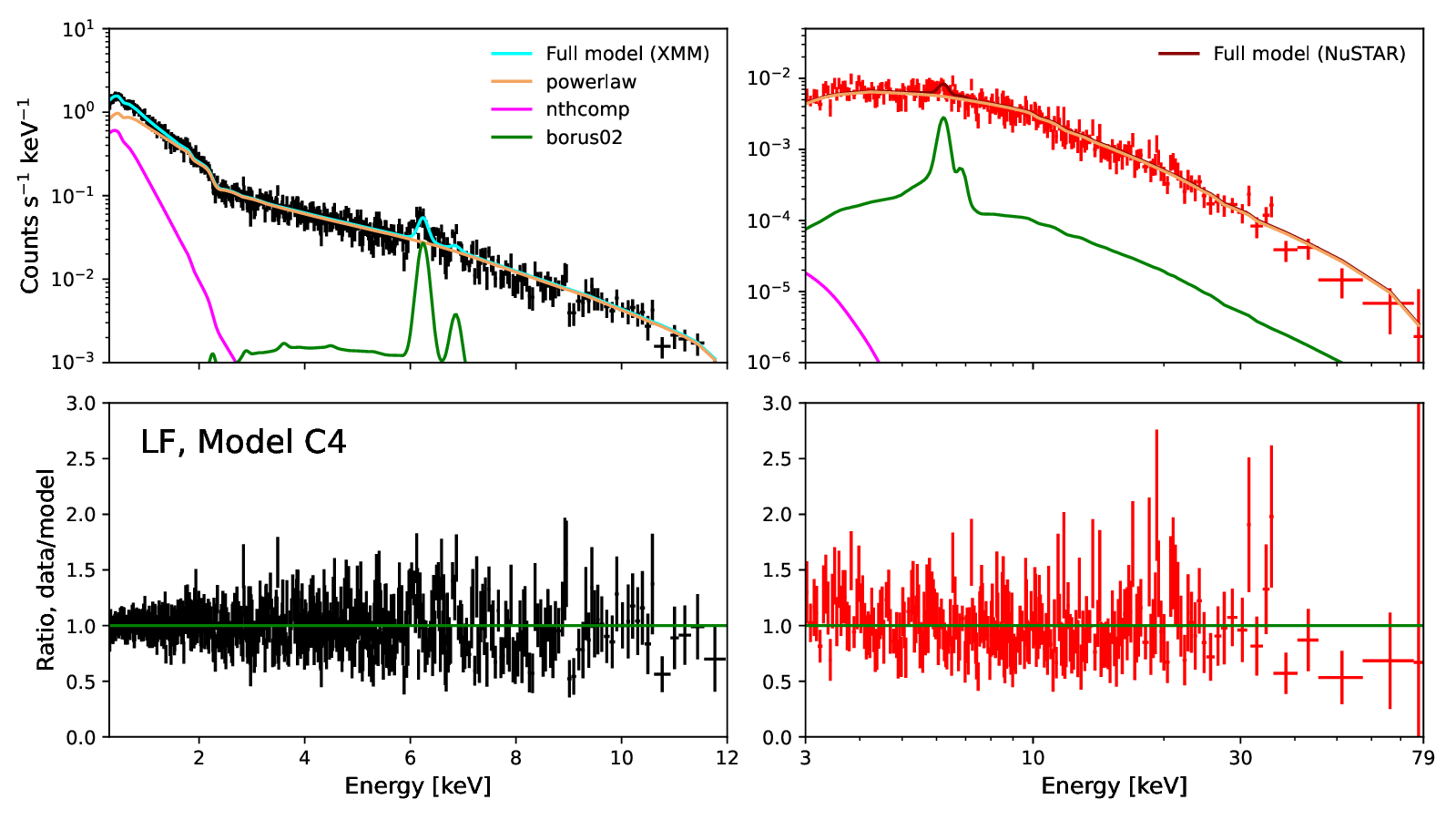}
    \caption{As Figure \ref{fig:appendixC_modelC_LF} but for Model C4, for the LF data set. \label{fig:appendixC_modelC4_LF}}
\end{figure*}

\begin{figure*}
    \includegraphics[scale=0.65,trim={15 0 0 0}]{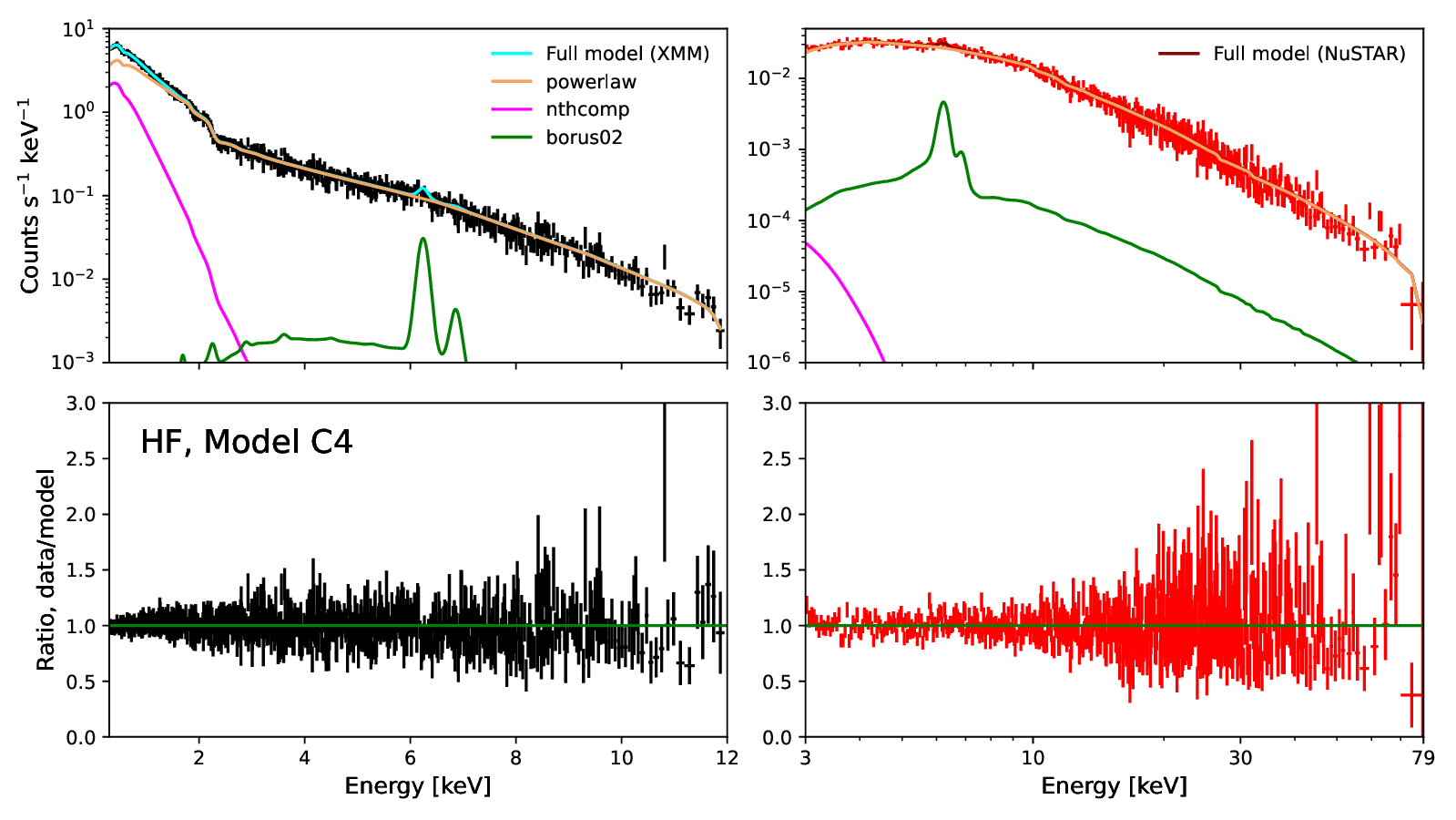}
    \caption{As Figure \ref{fig:appendixC_modelC_LF} but for Model C4, for the HF data set. \label{fig:appendixC_modelC4_HF}}
\end{figure*}

\begin{figure*}
    \includegraphics[scale=0.65,trim={15 0 0 0}]{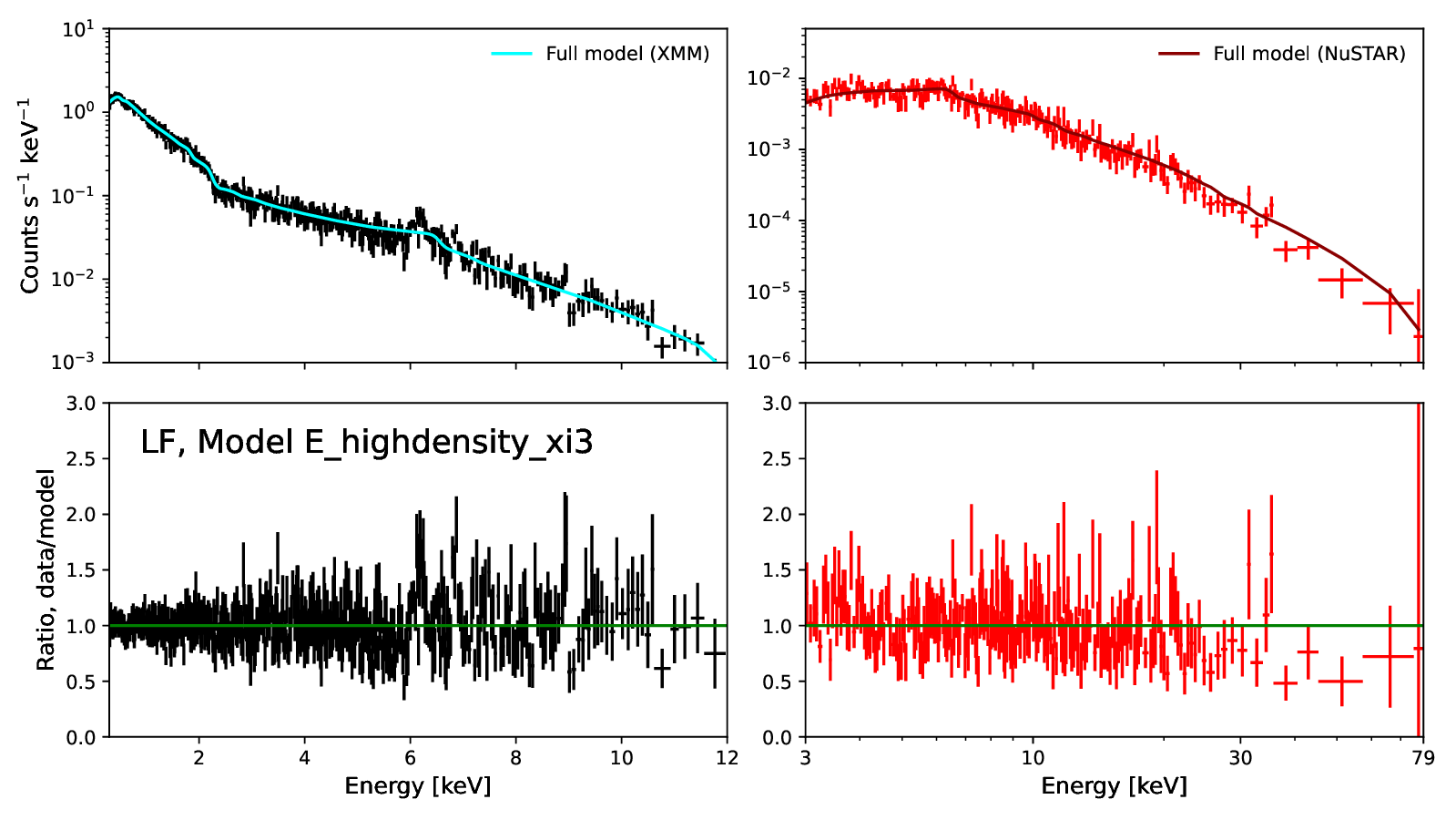}
    \caption{As Figure \ref{fig:appendixC_modelC_LF} but for Model E, for the LF data set. \label{fig:appendixC_modelE_LF}}
\end{figure*}

\begin{figure*}
    \includegraphics[scale=0.65,trim={15 0 0 0}]{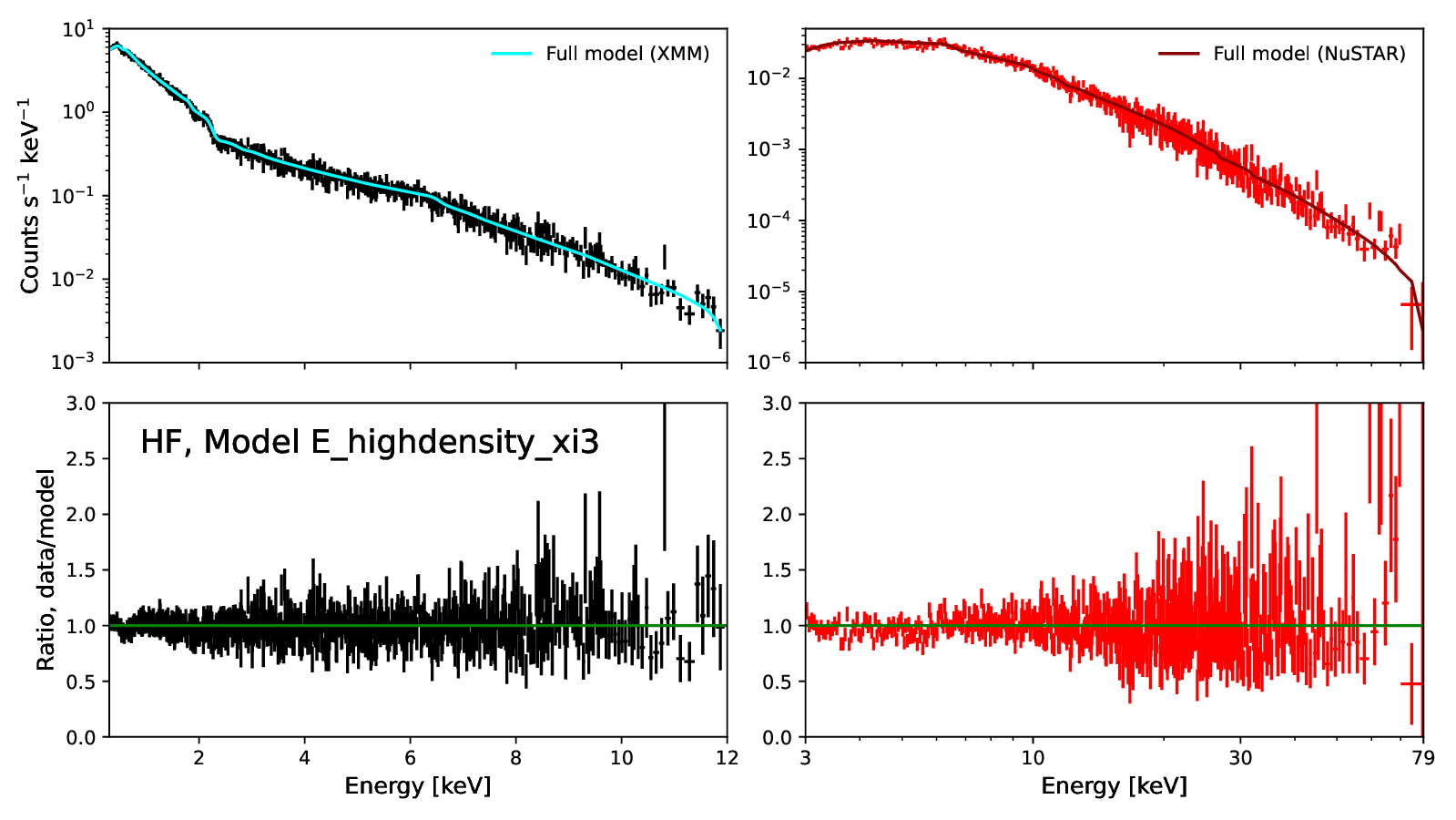}
    \caption{As Figure \ref{fig:appendixC_modelC_LF} but for Model E, for the LF data set. \label{fig:appendixC_modelE_LH}}
\end{figure*}

\begin{figure*}
    \includegraphics[scale=0.65,trim={15 0 0 0}]{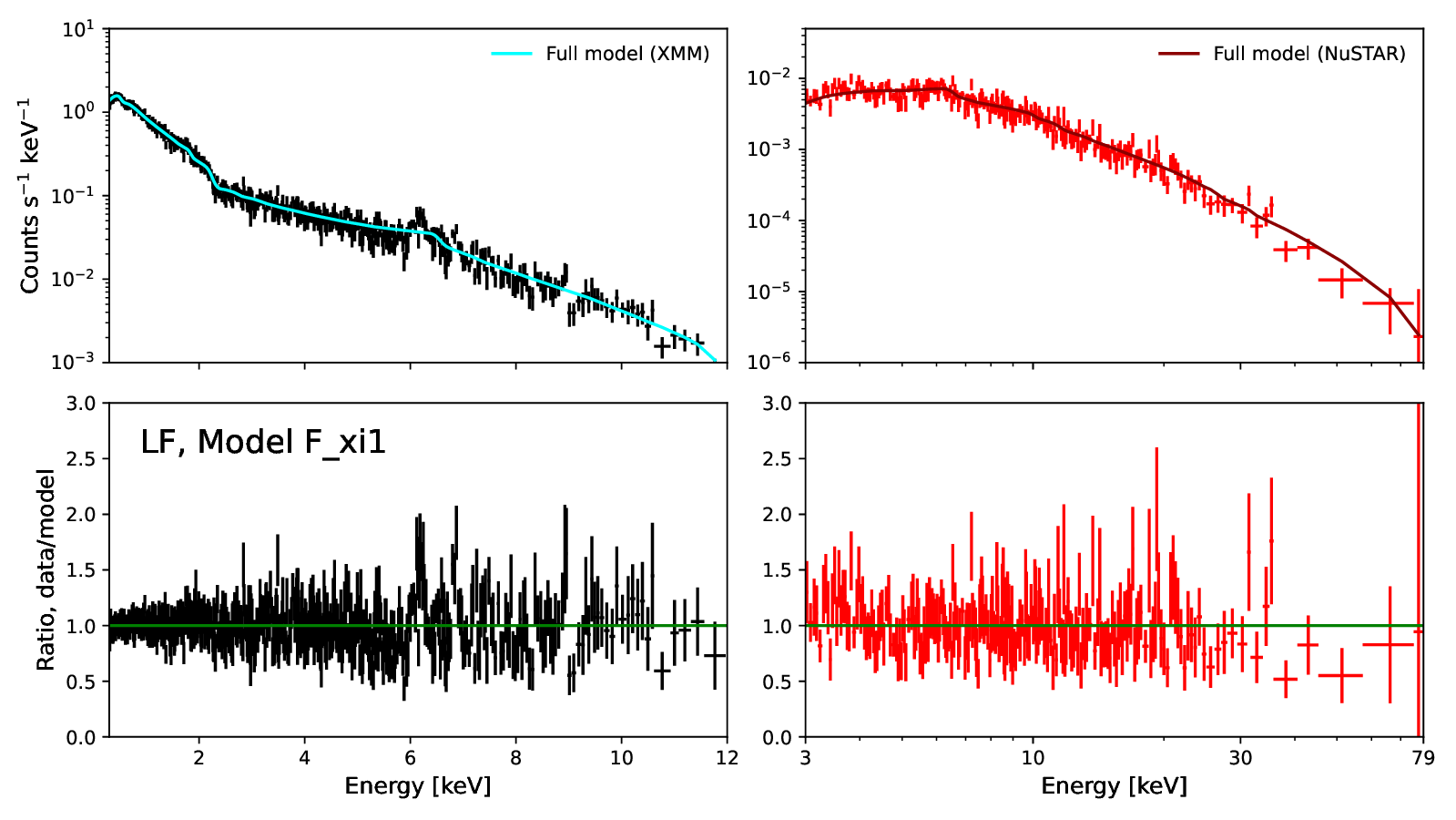}
    \caption{As Figure \ref{fig:appendixC_modelC_LF} but for Model F, for the LF data set. \label{fig:appendixC_modelF_LF}}
\end{figure*}

\begin{figure*}
    \includegraphics[scale=0.65,trim={15 0 0 0}]{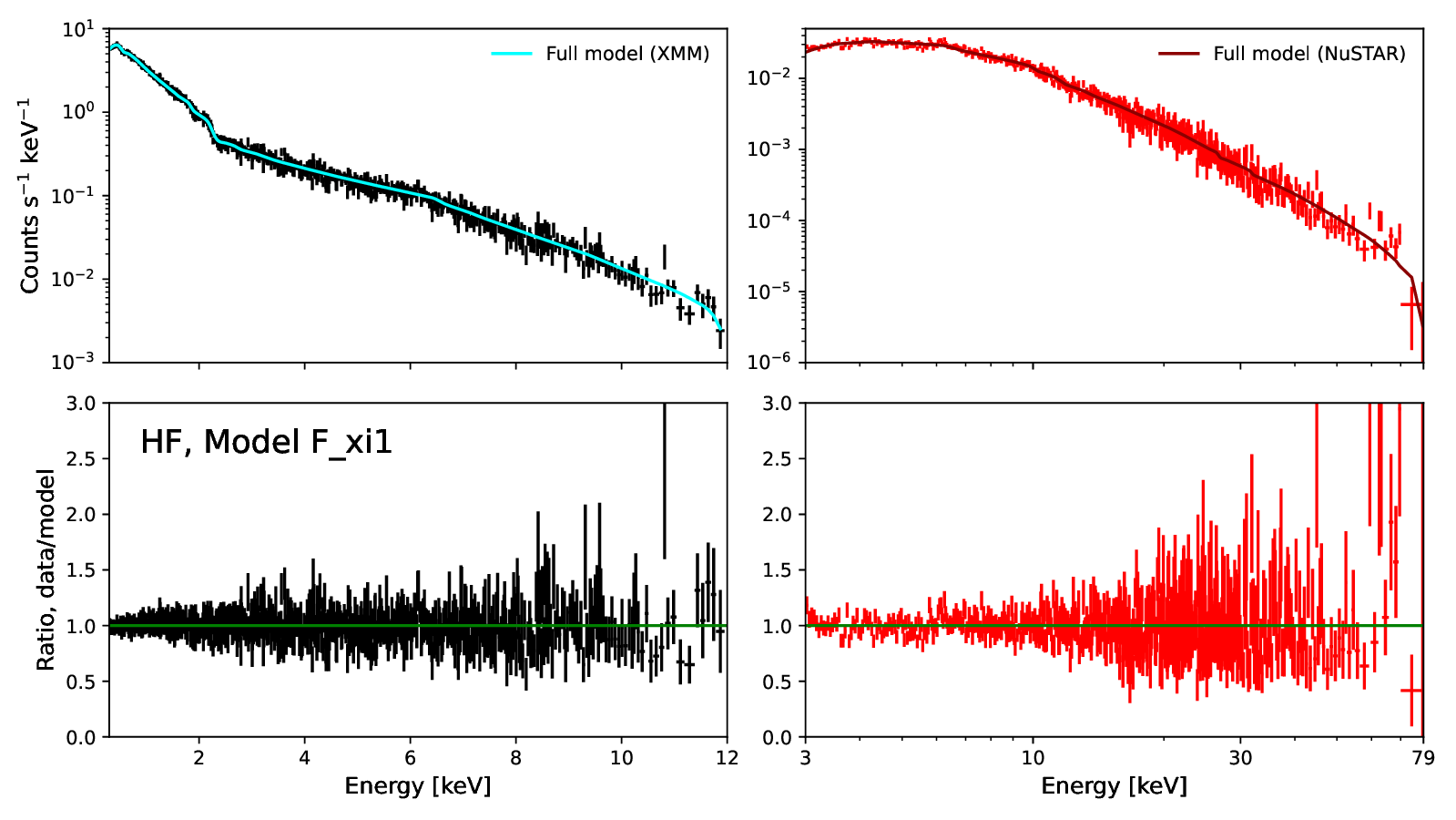}
    \caption{As Figure \ref{fig:appendixC_modelC_LF} but for Model F, for the HF data set. \label{fig:appendixC_modelF_HF}}
\end{figure*}

\begin{figure*}
    \includegraphics[scale=0.65,trim={15 0 0 0}]{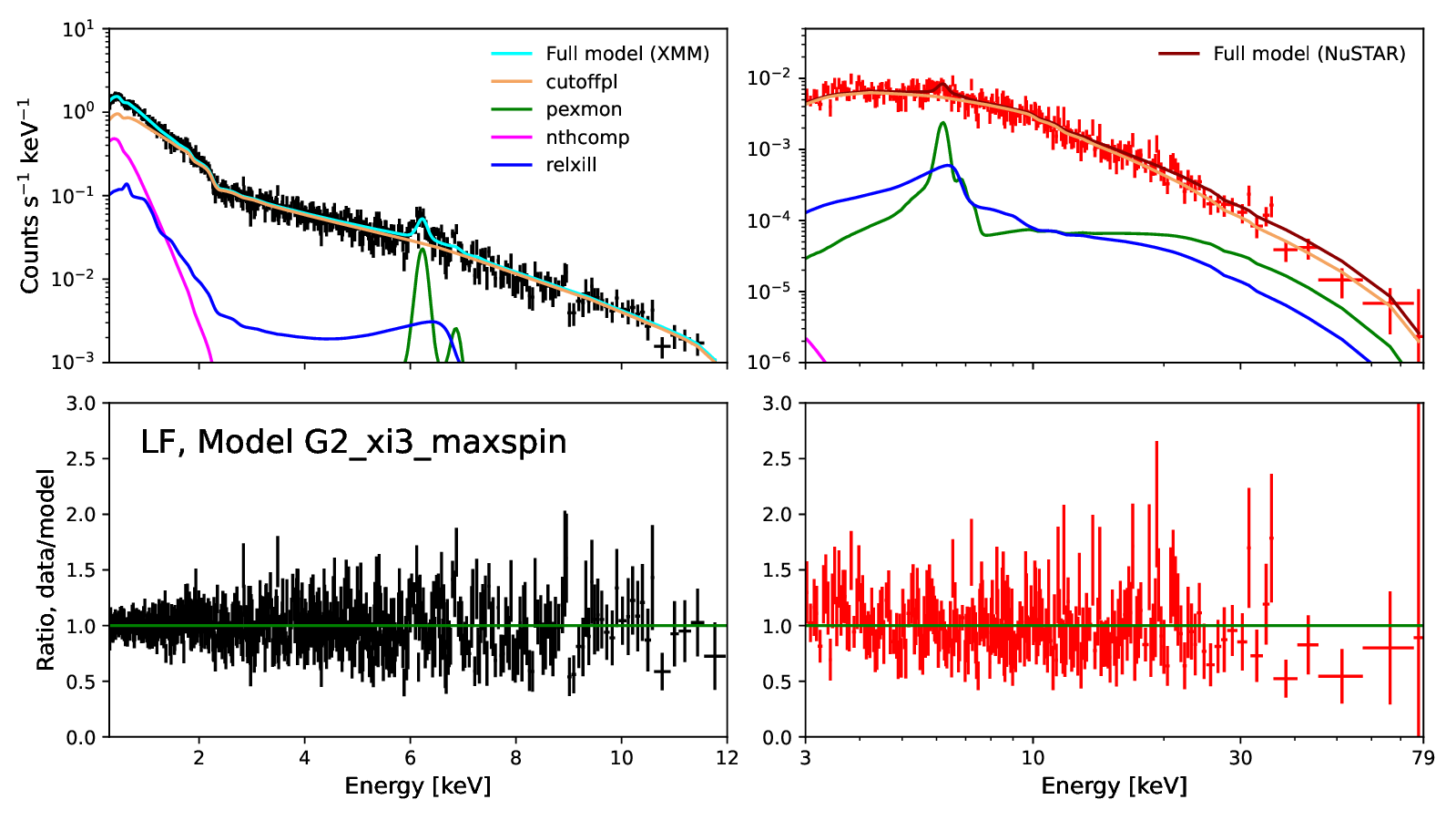}
    \caption{As Figure \ref{fig:appendixC_modelC_LF} but for Model G, for the LF data set. \label{fig:appendixC_modelG_LF}}
\end{figure*}

\begin{figure*}
    \includegraphics[scale=0.65,trim={15 0 0 0}]{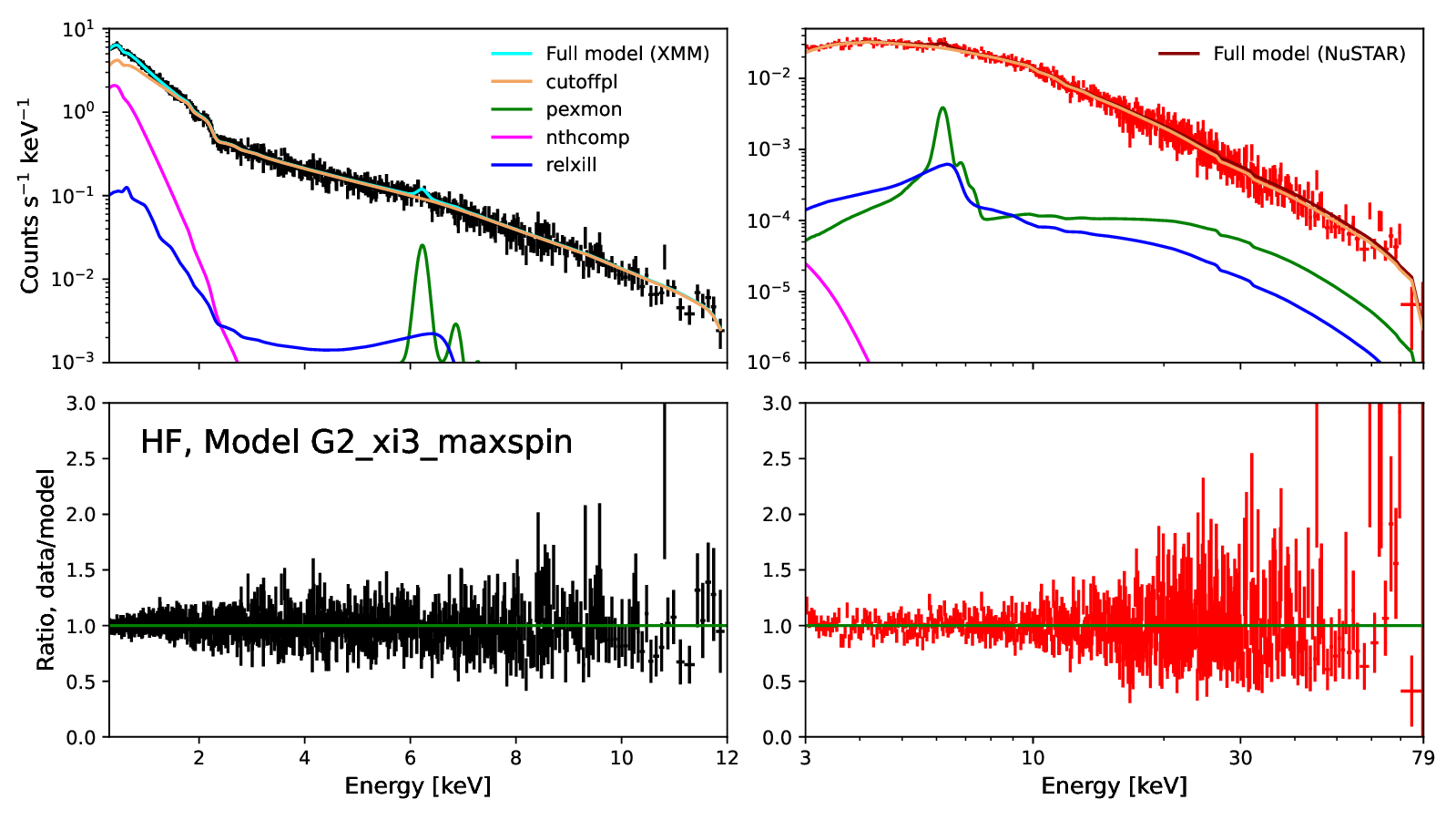}
    \caption{As Figure \ref{fig:appendixC_modelC_LF} but for Model G, for the HF data set. \label{fig:appendixC_modelG_HF}}
\end{figure*}

\begin{figure*}
    \includegraphics[scale=0.65,trim={15 0 0 0}]{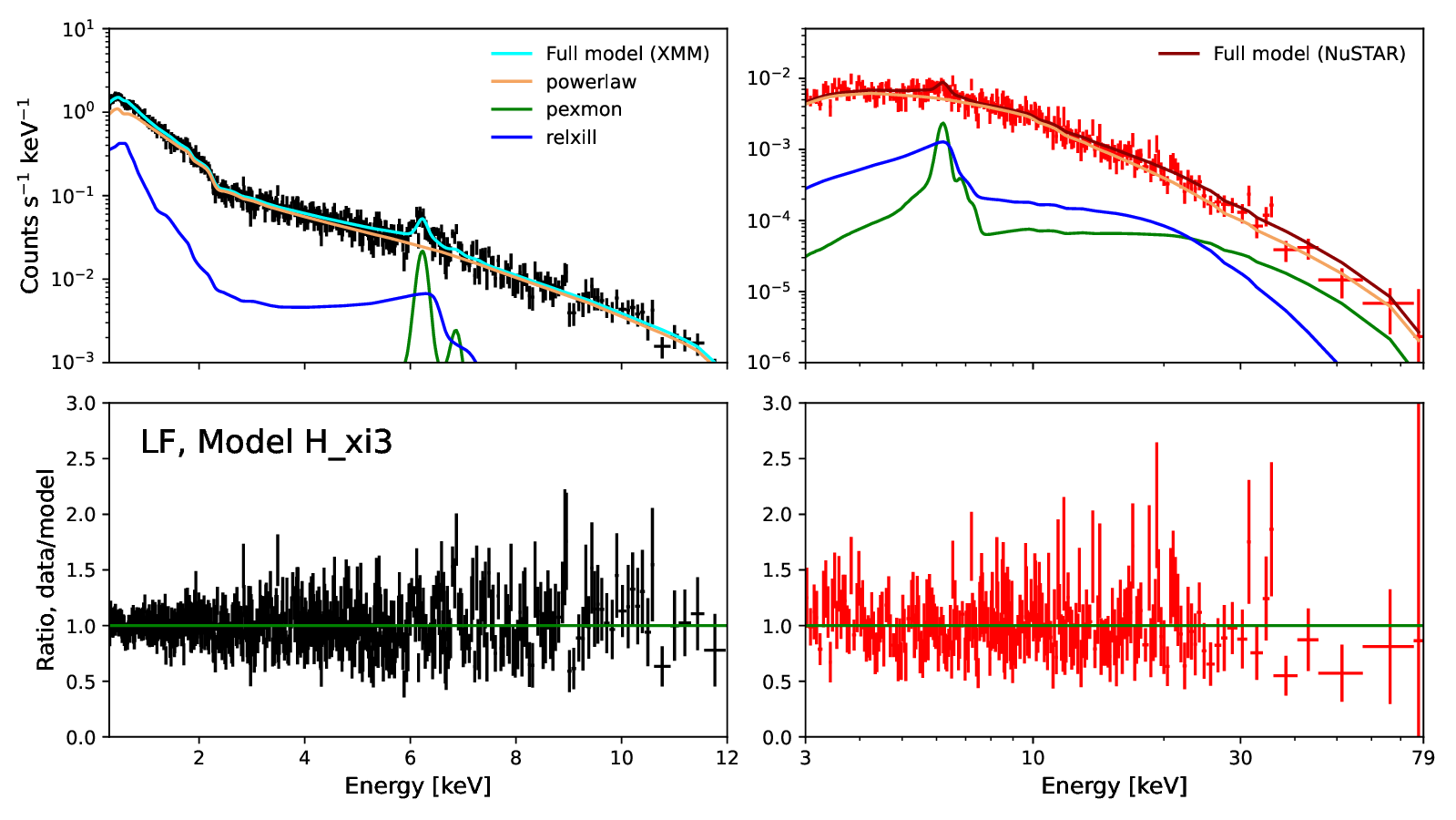}
    \caption{As Figure \ref{fig:appendixC_modelC_LF} but for Model H, for the LF data set. \label{fig:appendixC_modelH_LF}}
\end{figure*}

\begin{figure*}
    \includegraphics[scale=0.65,trim={15 0 0 0}]{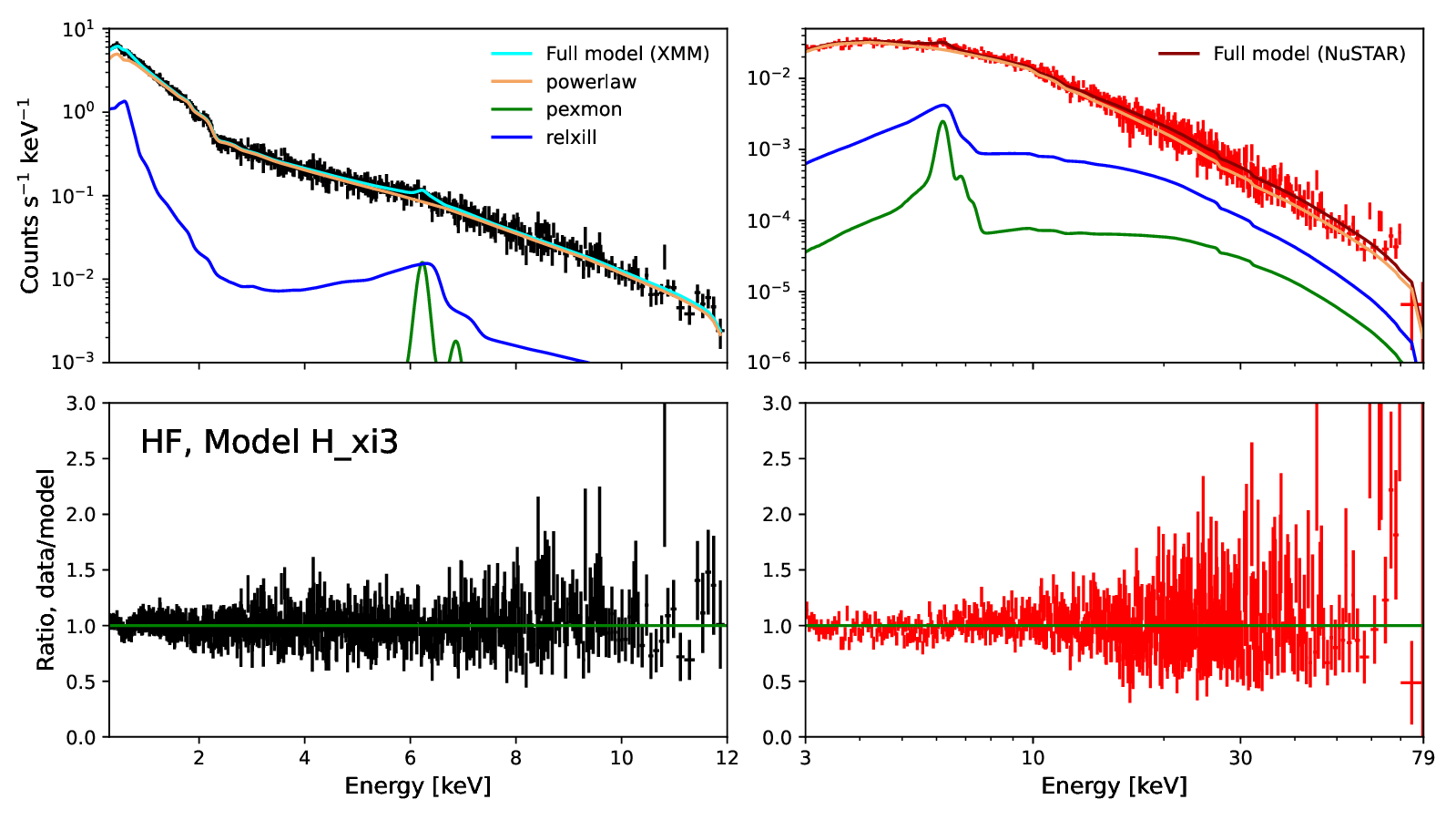}
    \caption{As Figure \ref{fig:appendixC_modelC_LF} but for Model H, for the HF data set. \label{fig:appendixC_modelH_HF}}
\end{figure*}

\begin{figure*}
    \includegraphics[scale=0.5]{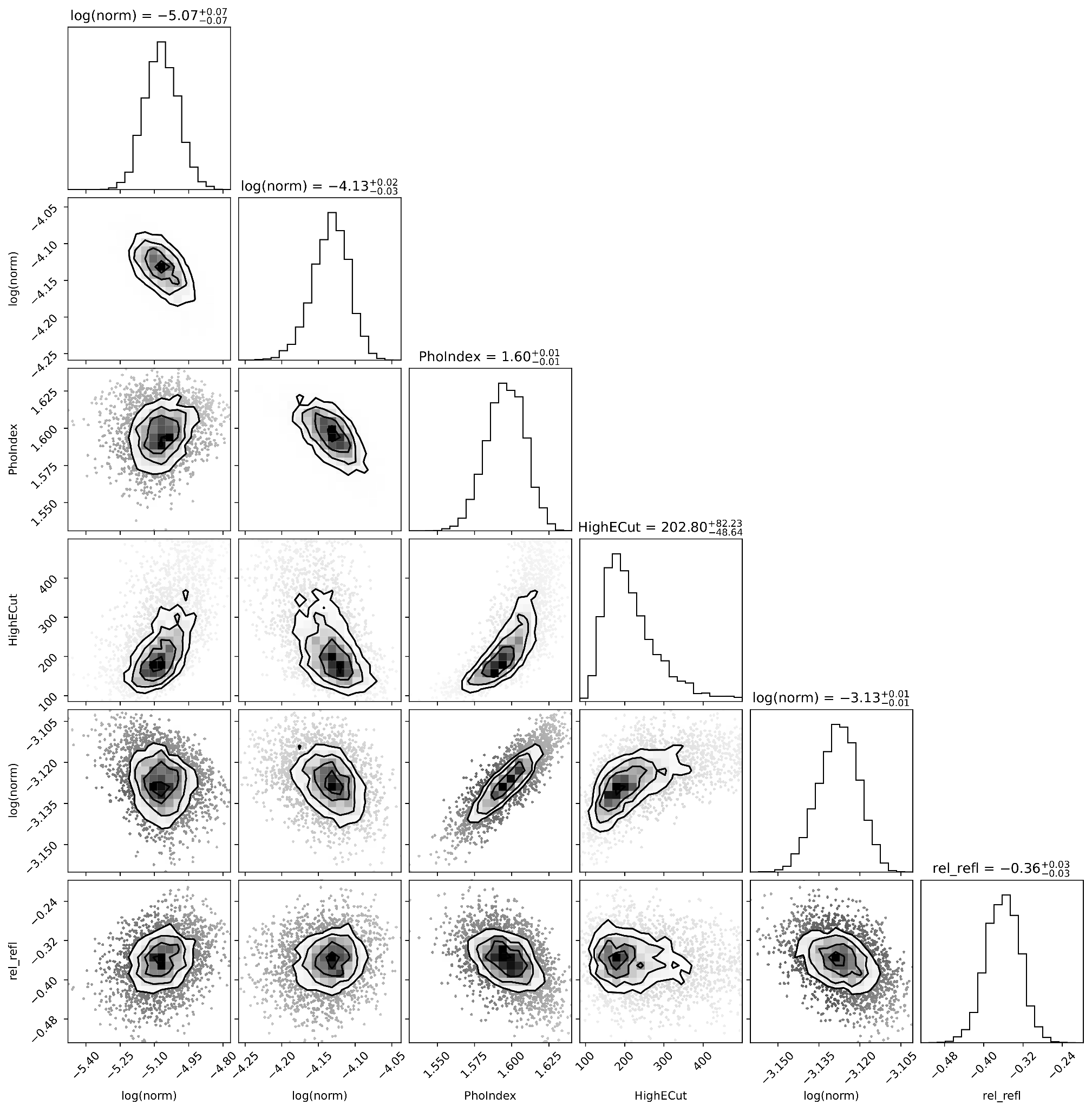}
    \caption{`Corner plot' for the posterior distribution of our BXA run for Model G with $\log(\xi)=3$ to the LF data set. The continuum spectral index $\Gamma$ is somewhat degenerate with both the continuum normalization and the cutoff energy $E_{\mathrm{cut}}$.  The \textsc{pexmon} $R$ parameter denotes the strength of the distant reflection component. It shows no strong degeneracy with the normalization of the relativistic reflection (or any other model parameters tested), and is rather weak: we find a posterior median $R=-0.36\pm0.03$, where $R=-1$ corresponds to a slab geometry extending to infinity. 
    \emph{The three normalizations are, ordered from left to right, for the \textsc{relxill}, \textsc{nthcomp} and \textsc{zcutoffpl} model components. A clearly labeled version of this Figure will be included in the final published manuscript; an issue with image file conversion is delaying its inclusion in the arxiv.org preprint.}\label{fig:corner_plot_G2_LF}}
\end{figure*}

\begin{figure*}
    \includegraphics[scale=0.5]{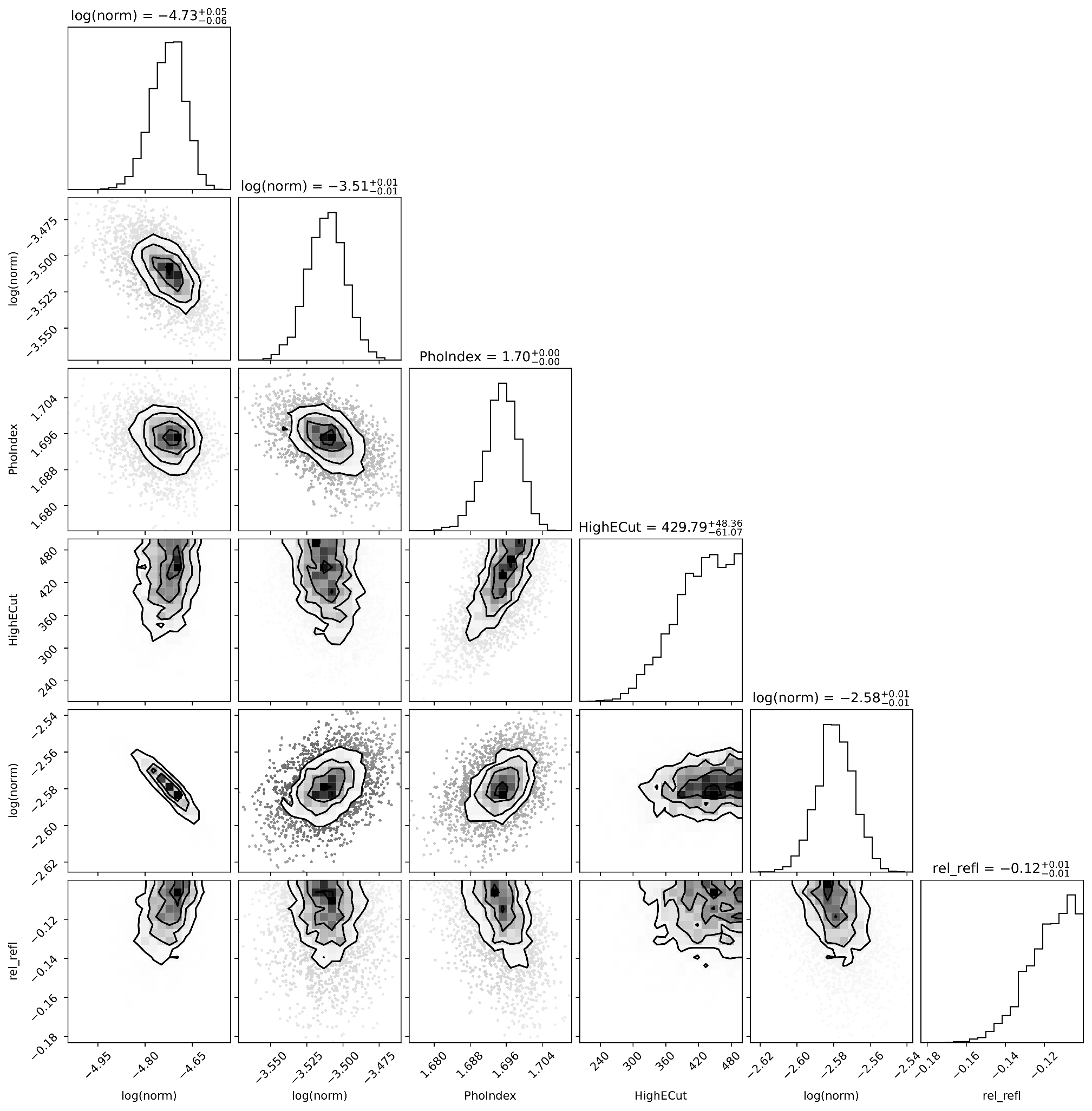}
    \caption{As Figure \ref{fig:corner_plot_G2_LF}, but for the HF data set. Here, the distant reflection is significantly weaker than in the low-flux state, with a posterior median reflection factor $R=-0.10_{-0.01}^{+0.02}$ for the \textsc{pexmon} component. \emph{The three normalizations are, ordered from left to right, for the \textsc{relxill}, \textsc{nthcomp} and \textsc{zcutoffpl} model components. A clearly labeled version of this Figure will be included in the final published manuscript; an issue with image file conversion is delaying its inclusion in the arxiv.org preprint.}\label{fig:corner_plot_G2_HF}}
\end{figure*}

We define models A--G (roughly in order of increasing complexity) below. All models include a \textsc{const} term representing the per-detector calibration factors $C_{\mathrm{inst}}$, and a \textsc{tbabs} term representing Galactic absorption with \mbox{$N_{\mathrm{H}}=2.77\times10^{20}$ cm$^{-2}$} \citep{Bekhti2016}. We present the parameter bounds for each model in Table \ref{tab:parameters}. In cases where different model components have parameters with the same physical meaning (\emph{e.g.}, $E_\mathrm{cut}$ for Model C2 should be roughly equivalent to $3kT_e$ for Model C3), we apply consistent parameter bounds; this avoids spurious differences in $Z$ due to choices of prior. We present the logarithmic Bayes factors $\Delta\log(Z)$ for each model, compared to the `best' model, in Table \ref{tab:bayesfactors}. We illustrate models C1 through H as C-stat optimized fits to the LF and HF data sets in Figures \ref{fig:appendixC_modelC_LF} through \ref{fig:appendixC_modelH_HF}. For Model G, we include `corner plots' of the BXA posterior distributions, to illustrate the parameter degeneracies at play when using two reflection regions (Figures \ref{fig:corner_plot_G2_LF} and \ref{fig:corner_plot_G2_HF}).\\

\noindent\textbf{Model A: continuum only} \\

\noindent\textsc{Xspec} model definition: \textsc{const$\times$tbabs$\times$powerlaw}\\

\noindent We begin our model comparison with a simple power law, with two free parameters: the normalization $N_{\mathrm{cont}}$ and photon index $\Gamma$. This model is clearly inadequate; we include it to quantify the improvement in evidence for the more complex Models B--H.\\

\noindent\textbf{Model B: warm Comptonization} \\

\noindent\textsc{const$\times$tbabs(powerlaw+nthcomp)}\\

\noindent We add a Comptonization component to model the soft excess, represented by the \textsc{nthcomp} model \citep{Zdziarski1996,Zycki1999}. This model is characterized by a photon index $\Gamma_{\mathrm{warm}}$, which depends physically on the optical depth and scattering geometry. For the BXA runs, we set a constant $\Gamma_{\mathrm{warm}}\equiv2.5$ , as typically found for Seyfert galaxies \citep{Petrucci2018}. For the seed photons we set a constant energy of 10 eV (\emph{i.e.}, UV seed photons). The resulting model has two additional free parameters relative to model A: the soft excess temperature $kT_e$, and normalization $N_{\mathrm{warm}}$. For all four data sets (LF, HF, J21 and F23) we see a substantial increase in evidence for Model B relative to Model A. \\

\noindent\textbf{Models C1--C4 (preferred!): warm Comptonization plus distant reflection} \\

\noindent C1: \textsc{const$\times$tbabs(powerlaw+nthcomp+pexmon)}

\noindent C2: \textsc{const$\times$tbabs(zcutoffpl+nthcomp+pexmon)}

\noindent C3: \textsc{const$\times$tbabs(nthcomp$_1$+nthcomp$_2$+pexmon)}

\noindent C4: \textsc{const$\times$tbabs(powerlaw+nthcomp} 

\textsc{+atable\{borus02\_v170323c.fits\})}\\

\noindent We now add a non-relativistic reflection component. The \textsc{pexmon} component \citep{Nandra1997} combines Compton reflection \citep{Magdziarz1995} with energetically consistent Iron and Nickel K-shell emission lines. The strength of the reflection features is parameterized by $R$, where $R=-1$ is the expected reflection from an infinite slab of neutral, Compton-thick gas\footnote{The negative sign causes the \textsc{pexmon} component to only return the reflected spectrum, such that the continuum can be modeled separately. We use negative $R$ in all \textsc{pexmon} modeling, as this allows us to test different continuum models, \emph{e.g.}, Model C3.}. We allow the reflection strength to vary between $-0.1>R>-5$; here, $R=-0.1$ is barely detectable in typical observing situations. We tie the \textsc{pexmon} incident continuum normalization and photon index to that of the \textsc{powerlaw} component. 

The reflected spectrum depends on the inclination $i$ of the line-of-sight to the reflecting surface (where $i=0^\circ$ is face-on). We impose $i\equiv30^\circ$ for all reflection components (Models C1 through H). This is a typical estimate for the accretion plane of Type 1 AGN \citep[\emph{e.g.},][]{Wu2001, Marin2016}, and is consistent with the estimate $i=47_{-47}^{+38}{^\circ}$ made by \citet{Bhayani2011} for Mrk 590 specifically. 

The elemental abundance $Z$ and Iron abundance $A_\mathrm{Fe}$ affect the shape of the reflected spectrum in non-trivial ways \citep[\emph{e.g.},][]{Garcia2014}. To ensure that the parameter space is of manageable size, we keep $Z$ and $A_\mathrm{Fe}$ constant during the BXA runs. After some initial experimentation, we set $A_\mathrm{Fe}\equiv5$; we find this is necessary to match the observed 6.4 keV line strength. We discuss the issue of super-Solar iron abundance further in \S \ref{sec:bestmodels}, where we include $A_\mathrm{Fe}$ as a free parameter in our preferred models.

In all cases, the evidence for Model C1 is higher than for Model B. However, Model C1 over-predicts the X-ray flux above \mbox{$\sim30$ keV}, especially for the LF data set (Figure \ref{fig:appendixC_modelC_LF}). We therefore test whether there is evidence for a high-energy cutoff of the X-ray continuum. For Model C2, we include an ad-hoc exponential cutoff in the continuum at energy $E_{\mathrm{cut}}$. For Model C3, we replace the power law continuum with a physically motivated hot Comptonization (\textsc{nthcomp}) component \citep[\emph{e.g.},][]{Zycki1999}, for which the high-energy curvature is parameterized by the electron temperature $kT_{e,\mathrm{hot}}$. For all four data sets, Model C2 is preferred over C1 and C3. Thus, while our data require some additional spectral curvature, its detailed shape is \emph{not} particularly well-described by the high-energy peak of a hot Comptonization component. 

Motivated by this finding, we experiment with a Compton-thin reflection component. This would drive down the flux near 30 keV by producing a weaker Compton reflection hump. For Model C4, we replace \textsc{pexmon} with the \textsc{borus02} torus-geometry model \citep{Balokovic2018}, which includes a torus opening angle $i_t$ and column density $N_{H,t}$ as additional parameters. We limit the torus opening angle to $i_t>30^\circ$ (\emph{i.e.}, the line of sight, $i\equiv30^\circ$, is not obscured by the torus). For the LF, HF and J21 data sets, the evidence for Model C4 is roughly equal to that of C2, while for F23, Model C2 is preferred. In summary, either a high-energy cutoff or a Compton-thin reflector are required; we use an exponential cutoff in Models E through H.\\

\noindent\textbf{Models D$_n$ and D$_i$: intrinsic absorption}\\

\noindent D$_n$: \textsc{const$\times$tbabs$\times$ztbabs(powerlaw+nthcomp}

\textsc{+pexmon)}

\noindent D$_i$: \textsc{const$\times$tbabs$\times$zxipcf(powerlaw+nthcomp}

\textsc{+pexmon)}\\

\noindent To test for any strong intrinsic absorption, we firstly add a neutral absorber component at the systemic redshift (Model D$_n$). Model C1 is preferred over Model D$_n$ for all data sets. Thus, the intrinsic neutral absorption is negligible for Mrk 590 in both low- and high-flux states. We also test for ionized and/or partially covering absorption  using the \textsc{zxipcf} component \citep{Reeves2008} (Model D$_i$), which includes the ionization strength $\xi$ and the absorber covering fraction $CF$ as additional parameters. For the BXA analysis, we set the \textsc{zxipcf} redshift to the systemic redshift. We test covering fractions of 1, 2/3, and 1/3, where a fraction $CF-1$ of the source flux bypasses the absorber entirely. For all data sets, Model C1 is preferred over D$_i$. Finally, we test whether an ionized absorber is in- or outflowing at a relativistic velocity, which would produce a shift in the energies of the absorption troughs. To avoid a large computational burden, we turn to \emph{C-stat} minimization for this test. We allow $0.33<CF<1$, and allow the absorber redshift to vary from the systemic by up to $\Delta v=\pm$0.5c. We find that the column density $N_{\mathrm{H,int}}$ converges at its lower limit of $10^{20}$ cm$^{-2}$, while both the absorber redshift and $\xi$ are very poorly constrained. This confirms the lack of strong intrinsic absorption in our data.\\ 

\noindent\textbf{Model E: relativistic reflection only} \\

\noindent\textsc{const$\times$tbabs(powerlaw+relxillLpCp)}\\

\noindent Reflection in an optically thick accretion disk, commonly invoked to explain broad Fe K emission lines \citep[\emph{e.g.},][]{Nandra2007,Fabian2009}, can also produce a soft excess due to relativistically blurred atomic transitions \citep[\emph{e.g.},][]{Crummy2006, Fabian2009, Mundo2020, Xu2021}. We use the \textsc{Relxill} suite of models \citep{Garcia2014,Dauser2014,Dauser2022} to explore whether the observed soft excess features can be fully explained by accretion disk reflection. Of the available \textsc{Relxill} variants, we found that only the variable-density model \textsc{RelxillLpCp} generates sufficient blurred soft reflection to roughly match the observed soft X-ray flux. This is broadly consistent with the findings of \citet{Mallick2022} and \citet{Yu2023} for low-mass AGN. The geometry of this model is a `lamp-post' irradiating source above and below the disk. The reflection strength\footnote{Subtle differences in the definitions of the reflection strengths between \textsc{relxill} and \textsc{pexmon} are discussed by \citet{Dauser2022}.} for \textsc{relxill} models is parameterized by $R_{\mathrm{rel}}$, where $R_{\mathrm{rel}}=-1$ is the theoretically predicted strength for disk reflection. The local ionization level of the reflecting material is radially stratified according to the thin-disk prediction, and is parameterized by its level $\xi_{\mathrm{rel}}$ at the innermost stable circular orbit. Typically, higher ionization produces a more featureless reflection spectrum \citep[\emph{e.g.},][]{Garcia2014}. We keep $\xi_{\mathrm{rel}}$ constant in our individual BXA runs, but `step through' five different values: $\log(\xi)=(0;1;2;3;4)$. The black hole spin parameter $a_*$ affects the reflection fraction largely by changing the radius of the inner disk. We test variants with a non-spinning ($a_*=0)$ and maximally spinning ($a_*=0.998$) black hole. Qualitatively, \textsc{RelxillLpCp} can produce sufficient soft X-rays to roughly match the data, but lacks the required strong narrow 6.4 keV line to match the LF data set (Figure \ref{fig:appendixC_modelE_LF}). We conclude that some contribution from distant reflection is necessary. In all cases, the evidence for model E is lower than for Model C.\\

\noindent\textbf{Model F: warm Comptonization and relativistic reflection} \\ 

\noindent\textsc{const$\times$tbabs(powerlaw+nthcomp+relxillLpCp)}\\

\noindent For completeness, we test a model including both warm Comptonization and relativistic reflection, but without distant reflection. Similarly to Model E, this fails to produce sufficient narrow Iron emission to match the data, and is disfavored relative to models including a distant reflection component. \\

\noindent\textbf{Model G (preferred!): warm Comptonization, distant reflection, and relativistic reflection}\\

\noindent\textsc{const$\times$tbabs(zcutoffpl+nthcomp}

\textsc{+pexmon+relxillLpCp)}\\

\noindent If Mrk 590 indeed harbors a standard `thin disk' extending to the innermost stable orbit, we would expect some contribution from relativistic reflection. To explore this, we now include both distant and relativistic reflection components. We tie the incident photon indices $\Gamma$ and high-energy cutoffs $E_{\mathrm{cut}}$ of the two reflection components to that of the power law continuum, but allow them to have different reflection contributions. For this dual-reflection model, we fix the relativistic component reflection strength to $R_{\mathrm{rel}}\equiv-1$ and allow the \textsc{RelxillLpCp} overall normalization to vary. While not strictly consistent (as the \emph{pexmon} reflecting geometry is a uniform plane, thus 'double-counting' the reflection at small radii), this model provides an estimate of the relative contributions of relativistic and  distant reflection. As for Model E, we step through a range of ionization parameters for the disk reflection component, and test both non-rotating and maximally rotating black holes. For the LF, HF and J21 data sets, model G is marginally preferred over C2. Thus, we cannot exclude a contribution from relativistic reflection at both high and low flux levels.\\

% We set the reflection factor $R=-1$ in \textsc{relxillLpCp} such that it only returns the reflected component, and allow its normalization to vary freely.

\noindent\textbf{Model H: distant and relativistic reflection only.}\\ 

\noindent H: \textsc{const$\times$tbabs(zcutoffpl+pexmon+relxillLpCp)}\\

\noindent Finally, to investigate whether a separate warm Comptonization region is \emph{required} in addition to two reflection components, we remove the \textsc{nthcomp} component from Model G. The soft excess is then entirely due to the disk reflection component, while the distant component contributes to the narrow 6.4 keV line. This scenario is in all cases strongly disfavored relative to models including a warm Comptonization region.

\subsection{On model comparison methodology} 

Several authors have compared the blurred reflection and warm Comptonization scenarios for individual AGN. Typically, both models produce similar quality fits to the observed data, based on the reduced-$\chi^2$ fit statistic \citep[\emph{e.g.},][]{Pal2016, Garcia2019, Xu2021, Chalise2022, Yu2023, Madathil2024}. As we are working with minimally binned data, we do not use the $\chi^2$ statistic in our analyses. We can instead consider the statistic $C/$DOF, \emph{i.e.}, the ratio of the Cash statistic \emph{C} \citep{Cash1979} to the model-fit degrees of freedom. For our minimal binning scheme (\S \ref{sec:method_xspec}), the expected value of this statistic (assuming that the model is correct and the uncertainties are accurate) is $C/$DOF=1.01 \citep[][their Equation (11)]{Kaastra2017}. Thus, in analogy to the reduced $\chi^2$ statistic, a value near 1 is consistent with a satisfactory model fit, given the uncertainties. Similar to other model comparison studies, we indeed only find small differences in $C/$DOF between models with warm Comptonization and models with relativistic reflection (Table \ref{tab:bayesfactors}). This illustrates the difference between the Bayesian approach we apply, which displays a clear preference for warm Comptonization models, and selection strategies that compare optimized (`best-fit') models.

However, we note that the `best-fit' models including warm Comptonization do perform \emph{slightly} better in terms of $C/$DOF for our data. For LF, the best Model G variant has $C/$DOF$=1.05$, while the best Model H variant (\emph{i.e.}, distant and relativistic reflection, but no Comptonization) displays $C/$DOF$=1.07$. For HF, the best-performing Model G variant yields $C/$DOF=1.08, while we obtain $C/$DOF=1.15 for Model H. Thus, the models including warm Comptonization do yield a `better fit' in the traditional sense. 

\onecolumn

\begin{deluxetable}{cccccccccc}
	\tabletypesize{\footnotesize}
	\tablewidth{0pt}
	\tablecaption{Model parameter ranges used for BXA runs \label{tab:parameters}}
	\tablehead{
    Col. & \colhead{Parameter} & \colhead{Model A} & \colhead{B} & \colhead{C, C2, C3, C4} & \colhead{D$_n$, D$_i$} & \colhead{E} & \colhead{F} & \colhead{G} & \colhead{H}\\
    }
    \startdata
    (1) & $N_{\mathrm{H,Gal}}$ & \textcolor{Gray}{2.77$\times10^{20}$} & \textcolor{Gray}{2.77$\times10^{20}$} & \textcolor{Gray}{2.77$\times10^{20}$} & \textcolor{Gray}{2.77$\times10^{20}$} & \textcolor{Gray}{2.77$\times10^{20}$} & \textcolor{Gray}{2.77$\times10^{20}$} & \textcolor{Gray}{2.77$\times10^{20}$} & \textcolor{Gray}{2.77$\times10^{20}$} \\
    (2) & N$_{\mathrm{cont}}$ & \textcolor{BurntOrange}{$10^{-5}$, $10^{-2}$}  & \textcolor{BurntOrange}{$10^{-5}$, $10^{-2}$} & \textcolor{BurntOrange}{$10^{-5}$, $10^{-2}$} & \textcolor{BurntOrange}{$10^{-5}$, $10^{-2}$} & \textcolor{BurntOrange}{$10^{-7}$, $10^{-4}$} & \textcolor{BurntOrange}{$10^{-5}$, $10^{-2}$} & \textcolor{BurntOrange}{$10^{-5}$, $10^{-2}$} & \textcolor{BurntOrange}{$10^{-5}$, $10^{-2}$}\\ 
    (3) & $\Gamma$ & 1.1, 2.4 & 1.1, 2.4 & 1.1, 2.4 & 1.1, 2.4 & 1.1, 2.4 & 1.1, 2.4 & 1.1, 2.4 & 1.1, 2.4 \\ 
    (4) & N$_{\mathrm{warm}}$ &  & \textcolor{BurntOrange}{$10^{-6}$, $10^{-2}$} & \textcolor{BurntOrange}{$10^{-6}$, $10^{-2}$} & \textcolor{BurntOrange}{$10^{-6}$, $10^{-2}$} &  & \textcolor{BurntOrange}{$10^{-6}$, $10^{-2}$} & \textcolor{BurntOrange}{$10^{-6}$, $10^{-2}$} & \\ 
    (5) & $\Gamma_{\mathrm{warm}}$ &  & \textcolor{Gray}{2.5} & \textcolor{Gray}{2.5} & \textcolor{Gray}{2.5} &  & \textcolor{Gray}{2.5} & \textcolor{Gray}{2.5} \\ 
    (6) & $kT_e$ &  & 0.05, 1 & 0.05, 1 & 0.05, 1 &  & 0.05, 1 & 0.05, 1 & \\ 
    (7) & $E_{\mathrm{cut}}$ &  &  & 20, 300 &  &  &  & 20, 300 & 20, 300 \\
    (8) & $kT_{e,\mathrm{hot}}$ &  &  & 6, 100 &  &  &  & $\equiv E_{\mathrm{cut}}/3$ & $\equiv E_{\mathrm{cut}}/3$ \\
    (9) & $i$ &  &  & \textcolor{Gray}{30$^\circ$} & \textcolor{Gray}{30$^\circ$} & \textcolor{Gray}{30$^\circ$} & \textcolor{Gray}{30$^\circ$} & \textcolor{Gray}{30$^\circ$} & \textcolor{Gray}{30$^\circ$} \\ 
    (10) & $Z$ &  &  & \textcolor{Gray}{1} & \textcolor{Gray}{1} & \textcolor{Gray}{1} & \textcolor{Gray}{1} & \textcolor{Gray}{1} & \textcolor{Gray}{1} \\ 
    (11) & $A_{\mathrm{Fe}}$ &  &  & \textcolor{Gray}{5} & \textcolor{Gray}{5} & \textcolor{Gray}{5} & \textcolor{Gray}{5} & \textcolor{Gray}{5} & \textcolor{Gray}{5} \\ 
    (12) & $R$ &  &  &  -5, -0.05 & -5, -0.05 &  &  & -5, -0.05 & -5, -0.05 \\
    (13) & $N_{\mathrm{H,int}}$ &  &  &  & \textcolor{BurntOrange}{$10^{20}$, $10^{24}$} \\ 
    (14) & $\xi_{\mathrm{abs}}$ &  &  &  & \textcolor{BurntOrange}{$10^{-3}$, $10^{5}$} \\ 
    (15) & $z_{\mathrm{abs}}$ &  &  &  & -0.48, 0.52 &  &  &  &  \\ 
    (16) & $CF$ &  &  &  & \textcolor{PineGreen}{1, $\frac{2}{3}$, $\frac{1}{3}$} & \\
    (17) & $n_\mathrm{rel}$ &  &  &  &  & \textcolor{BurntOrange}{$10^{15}$, $10^{20}$} & \textcolor{BurntOrange}{$10^{15}$, $10^{20}$} & \textcolor{Gray}{$10^{15}$} & \textcolor{BurntOrange}{$10^{15}$, $10^{20}$} \\ 
    (18) & $\log(\xi)_\mathrm{rel}$ &  &  &  &  & \textcolor{PineGreen}{0,1,2,3,4} & \textcolor{PineGreen}{0,1,2,3,4} & \textcolor{PineGreen}{0,1,2,3,4} & \textcolor{PineGreen}{0,1,2,3,4} \\ 
    (19) & $a_\mathrm{*,rel}$ &  &  &  &  & \textcolor{PineGreen}{0, 0.998} & \textcolor{PineGreen}{0, 0.998} & \textcolor{PineGreen}{0, 0.998} & \textcolor{PineGreen}{0, 0.998} \\ 
    (20) & $R_{\mathrm{rel}}$ &  &  &  &  & -5, -0.05 & -5, -0.05 & \textcolor{Gray}{1} & \textcolor{Gray}{1} \\ 
    (21) & $h_{\mathrm{rel}}$ &  &  &  &  & \textcolor{Gray}{10} & \textcolor{Gray}{10} & \textcolor{Gray}{10} & \textcolor{Gray}{10} \\ 
    (22) & $\beta_{\mathrm{rel}}$ &  &  &  &  & \textcolor{Gray}{0} & \textcolor{Gray}{0} & \textcolor{Gray}{0} & \textcolor{Gray}{0} \\
    \enddata
    \tablecomments{Allowed parameter ranges for the different models tested in our BXA analysis (\ref{sec:analysis_joint}). Parameters that are held constant during our BXA runs are listed in \textcolor{Gray}{gray}. Parameters that are `stepped through' (\emph{i.e.}, we perform multiple BXA runs with different constant values of that parameter) are highlighted in \textcolor{PineGreen}{green}. For parameters that may plausibly range over several orders of magnitude we adopt a log-uniform prior; these are listed in \textcolor{BurntOrange}{orange}. For all other parameters we adopt a uniform prior between the lower and upper limits listed. We note that some parameters are held constant or `stepped through' here, but instead treated as free parameters in our final analysis (\S \ref{sec:bestmodels}). This is due to the high computational demands of the BXA runs.\\
    \emph{Remarks on individual parameters:} (1) Galactic column density in cm$^{-2}$. (2) The normalization of the primary continuum component. For \textsc{powerlaw} or \textsc{zcutoffpl} continua, this is given as photons keV$^{-1}$ cm$^{-2}$ measured at 1 keV. For \textsc{relxill} models the normalization is as defined by \citet{Dauser2022}.(3) The photon index of the primary (observed) continuum; for reflection models we tie this to the $\Gamma$ of the incident continuum. (4) The normalization of the warm Comptonization (\textsc{nthcomp}) component. (5) The photon index of the warm Comptonization component. (6) The electron temperature of the warm Comptonization component, in keV. (7) The cutoff energy for \textsc{zcutoffpl} components, in keV. (8) The electron temperature of the hot Comptonization component, in keV. This is treated as a free parameter for Model C3 only.(9) The reflection plane inclination, where $i=0$ corresponds to face-on, is held constant for all reflection models. (10) The metallicity, assumed Solar for all reflection models. (11) The Iron abundance, set to five times Solar for all reflection models, for reasons discussed in Appendix \ref{AppendixB_modeldefs}. (12) The reflection factor for \textsc{pexmon} components. (13) The column density of intrinsic absorption components (neural or ionized), in cm$^{-2}$. (14) The ionization parameter of an ionized absorber. (15) The in- or outflow velocity of an ionized absorber, expressed as a redshift. (16) The covering fraction of an ionized absorber, where $CF=1$ denotes full coverage of the continuum source. (17) The disk particle density for \textsc{relxillCpLp} models, in cm$^{-3}$. (18) The ionization parameter of \textsc{relxillLp} or \textsc{relxillCpLp} components, parameterized at the innermost stable circular orbit. (19) The black hole spin parameter, for  \textsc{relxillLp} or \textsc{relxillCpLp} components. (20) The reflection strength of relativistic reflection, as defined by \citet{Dauser2022}. (21) The height of the X-ray continuum source above the disk, in gravitational radii. (22) The velocity of the continuum source relative to the disk.}
\end{deluxetable}

\begin{deluxetable}{cccccccccc}
	\tabletypesize{\small}
	\tablewidth{0pt}
	\tablecaption{Model comparison for joint data sets \label{tab:bayesfactors}}
	\tablehead{
    \colhead{Model} & \colhead{Variant} & \colhead{LF} & \colhead{HF} & \colhead{J21} & \colhead{F23} & \colhead{LF} & \colhead{HF} & \colhead{J21} & \colhead{F23}\\
    \colhead{} & \colhead{} & \colhead{$\Delta\log(Z)$} & \colhead{$\Delta\log(Z)$} & \colhead{$\Delta\log(Z)$} & \colhead{$\Delta\log(Z)$} & \colhead{\emph{C}/DOF} & \colhead{\emph{C}/DOF} & \colhead{\emph{C}/DOF} & \colhead{\emph{C}/DOF} \\
    \colhead{(1)} & \colhead{(2)} & \colhead{(3)} & \colhead{(4)} & \colhead{(5)} & \colhead{(6)} & \colhead{(7)} & \colhead{(8)} & \colhead{(9)} & \colhead{(10)} \\
    }
    \startdata
    A  & & 608.8 & 1309.6 & 160.4 & 268.4 & 5936/4674 & 8266/5764 & 3019/2593 & 3056/2671 \\ 
    B  & & 158.8 & 75.2 & 62.7 & 89.6 & 5247/4672 & 6372/5762 & 2747/2591 & 2776/2669 \\ 
    C  & & 12.8 & 3.3 & 14.5 & 13.6 & 4948/4671 & 6218/5761 & 2640/2590 &  2619/2668 \\ 
    C2 & cutoff PL & \textcolor{BurntOrange}{1.9} & \textcolor{BurntOrange}{0.6} & \textcolor{PineGreen}{$\bigstar$} & \textcolor{PineGreen}{$\bigstar$} & 4935/4670 & 6215/5760 & 2610/2589 & 2589/2667  \\ 
    C3 & \textsc{nthcomp} cont. &  9.8 & 10.8 & 4.1 & 7.9 & 4936/4670 & 6231/5760 & 2609/2589 & 2596/2667 \\ 
    C4 & Torus & \textcolor{BurntOrange}{2.6} & \textcolor{BurntOrange}{0.4} & \textcolor{BurntOrange}{2.4} & 5.0 & 4942/4669 & 6197/5759 & 2622/2588 & 2600/2666 \\
    D$_n$ & & 13.9 & 18.0 & 12.9 & 10.8 & 4933/4670 & 6278/5760 &  2631/2589 & 2604/2667 \\ 
    D$_i$ & CF=1 & 31.9 & 36.4 & 15.2 & 17.7 & 4985/4669 & 6245/5759 & 2693/2588 &  2641/2666 \\ 
    D$_i$ & CF=0.66 & 31.9 & 25.2 & 12.6 & 17.7 & 4985/4669 & 6444/5759 & 2751/2588 & 2634/2666 \\ 
    D$_i$ & CF=0.33 & 4.9 & 35.4 & 11.6 & \textcolor{BurntOrange}{0.1} & 4998/4669 & 6438/5759 & 2656/2588 & 2637/2666 \\ 
    E & $\log\xi=0$, $a_*=0$ & 100.4 & 208.5 & 46.1 & 58.6 & 5192/4672 & 6680/5761 & 2719/2590 & 2695/2668 \\ 
    E & $\log\xi=1$, $a_*=0$ & 99.2 & 202.0 & 43.6 & 55.8 & 5129/4672 & 6656/5761 & 2702/2590 & 2691/2668 \\ 
    E & $\log\xi=2$, $a_*=0$ & 89.0 & 188.0 & 41.6 & 52.6 & 5105/4672 & 7759/5671 & 2702/2590 & 2685/2668 \\ 
    E & $\log\xi=3$, $a_*=0$ & 83.6 & 194.4 & 40.4 & 55.1 & 5094/4672 & 6618/5761 & 2690/2590 & 2680/2668 \\ 
    E & $\log\xi=4$, $a_*=0$ & 116.8 & 200.2 & 51.4 & 75.3 & 5164/4672 & 6751/5761 & 2708/2590 & 2700/2668 \\ 
    E & $\log\xi=0$, $a_*=0.98$ & 101.8 & 208.4 & 47.0 & 59.4 & 5154/4672 & 6850/5761 & 2741/2590 & 2710/2668 \\ 
    F & $\log\xi=0$, $a_*=0$ & 52.8 & 52.8 & 39.6 & 40.1 & 5016/4670 & 6261/5760 & 2664/2589 & 2642/2667 \\
    F & $\log\xi=1$, $a_*=0$ & 55.7 & 29.1 & 38.5 & 40.9 & 5022/4670 & 6266/5760 & 2664/2589 & 2643/2667 \\
    F & $\log\xi=2$, $a_*=0$ & 59.2 & 39.7 & 38.8 & 43.6 & 5026/4670 & 6286/5760 & 2662/2589 & 2653/2667 \\
    F & $\log\xi=3$, $a_*=0$ & 80.5 & 29.5 & 43.4 & 57.3 & 5055/4670 & 6266/5670 & 2674/2589 & 2672/2667 \\
    F & $\log\xi=4$, $a_*=0$ & 114.1 & 33.2 & 37.8 & 56.7 & 5141/4670 & 6283/5760 & 2695/2589 & 2689/2667 \\
    F & $\log\xi=0$, $a_*=0.98$ & 56.6 & 31.2 & 41.0 & 42.7 & 5023/4670 & 6269/5760 & 2666/2589 & 2645/2667 \\
    F & $\log\xi=3$, $a_*=0.98$ & 81.8 & 33.0 & 43.9 & 57.8 & 5055/4570 & 6272/5760 & 2675/2589 & 2672/2667 \\
    %G & $\log\xi=0$, $a_*=0$ & -9.8 & -4.7 & -15.9 & -15.2 & 4937/4670 & 6218/5760 & 2646/2589 & 2618/2667 \\ 
    G & $\log\xi=0$, $a_*=0$ & \textcolor{BurntOrange}{3.0} & \textcolor{PineGreen}{$\bigstar$} & 12.7 & 8.6 & 4918/4668 & 6209/5759 & 2610/2588 & 2586/2666 \\ 
    G & $\log\xi=1$, $a_*=0$ & 22.9 & 39.9 & 7.1 & 9.2 & 4923/4668 & 6210/5759 & 2610/2588 & 2586/2666 \\
    G & $\log\xi=2$, $a_*=0$ & 14.2 & 38.4 & 4.7 & 6.3 & 4918/4668 & 6210/5759 & 2608/2588 & 2584/2666 \\
    G & $\log\xi=3$, $a_*=0$ & \textcolor{BurntOrange}{0.9} & 7.3 & \textcolor{BurntOrange}{2.5} & \textcolor{BurntOrange}{2.9} & 4913/4668 & 6206/5759 & 2608/2588 & 2585/2666 \\
    G & $\log\xi=4$, $a_*=0$ & 16.8 & 39.9 & 23.2 & 6.9 & 4921/4668 & 6209/5759 & 2610/2588 & 2586/2666 \\
    G & $\log\xi=0$, $a_*=0.98$ & 21.0 & 39.7 & 7.4 & 9.4 & 4921/4668 & 6210/5759 & 2611/2588 & 2586/2666 \\
    G & $\log\xi=1$, $a_*=0.98$ & 23.3 & 40.8 & 7.4 & 9.7 & 4923/4668 & 6210/5759 & 2610/2588 & 2585/2666 \\
    G & $\log\xi=2$, $a_*=0.98$ & 15.2 & 38.7 & 4.9 & 6.4 & 4919/4668 & 6211/5759 & 2609/2588 & 2584/2666 \\
    G & $\log\xi=3$, $a_*=0.98$ & \textcolor{PineGreen}{$\bigstar$} & 9.2 & \textcolor{BurntOrange}{2.2} & \textcolor{BurntOrange}{2.6} & 4912/4668 & 6207/5759 & 2608/2588 & 2585/2666 \\
    G & $\log\xi=4$, $a_*=0.98$ & 17.1 & 40.1 & 6.6 & 6.8 & 4921/4668 & 6209/5759 & 2610/2588 & 2586/2666 \\
    H & $\log\xi=0$, $a_*=0$ & 79.8 & 182.4 & 27.4 & 38.3 & 5254/4670 & 7302/5760 & 2758/2589 & 2792/2667 \\
    H & $\log\xi=1$, $a_*=0$ & 74.0 & 208.9 & 29.8 & 40.5 & 5037/4670 & 6666/5760 & 2670/2589 & 2637/2667 \\
    H & $\log\xi=2$, $a_*=0$ & 57.9 & 208.1 & 24.3 & 36.2 & 5046/4670 & 6618/5760 & 2679/2589 & 2633/2667 \\
    H & $\log\xi=3$, $a_*=0$ & 52.7 & 201.7 & 19.2 & 36.0 & 4994/4670 & 6603/5760 & 2659/2589 & 2633/2667 \\
    H & $\log\xi=4$, $a_*=0$ & 75.6 & 307.9 & 29.0 & 48.0 & 5021/4670 & 6776/5760 & 2681/2589 & 2652/2667 \\
    \enddata
    \tablecomments{Comparison of Bayesian evidence $Z$ (columns 3--6) for different models (column 1) fitted to our joint data sets (\ref{sec:analysis_joint}). Models A through H are described in Appendix \ref{AppendixB_modeldefs}. For models where we `step through' discrete values of certain parameters, the relevant values are listed in column (2). The evidence (\emph{i.e.}, the marginal likelihood) is calculated using the BXA software \citep{Buchner2014}, over the parameter space defined by the bounds in Table \ref{tab:parameters}. The model with the highest evidence, for a given data set, is indicated with a green star symbol (\textcolor{PineGreen}{$\bigstar$}). For all other models, we tabulate the \emph{difference} in log-evidence, $\Delta\log(Z)=\log(Z_{\mathrm{best}})-\log(Z_{\mathrm{model}})$, between that model and the best (\textcolor{PineGreen}{$\bigstar$}) model. Models colored \textcolor{BurntOrange}{orange} have $\Delta\log(Z)<3$; adopting a rather conservative threshold (\S \ref{sec:modelcomparison}), we do not consider such models to be decisively disfavored compared to the best model. We also tabulate the Cash statistic and degrees of freedom (\emph{C}/DOF, columns 7--10) for an \textsc{Xspec} C-stat optimization, initiated at the posterior peak-likelihood parameter values obtained during the relevant BXA run, to provide an idea of the goodness of fit for that model.}
\end{deluxetable}

\twocolumn

%% file: appendixC/appendixC.tex
\section{Comparison between near-contemporaneous \emph{Swift} XRT and \emph{XMM-Newton} spectra}\label{AppendixC}

\begin{figure}
\includegraphics[scale=0.53]{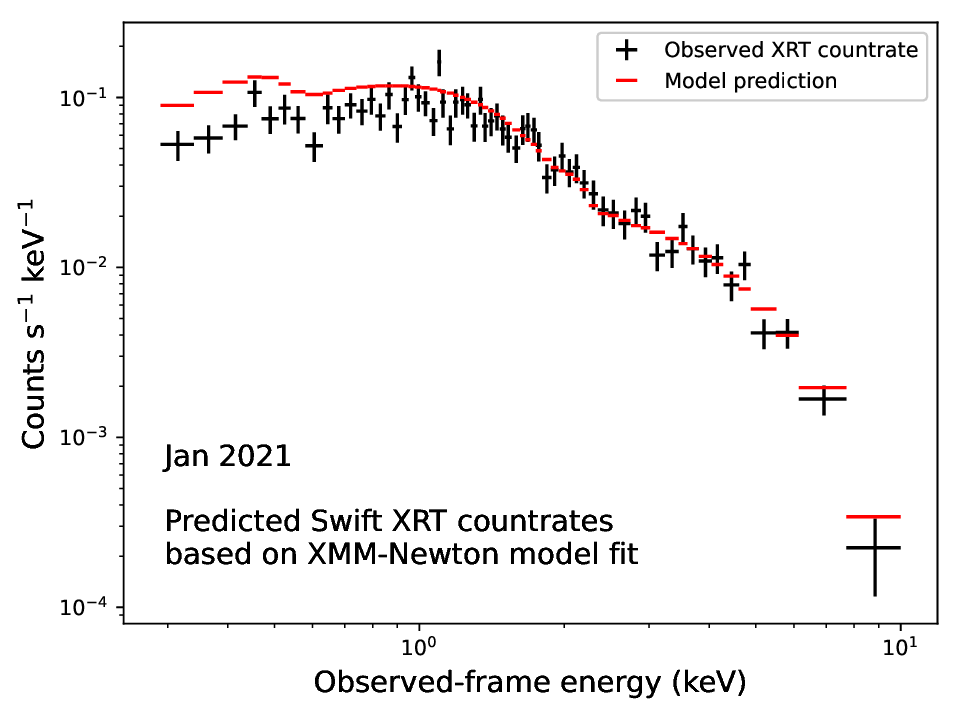}
\caption{The predicted \emph{Swift} XRT count rates (red lines), based on folding the best-fit phenomenological model (\S \ref{sec:analysis_bbody} for the 3rd January 2021 \emph{XMM-Newton} observation with the XRT instrumental response, is significantly higher in soft X-rays than the observed XRT data (black crosses; January 10th 2021). The predicted data are rescaled in flux to match the 2--10 keV flux level measured by XRT. Thus, either the source spectral index varied significantly between 3rd and 10th January 2021, or XRT was less sensitive to soft X-rays than \emph{XMM-Newton} during these observations. \label{fig:xrt_comparison}}
\end{figure}

We consistently detect a soft X-ray excess in our \emph{XMM-Newton} observations (\S \ref{sec:analysis_bbody}), even at the lowest observed continuum flux levels. In fact, after accounting for the X-ray continuum flux dependence of the soft excess, we see no indication that the soft excess behavior changed after the 2010s changing-look event, compared to archival spectra observed 2002--2004. This is at odds with the findings of \citet{Ghosh2022}, who analyze \emph{Swift} XRT data and report a disappearance of the soft excess at low flux levels during 2016--2022. To investigate this discrepancy, we retrieve the 10th January 2021 XRT observation of Mrk 590 (obsid: 00095662033), which is the longest-duration (10 ks) XRT pointing occurring within a few days of a \emph{XMM-Newton} observations. We firstly confirm that the XRT data are consistent with no soft excess by fitting our phenomenological model (\S \ref{sec:analysis_bbody}) using \textsc{Xspec}; the blackbody normalization for this model fit is consistent with zero within its 90\% confidence interval. Thus, we obtain the same qualitative results as \citet{Ghosh2022} when modeling the XRT data.

While it is in principle possible that the soft excess component weakened significantly between the \emph{XMM-Newton} and XRT observations, this seems unlikely given its tight correlation with continuum flux (Figure \ref{fig:bbody}). Assuming that the soft excess strength \emph{relative to the continuum flux} does not vary between the observations, we generate a simulated XRT spectrum based on the best-fit phenomenological model for the 3rd January \emph{XMM-Newton} data. We apply a flux scaling term (\textsc{const}) to the best-fit model, such that its integrated 2--10 keV flux is equal to that of the XRT data (measured between 2--10 keV using a simple power law model). We use the \textsc{Xspec} task \emph{`fakeit'} to generate the simulated data, applying the appropriate ARF and RMF files for the XRT detector. The resulting simulated data (red lines, Figure \ref{fig:xrt_comparison}) display a significantly higher count rate below $\sim1$ keV than do the observed XRT data (black crosses). The integrated count rate of the simulated spectrum between 0.3--1 keV is 0.11 counts keV$^{-1}$ s$^{-1}$, while the XRT observation displays 0.08 counts keV$^{-1}$ s$^{-1}$. Based on the naive assumption of a constant soft excess contribution, it appears that the sensitivities of \emph{XMM-Newton} and \emph{Swift} XRT may differ by $\sim27\%$ below 1 keV. Alternatively, it is possible that the 10 ks XRT observation was too short to detect the soft excess at this flux level, and the offsets we are seeing are purely due to statistical fluctuations. We cannot make any definitive conclusions regarding instrumental sensitivity based on a single comparison spectrum, especially as the observations are not fully contemporaneous. However, we note that \citet{Hagen2023} report a similar $\sim$30\% deficit of X-ray counts below 1 keV for XRT observations of the AGN Fairall 9, compared to their \emph{XMM-Newton pn} spectra; see their Appendix D for details. 

Given this significant count-rate offset, our results regarding the soft excess in Mrk 590 will depend strongly on whether we base our work on \emph{XMM-Newton} or \emph{Swift} XRT data. We choose to rely on \emph{XMM-Newton} for our spectral modeling in this work, due to its higher sensitivity, the longer exposure times of the available \emph{XMM-Newton} observations of Mrk 590, and that \emph{XMM-Newton} is more commonly used for soft X-ray excess studies.

%% file: appendixD/appendixD.tex
\section{Alternative configurations of the three-component \textsc{agnsed} model}\label{AppendixD}

Extending our preferred interpretation of the X-ray data (\emph{i.e.}, a substantial contribution from warm Comptonized emission) into the UV--optical regime, we find a reasonable overall match to the observed \emph{Swift} photometry using the \textsc{agnsed} \citep{Kubota2018} model with no thermal emission from an outer `thin disk' (\S \ref{sec:discussion_SED}). We also find a large inner truncation for the warm Comptonizing region, with $R_{\mathrm{hot}}\sim100$ $r_g$. In this Appendix, we present modeling results for alternative configurations of the \textsc{agnsed} model, which support our choice of the variant with no outer `thin disk' for the main analysis. We test four variants: \emph{1)} a model with an upper parameter bound $R_{\textsc{warm}}<10^3$ $r_g$ such that at least some outer disk is present, \emph{2)} a model with $R_{\mathrm{hot}}\equiv10$ $r_g$ (\emph{i.e.}, a compact X-ray continuum source), \emph{3)} a model where no warm-Comptonized emission is present, and \emph{4)} a model with the hot corona temperature $kT_{e\mathrm{,hot}}=50$ keV, with a corresponding reduction in the energy emitted as hard X-rays. We show model fits of these variants, for each joint data set, in Figures \ref{fig:HF_uvmodel} through \ref{fig:F23_uvmodel}.

\paragraph*{The thin-disk contribution appears negligible.} While the `hybrid' (\emph{i.e.}, $R_{\mathrm{warm}}<10^3$ $r_g$) variant produces a reasonable match to the UV photometry, it typically does so by pegging $R_{\textsc{warm}}$ at its upper limit, such that the outer disk emission is minimized. For the LF data set, \textsc{agnsed} cannot fully account for the observed UV--optical emission, as discussed in \S \ref{sec:discussion_SED} . It appears too bright to be solely due to the Comptonized flux component, and too blue to be attributed to a truncated outer disk. We defer further investigation of this issue to future work, as additional spectroscopic data will help distinguish between broadband SED models, and provide additional constraints on the all-important host galaxy stellar contribution. For the HF data set (Figure \ref{fig:HF_uvmodel}, top left), even the small contribution from the outer disk with $R_{\textsc{warm}}=10^3$ $r_g$ overestimates the flux in the optical bands. Removing the cool outer disk entirely provides a better match to observations.

\paragraph*{Models without warm-Comptonized emission predict a too-red UV--optical slope.} While the `no warm-Comptonized emission' model can very roughly match the overall UV-optical brightness, it fails to predict the shape of the optical--UV spectral slope (Figures \ref{fig:HF_uvmodel} through \ref{fig:F23_uvmodel}, top right panels). This can, in a sense, be attributed to the requirement of energetic consistency in \textsc{agnsed}, as follows. Firstly, $R_{\mathrm{hot}}$ must be large, in order to produce sufficient X-ray continuum emission to match the data. Then, due to the large $R_{\mathrm{hot}}$, the inner edge of any `thin disk' is distant from the central black hole; it is therefore cooler than a non-truncated disk, for a given black hole mass. The intrinsically cool disk emission must therefore be reprocessed by a warm Comptonization region to produce a sufficiently blue continuum to match our observations.

\paragraph*{Compact-corona variants produce insufficient X-rays at a given UV flux level.} The $R_{\mathrm{hot}}\equiv10$ $r_g$ variant significantly over-predicts the optical--UV flux levels, resulting in obviously poor model fits (Figures \ref{fig:HF_uvmodel} through \ref{fig:F23_uvmodel}, bottom left panels). In practice, this is due to the X-ray data having more `weight' in our modeling procedure: because we have very deep X-ray observations, a better fit statistic is obtained by matching the X-ray flux at the expense of the UV, rather than vice versa. A more physically meaningful inference is that, if the accretion energy really is dissipated according to the thin-disk prescription as a function of radius (as assumed by \textsc{agnsed}), the corona must have a much larger radius, $R_{\mathrm{hot}}\sim100$ $r_g$, to produce the observed UV to X-ray slope.

\paragraph*{Lowering the hot coronal temperature does not significantly alter the energy budget.} By reducing the temperature of the hot corona such that the high-energy cutoff is lowered, we can reduce the amount of energy emitted as X-ray continuum, for a given (\emph{e.g.},) 10 keV brightness. As \textsc{agnsed} is an energy-conserving model, this corresponds to a smaller fraction of the overall accretion energy being deposited in the corona. In the context of the assumed \textsc{agnsed} geometry, this should lead to a smaller $R_{\mathrm{hot}}$ at lower coronal temperatures $kT_{e,\mathrm{hot}}$. To test to what degree this would affect our qualitative results, we fit an \textsc{agnsed} model with $kT_{e,\mathrm{hot}}\equiv50$ keV. We choose this value as the high-energy cutoff in the X-ray spectra is in all cases 150 keV or higher (\S \ref{sec:bestmodels}); the high-energy cutoff is thought to be roughly 2--3 times the Comptonization temperature \citep{Zycki1999}, thus 50 keV is the lowest temperature warranted by our \emph{NuSTAR} data. Inspection of the best-fit models with  $kT_{e,\mathrm{hot}}\equiv50$ keV reveals that the overall picture is unchanged: the warm-Comptonized emission is the dominant contribution to the UV. We find that the best-fit $R_{\mathrm{hot}}$ values are indeed $\sim20$\% smaller for these models. However, they are still of order 100 $r_g$, and the difference is not highly significant given the substantial uncertainties on $R_{\mathrm{hot}}$ for the best-fit \textsc{agnsed} models. Neither do we see significantly reduced $R_{\mathrm{warm}}$ values, despite the reduction in $R_{\mathrm{hot}}$. We conclude that this effect is too subtle to accurately measure for the data presented in this work, and that it does not affect our qualitative result that the UV spectral energy distribution is more consistent with warm-Comptonized emission.

\begin{figure*}
    \centering
    \includegraphics[scale=0.395, trim={5 0 10 0}, clip]{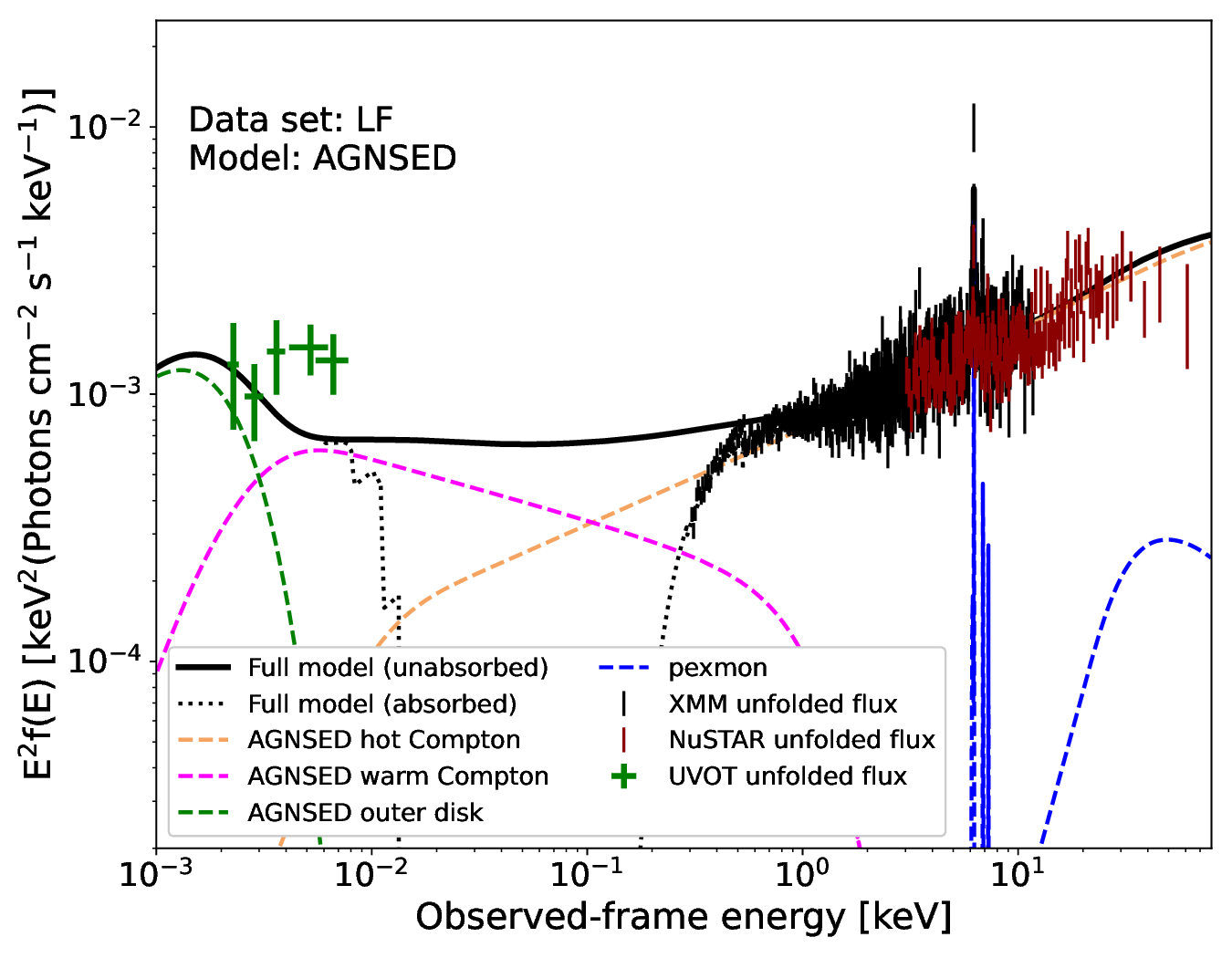}
    \includegraphics[scale=0.395, trim={30 0 10 0}, clip]{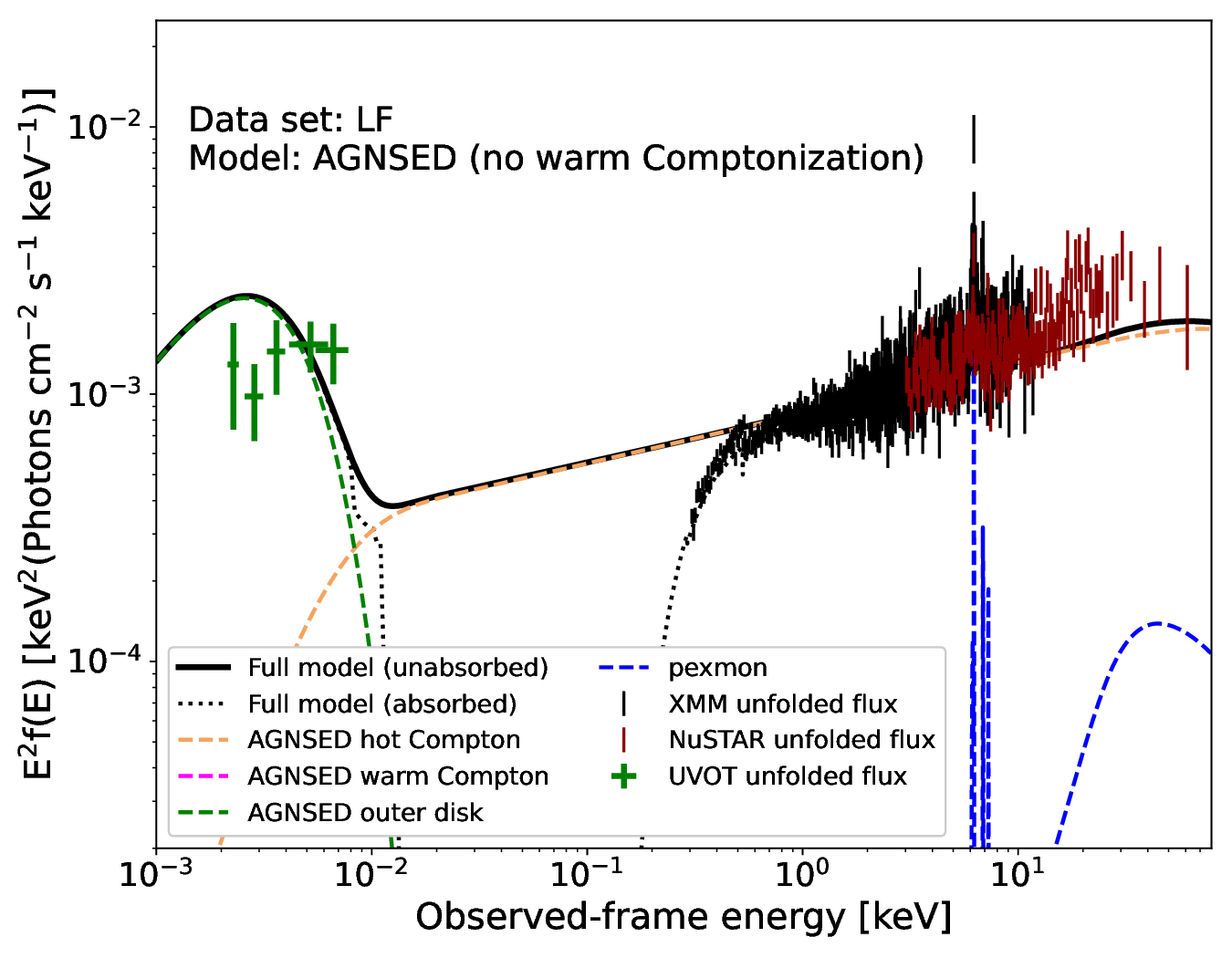}
    \includegraphics[scale=0.395, trim={5 0 10 0}, clip]{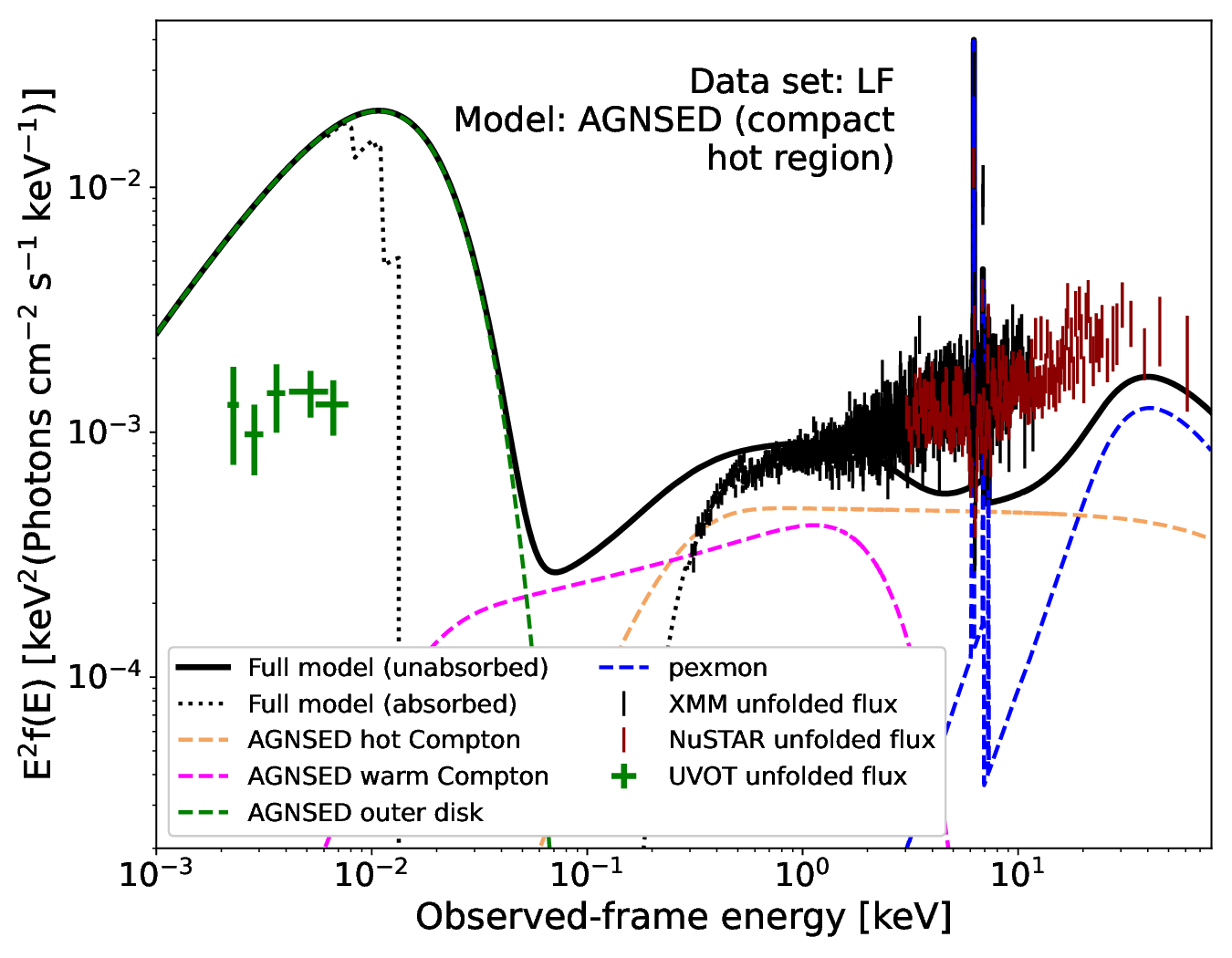}
    \includegraphics[scale=0.395, trim={10 0 10 0}, clip]{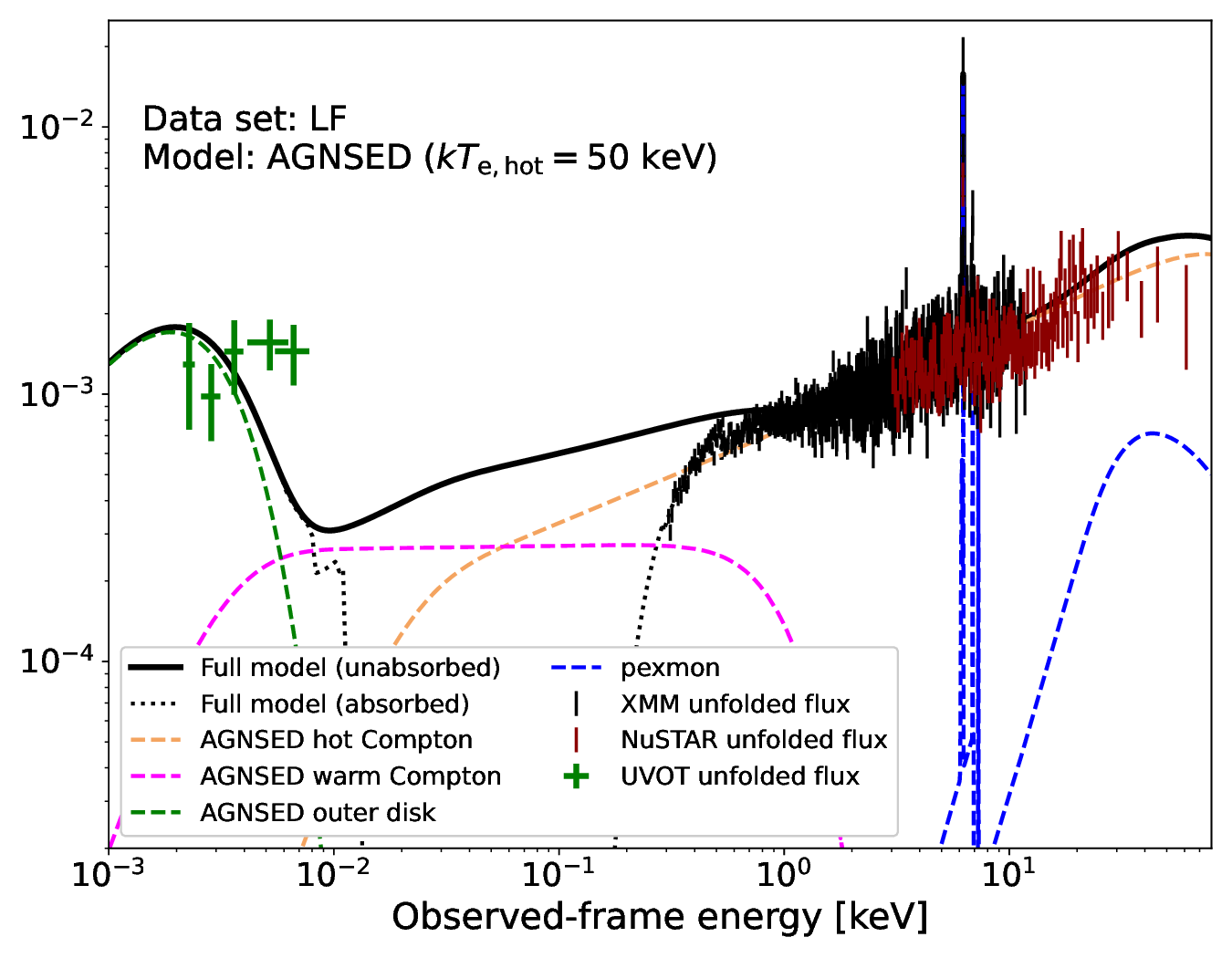}
    \caption{Alternative \textsc{agnsed + pexmon} model configurations for the LF data set. Here, we include the median-combined and host galaxy starlight-subtracted UVOT photometry; see text of \S \ref{sec:discussion_SED} for details. We display the unfolded \emph{Swift} UVOT, \emph{XMM-Newton pn}, and \emph{NuSTAR} FPMA spectra; data from \emph{XMM-Newton} MOS and \emph{NuSTAR} FPMB are included in the modeling but not shown. \emph{Top left:} We include both an outer disk, warm Comptonization region, and hot Comptonization region, requiring only that $R_\mathrm{warm}<10^3$ $r_g$. \emph{Top right:} Here, we apply a two-component \textsc{agnsed} model, with no warm Comptonization region. \emph{Bottom left:} Here, we constrain the hot Comptonization region to a compact size, with $R_\mathrm{hot}\equiv10r_g$. This yields a very poor fit even to the X-ray spectra, due to the trade-off between matching the UV and the X-ray flux levels while diverting only a small fraction of the accretion energy into the hot corona. \emph{Bottom right:} Here, we set the hot region of \textsc{agnsed} to an electron temperature of 50 keV, as opposed to 300 keV in the other cases. We note that the LF UV-optical data are not satisfactorily modeled by any \textsc{agnsed} variant; however, the UV flux level is high enough to allow a strong contribution from the warm-Comptonized emission component identified in our X-ray modeling. \label{fig:LF_uvmodel}}
\end{figure*}

\begin{figure*}
    \centering
    \includegraphics[scale=0.395, trim={5 0 10 0}, clip]{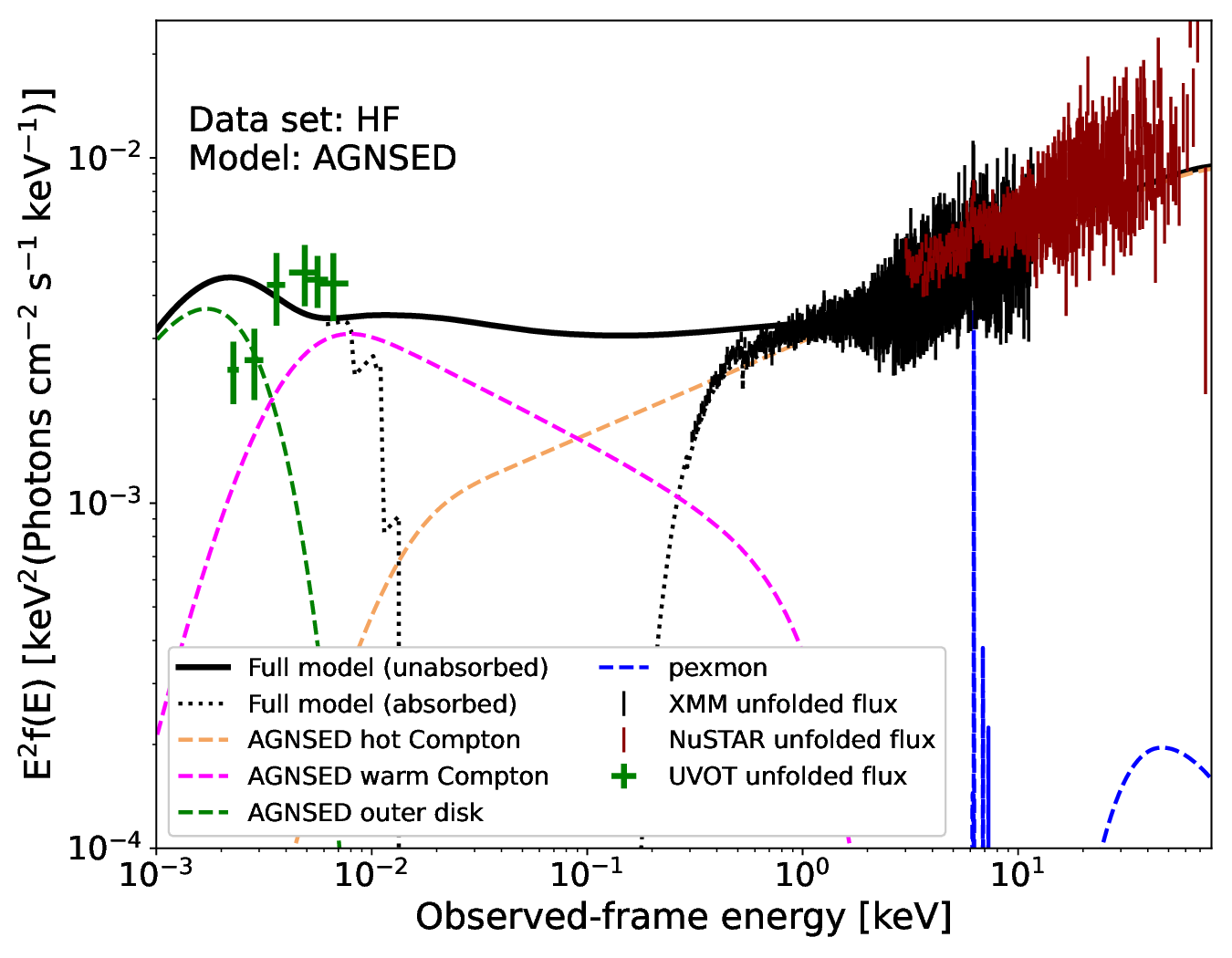}
    \includegraphics[scale=0.395, trim={30 0 10 0}, clip]{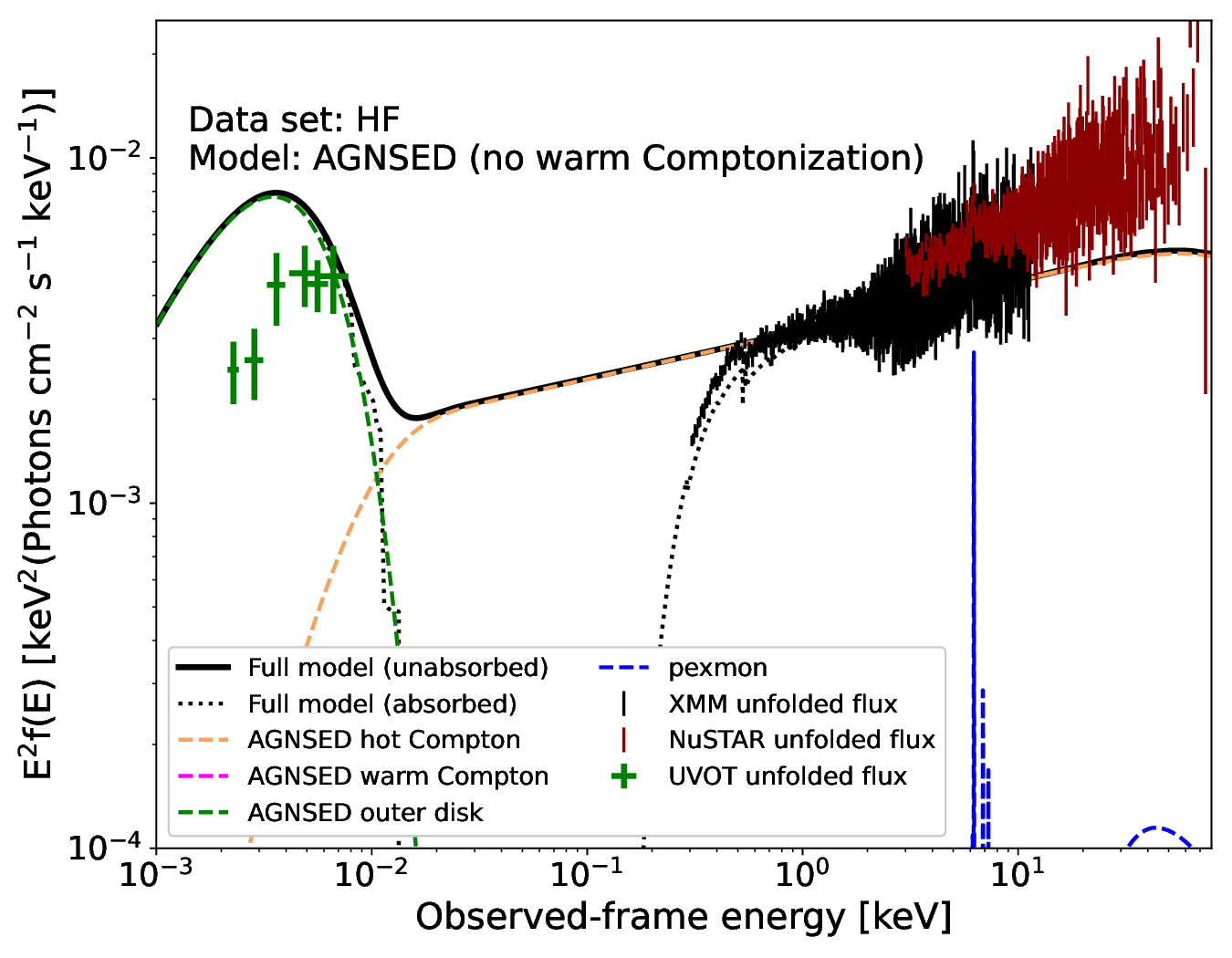}
    \includegraphics[scale=0.395, trim={5 0 10 0}, clip]{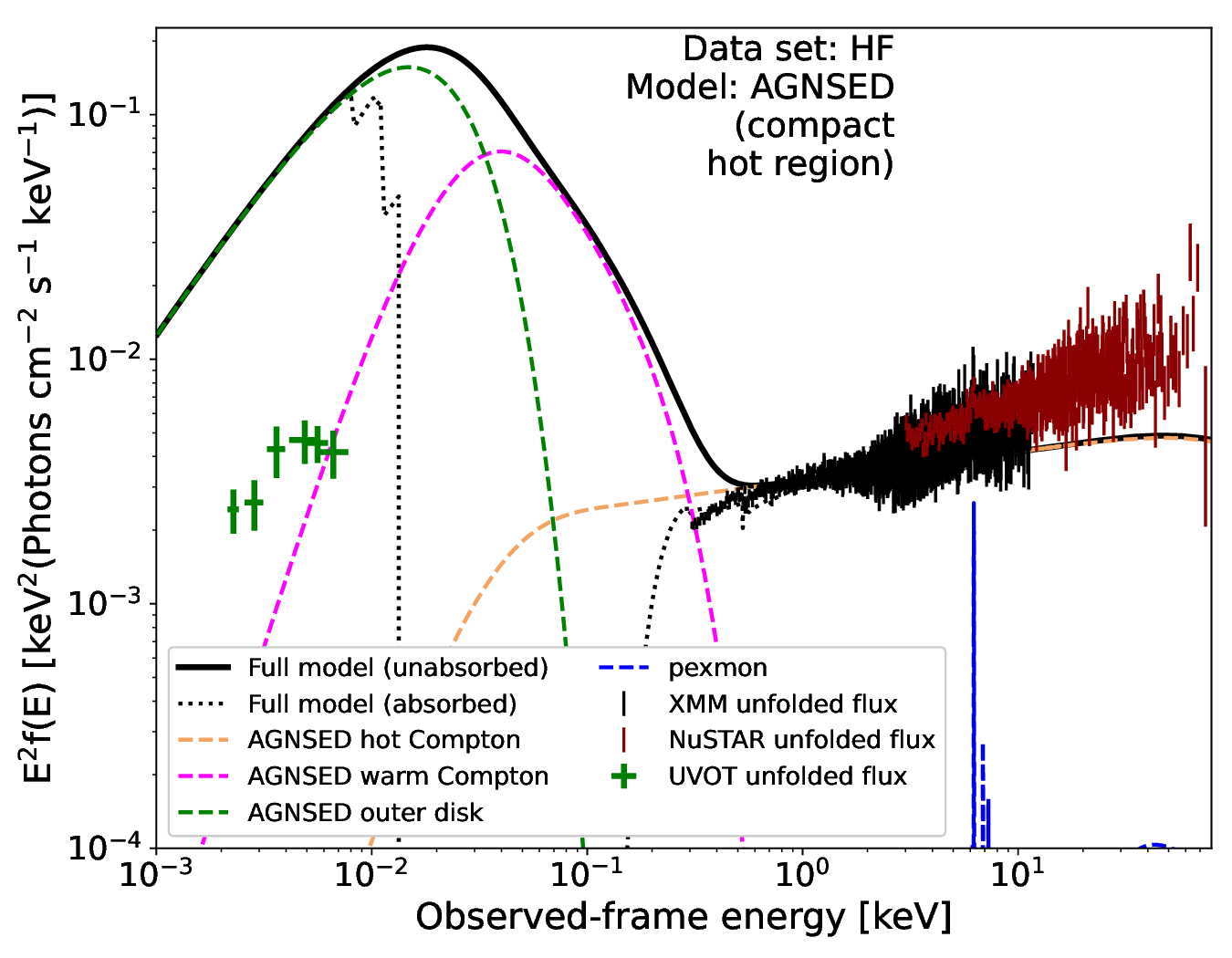}
    \includegraphics[scale=0.395, trim={10 0 10 0}, clip]{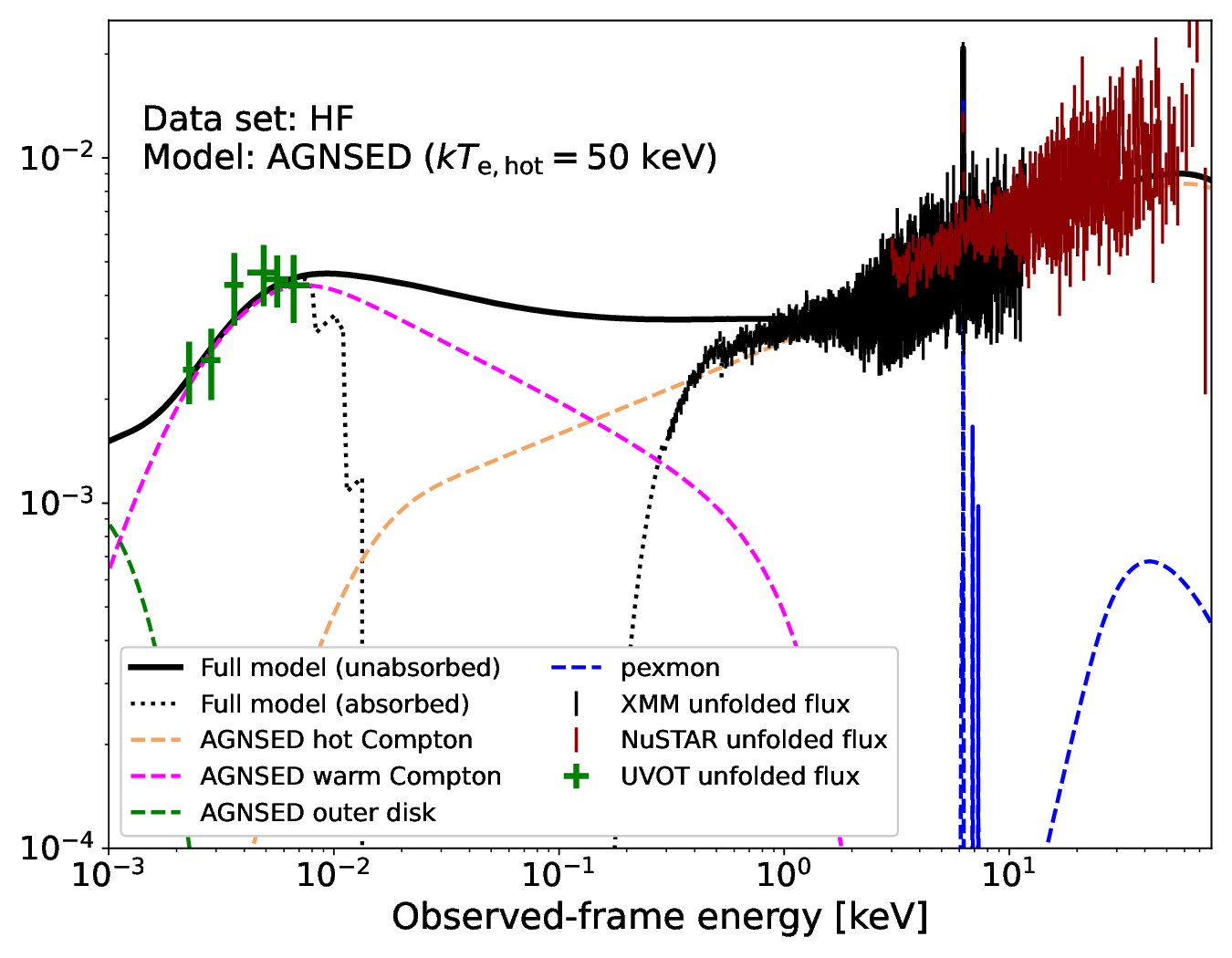}
    \caption{Alternative \textsc{agnsed + pexmon} model configurations for the HF data set. The model variants are defined as in Figure \ref{fig:LF_uvmodel}. For the model variant with $kT_{e,\mathrm{hot}}=50$ keV, the cool outer disk component is minimized in the fit, yielding very similar overall results to the `no cool disk' model presented in the main text (\S \ref{sec:discussion_SED}). We note that the `Full Model' spectra in these Figures (black curves) are normalized relative to the \emph{XMM-Newton} spectrum. Because the  \emph{NuSTAR} combined spectrum in the HF data set has a $\sim20$\% higher overall flux level compared to the \emph{XMM-Newton} spectrum, an offset is visible between \emph{NuSTAR} and \emph{XMM-Newton}. As discussed in \S \ref{sec:analysis_joint}, we account for flux offsets via multiplicative scaling constants in our modeling.  \label{fig:HF_uvmodel}}
\end{figure*}

\begin{figure*}
    \centering
    \includegraphics[scale=0.395, trim={5 0 10 0}, clip]{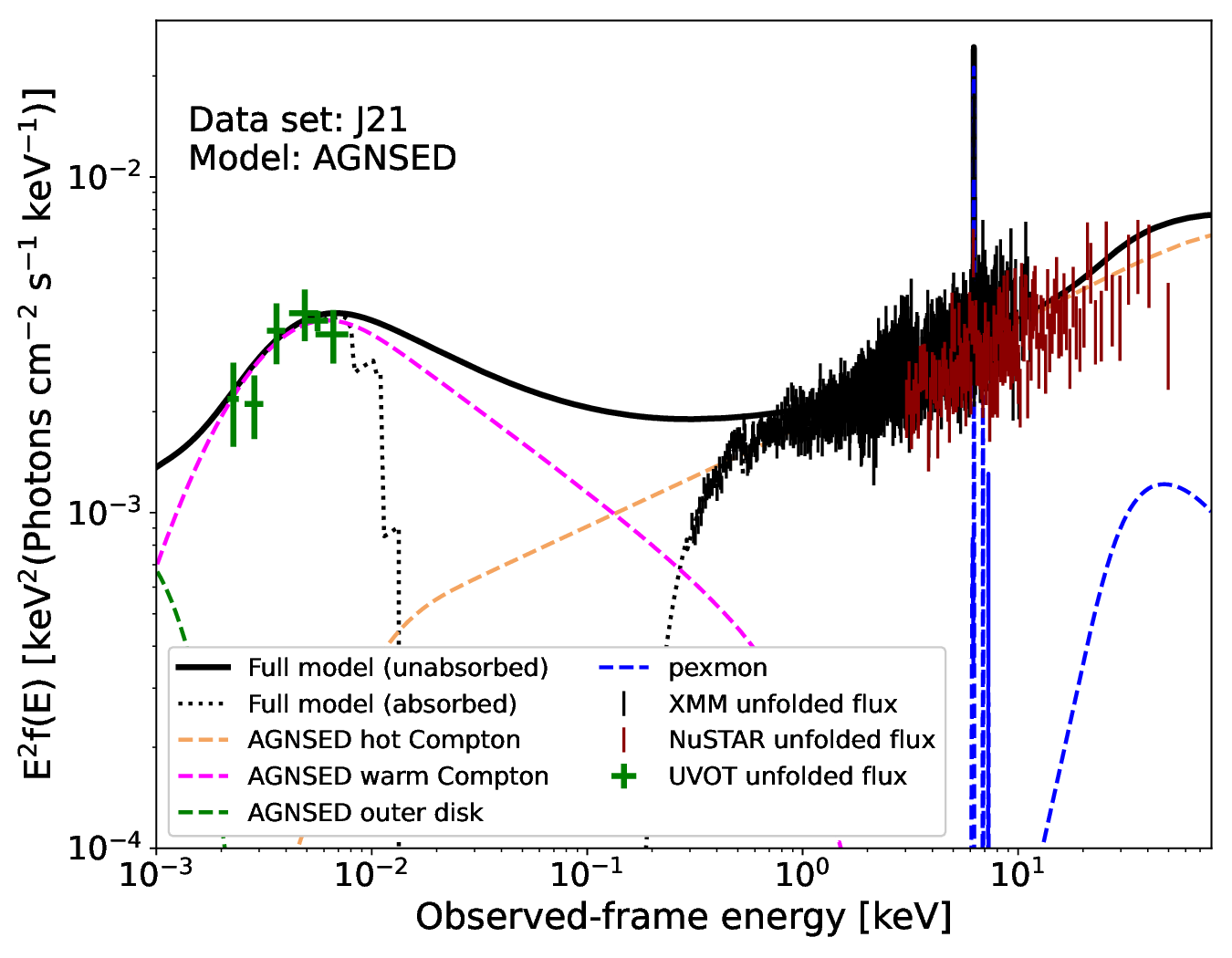}
    \includegraphics[scale=0.395, trim={30 0 10 0}, clip]{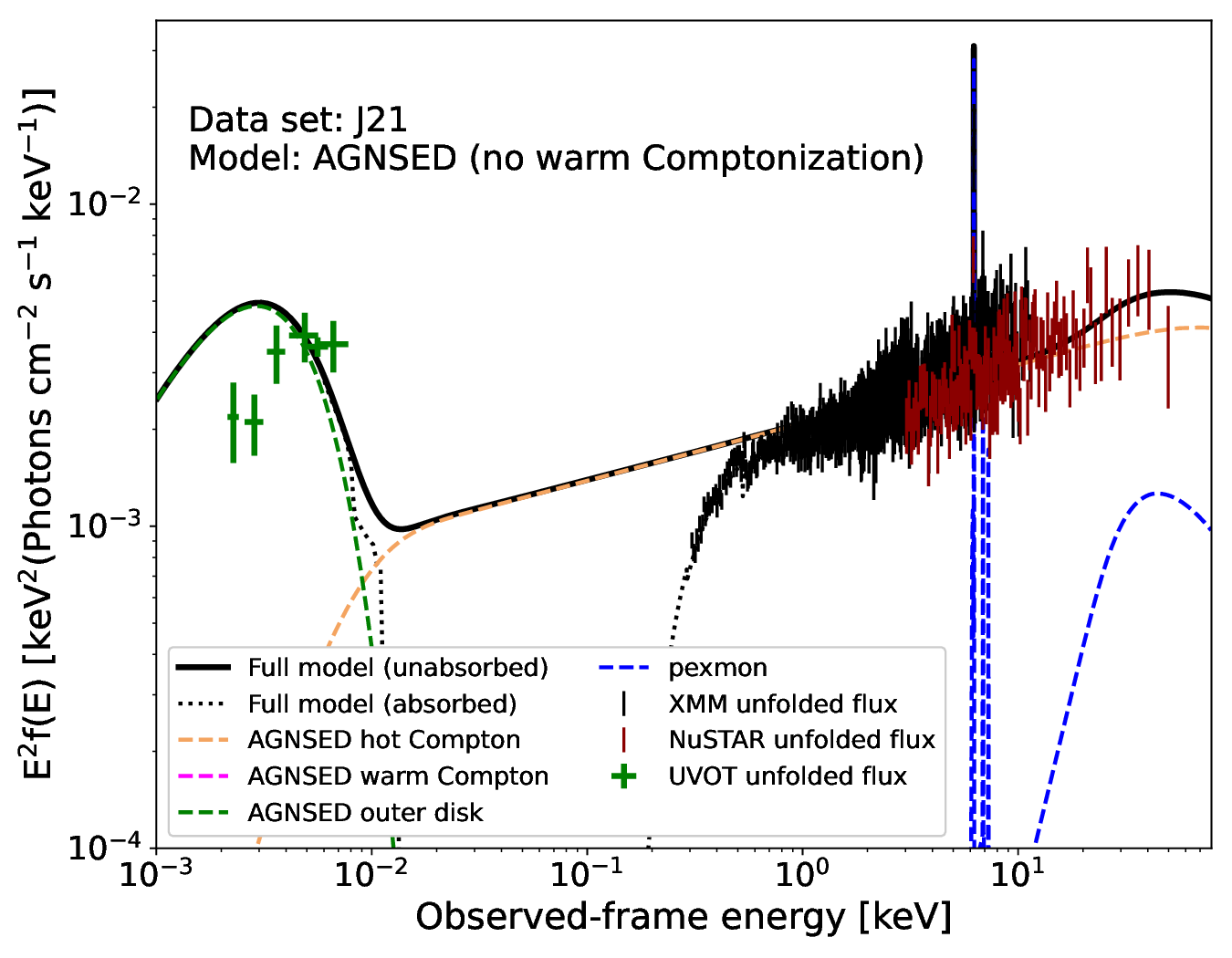}
    \includegraphics[scale=0.395, trim={5 0 10 0}, clip]{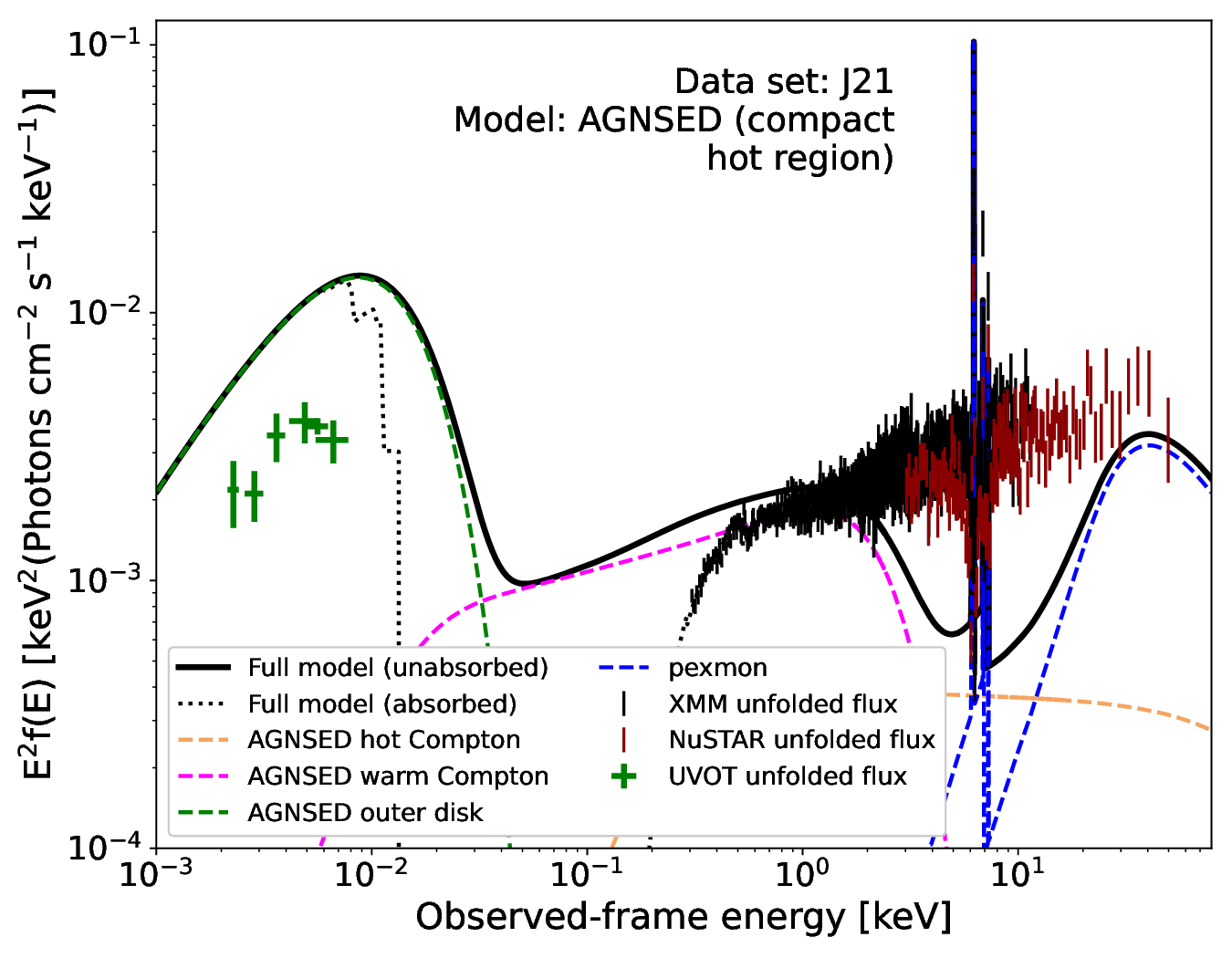}
    \includegraphics[scale=0.395, trim={10 0 10 0}, clip]{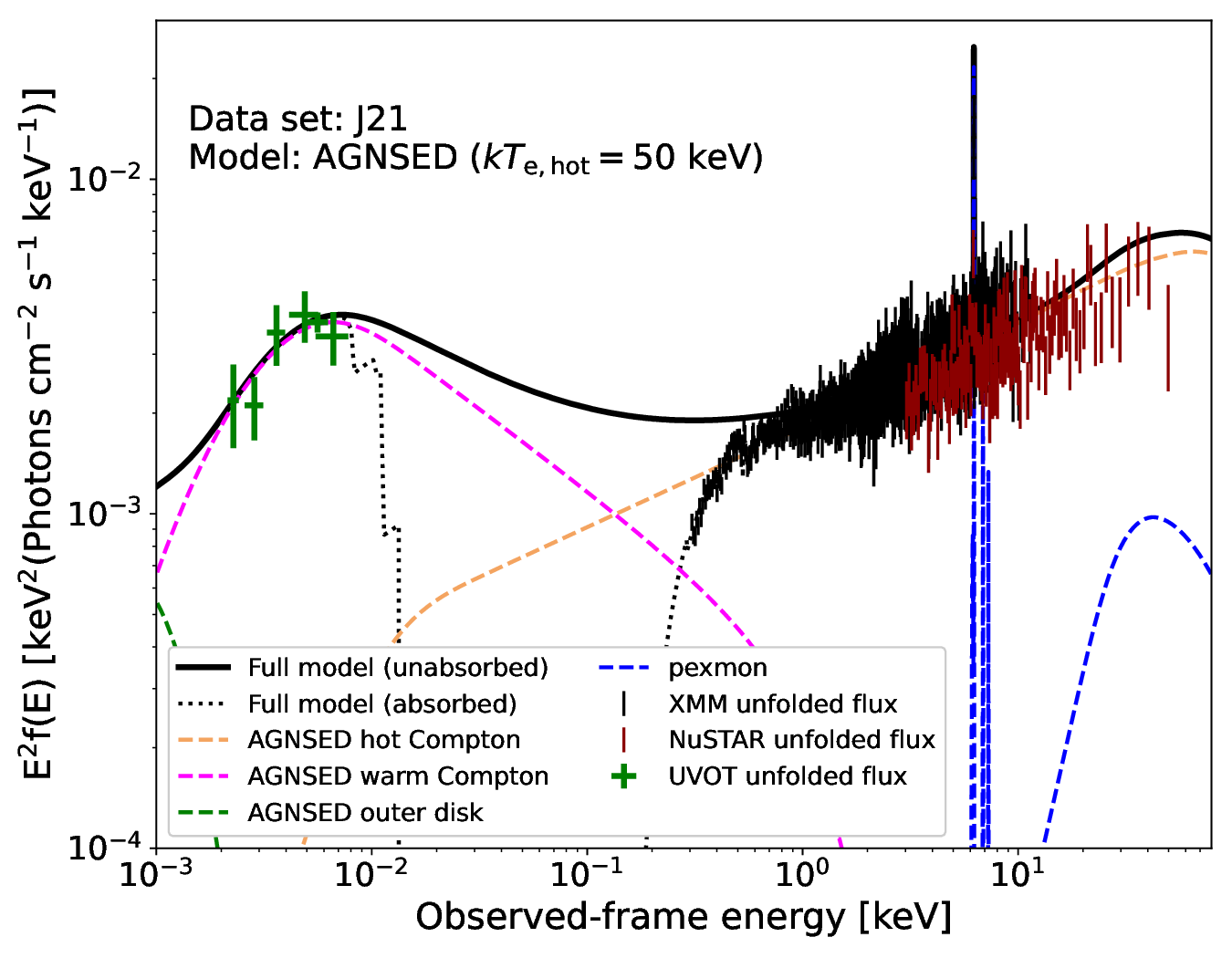}
    \caption{Alternative \textsc{agnsed + pexmon} model configurations for the J21 data set. The model variants are defined as in Figure \ref{fig:LF_uvmodel}.  \label{fig:J21_uvmodel}}
\end{figure*}

\begin{figure*}
    \centering
    \includegraphics[scale=0.395, trim={5 0 10 0}, clip]{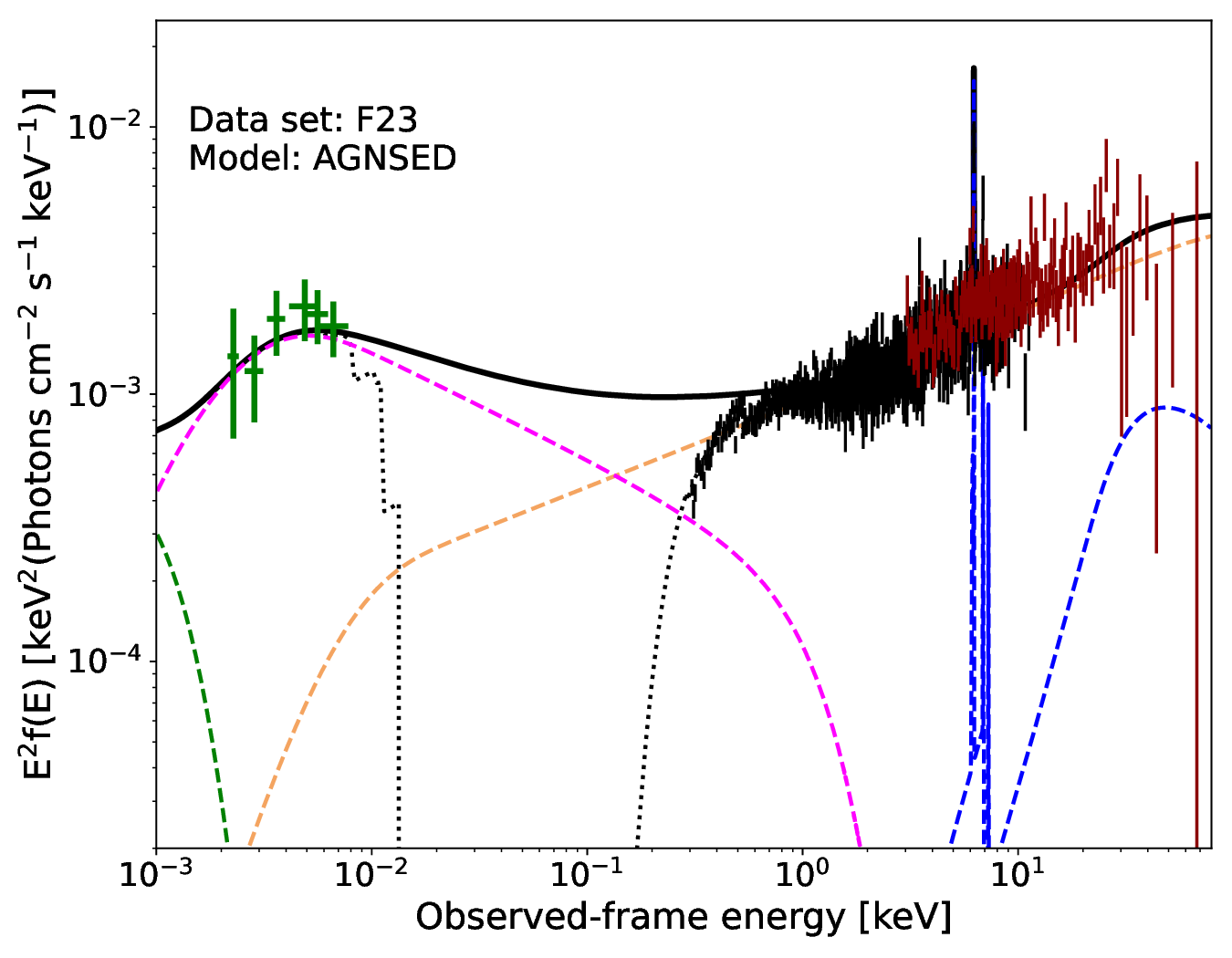}
    \includegraphics[scale=0.395, trim={30 0 10 0}, clip]{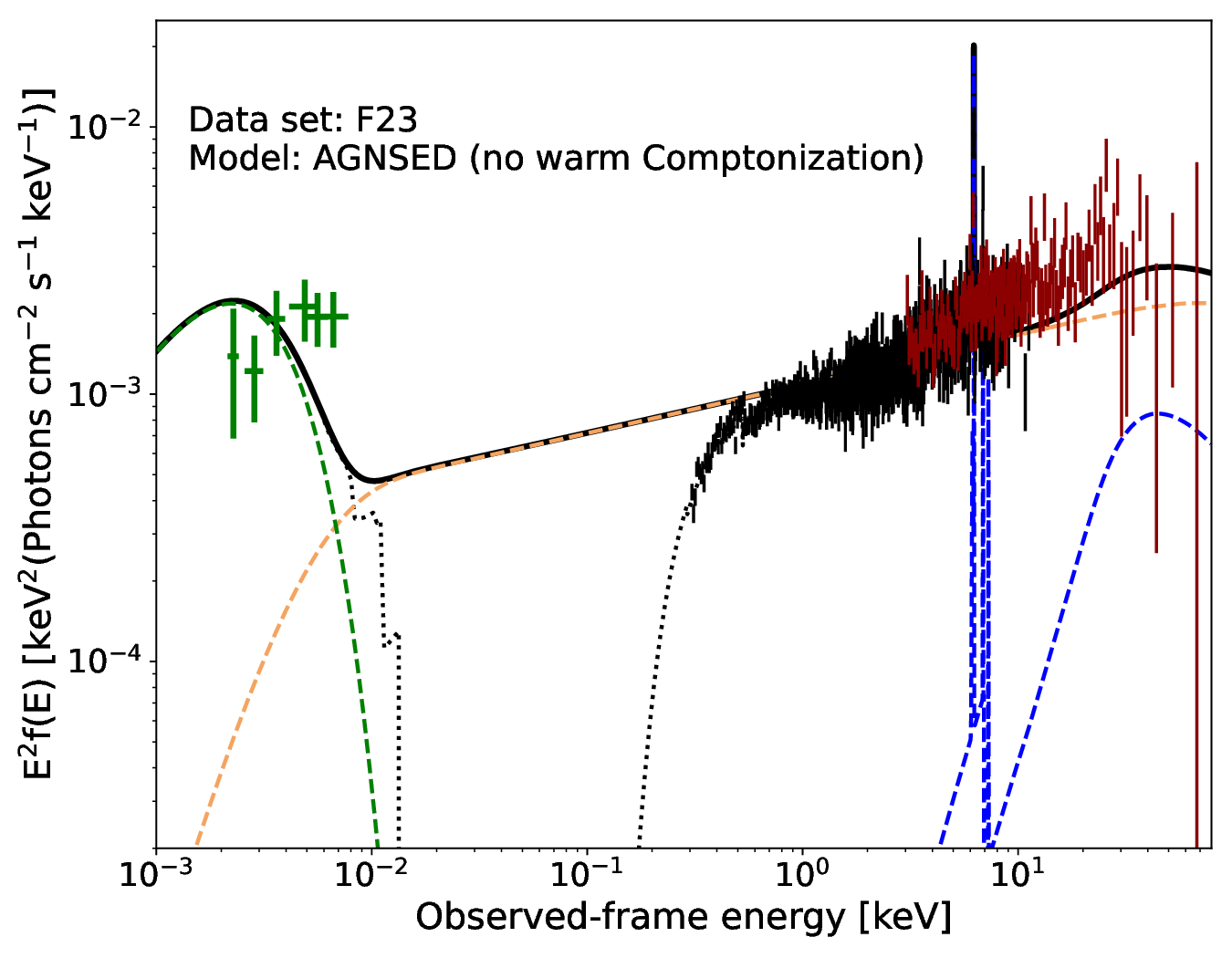}
    \includegraphics[scale=0.395, trim={5 0 10 0}, clip]{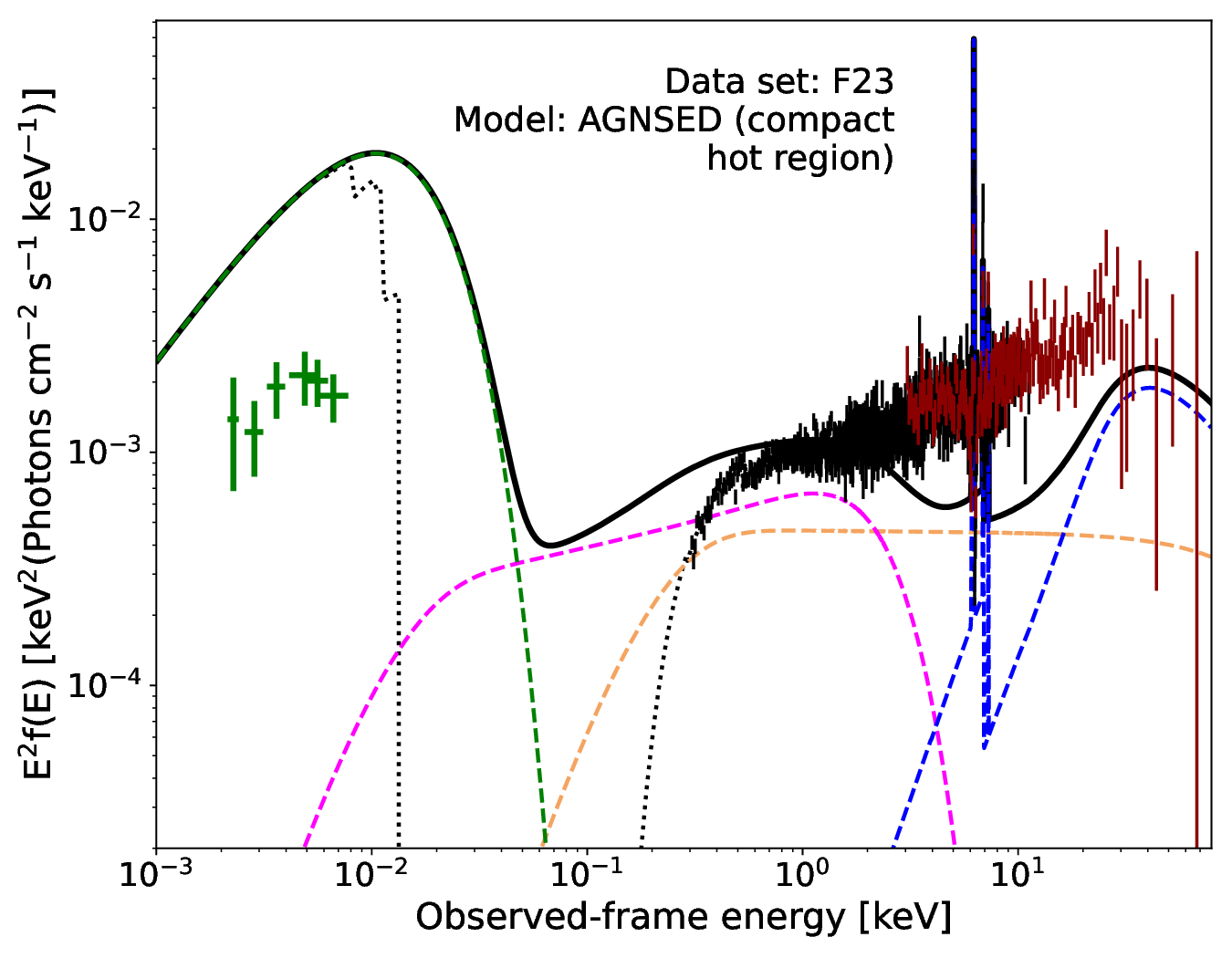}
    \includegraphics[scale=0.395, trim={10 0 10 0}, clip]{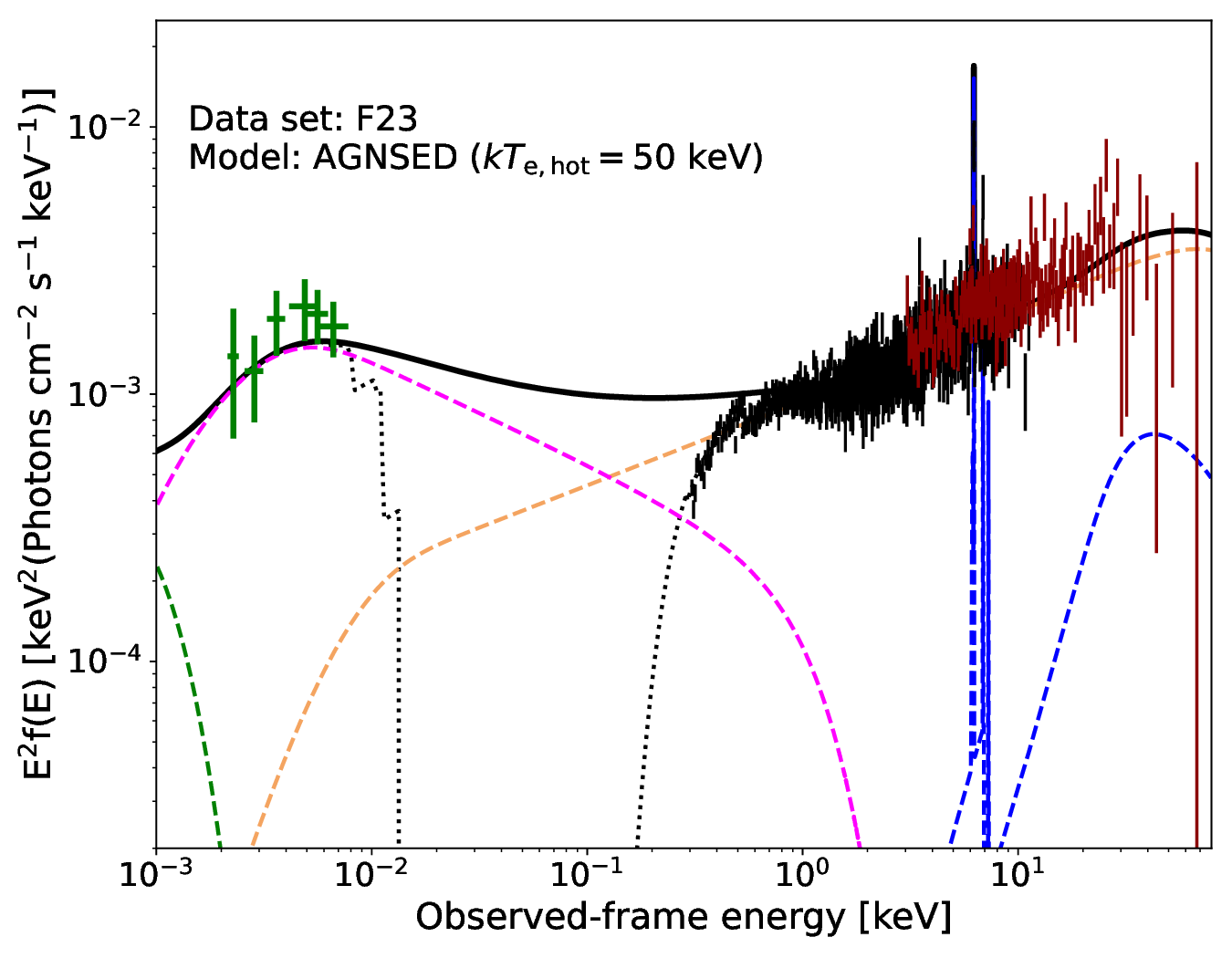}
    \caption{Alternative \textsc{agnsed + pexmon} model configurations for the F23 data set. The model variants are defined as in Figure \ref{fig:LF_uvmodel}.  \label{fig:F23_uvmodel}}
\end{figure*}